**Decarbonization pathways for liquid fuels: A multi-sector energy system perspective**

Jun Wen Law[1], Bryan K. Mignone[2], Dharik S. Mallapragada[3*]

1. MIT Energy Initiative, Massachusetts Institute of Technology, Cambridge, MA 02139
2. ExxonMobil Technology and Engineering Company, Annandale, NJ 08801
3. Chemical and Biomolecular Engineering Department, Tandon School of Engineering, New York University, Brooklyn, NY 11201

*Correspondence: Dharik S. Mallapragada (dharik.mallapragada@nyu.edu)

## Abstract

Low-carbon liquid fuels play a key role in energy system decarbonization scenarios. This study uses a multi-sector capacity expansion model of the contiguous United States to examine fuels production in deeply decarbonized energy systems. Our analysis evaluates how the shares of biofuels, synthetic fuels, and fossil liquid fuels change under varying assumptions about resource constraints (biomass and $CO_2$ sequestration availability), fuel demand distributions, and supply flexibility to produce different fuel products. Across all scenarios examined, biofuels provide a substantial share of liquid fuel supply, while synthetic fuels deploy only when biomass or $CO_2$ sequestration is assumed to be more limited. Fossil liquid fuels remain in all scenarios examined, primarily driven by the extent to which their emissions can be offset with removals. Limiting biomass increases biogenic $CO_2$ capture within biofuel pathways, while limiting sequestration availability increases the share of captured atmospheric (including biogenic) carbon directed toward utilization for synthetic fuel production. While varying assumptions about liquid fuel demand distributions and fuel product supply flexibility alter competition among individual fuel production technologies, broader energy system outcomes are robust to these assumptions. Biomass and $CO_2$ sequestration availability are key drivers of energy system outcomes in deeply decarbonized energy systems.

## 1. Introduction

Liquid and gaseous fuels dominate final energy demand in current energy systems. For example, liquid and gaseous fuels represented 70% of final energy use in the U.S. in 2023 [1]. While electrification of building, industrial, and transport end uses is projected to increase over time, particularly under deep decarbonization, energy system modeling consistently shows a continued role for gaseous fuels such as methane and liquid hydrocarbon fuels in net-zero energy systems [2], [3], [4], [5], [6], [7]. This persistence is largely due to the prevailing technological challenges of electrifying certain end uses such as heavy-duty transport, aviation, marine shipping, and high-temperature industrial processes [8], [9], [10]. Moreover, because fossil-derived liquid fuels are generally more carbon-intensive and costly on a per-energy basis than fossil gaseous fuels (e.g., methane) [3], there are stronger economic incentives under carbon pricing to develop and adopt alternative production routes for liquid fuels than to deploy analogous options for gaseous fuels.

Multiple technology options can reduce emissions associated with the use of conventional liquid fuels. These include: (1) deployment of carbon dioxide ($CO_2$) removal (CDR) technologies, such as direct air carbon capture and sequestration (DACCS) or bioenergy with carbon capture followed by sequestration (BECCS), to offset emissions from continued use of fossil liquids [11], [12], [13], (2) the conversion of biomass to liquid fuels (biofuels), and (3) the production of synthetic fuels using hydrogen ($H_2$) and captured atmospheric $CO_2$, including captured biogenic $CO_2$ [14]. Assessing the relative competitiveness of these liquid fuel pathways is challenging for several reasons. First, liquid fuel production pathways span a range of technological processes (e.g., refining, pyrolysis, and gasification followed by Fischer-Tropsch (FT) or methanol synthesis, among others), feeds (e.g., oil, multiple forms of biomass, hydrogen, and $CO_2$), and energy inputs (e.g., electricity and natural gas). Second, large-scale deployment of alternative fuel pathways would create new cross-sectoral interactions within the energy system – for instance, the increased demand for hydrogen and captured $CO_2$ associated with synthetic fuel production, related demands for electricity, and corresponding infrastructure requirements.

Several prior studies have undertaken techno-economic analysis (TEA) of biofuel and synthetic fuel production routes, with a focus on comparing process cost outcomes under various technology and input assumptions [15], [16], [17], [18], [19], [20]. These studies provide an engineering basis to quantify performance of pathways using different processes or feeds, including carbon and energy flows, feed requirements, and product distributions. These studies also highlight the sensitivity of process cost to feed and energy prices (e.g., electricity, $H_2$, $CO_2$, biomass) and co-product revenues (e.g., captured $CO_2$, light hydrocarbons), with one example being the importance of $H_2$ prices on synthetic fuel production costs [21]. However, TEA studies are unable to assess how resource constraints, such as limited biomass availability, coupled with cross-sectoral interactions, could impact process economics and in turn technology deployment. For example, under deep decarbonization scenarios, biomass used for liquid fuel production may face competition from other sectors such as power generation, driving up its price and consequently the cost of biofuels. Similarly, the price of $H_2$ for synthetic fuel production depends on the mix of $H_2$ production technologies deployed and may be weakly or strongly affected by electricity price dynamics, depending on the extent of electrolyzer deployment [22], [23], [24].

In this context, energy system models that account for competing uses of resource-constrained feedstocks and energy commodities across multiple sectors can complement TEAs in understanding fuel technology pathways under various technology and policy assumptions. Many different analysis frameworks have been used to evaluate the role of liquid fuel production technologies in regional or national decarbonization scenarios, such as system dynamic models (SDM) [25], [26], [27], integrated assessment models (IAMs) [28], [29], [30], [31], [32], [33], and capacity expansion models (CEMs) [34], [35], [36], [37], [38], [39]. System dynamics models (SDMs) capture feedback-driven dynamics and industry development at a sectoral or regional scale, while integrated assessment models (IAMs) span multiple regions and sectors, linking energy, land, and climate systems with the economy to explore long-term global mitigation pathways. In contrast, CEMs focus on cost-optimal investment and operation of infrastructure, typically at high temporal and technological resolution, making them well suited for analyzing sectoral interactions (e.g., linkage between power and other sectors) and the detailed pathways required for alternative fuel production.

Recent energy system studies [2], [3], [25], [28], [34], [38], [40] of liquid fuel decarbonization pathways illustrate how the competitiveness of alternative fuel production technologies is affected by assumptions about biomass and $CO_2$ sequestration availability as well as by the stringency of the emissions constraint. Key insights arising from these studies in the context of deep decarbonization include: (a) the potential for competing uses of biomass across fuels and other sectors (e.g., power) depending on technology, demand, and resource availability assumptions [25], [28], [34], [38], [40], (b) the sensitivity of synthetic fuel deployment to assumptions about $CO_2$ sequestration availability, with greater deployment of such fuels observed in scenarios with limited sequestration availability [34], [38], [40], and (c) the continued role of fossil liquid supply to meet demand, along with deployment of CDR technologies to offset remaining emissions [2], [3], [34], [40].

Despite these advancements, understanding fuel decarbonization pathways and their energy system drivers is limited by several gaps. First, most studies treat liquid fuels as a lumped commodity, even though demand for liquid fuels involves multiple fuel types – primarily gasoline, diesel, and jet fuel, with distinct demand drivers and competing technologies. For instance, although gasoline currently dominates liquid fuel demand in many regions, a combination of technological trends, including growing electrification of light duty vehicles (LDV), could lower the share of gasoline in total liquid fuel demand in the future. This raises a question about how changes in the liquid fuel demand distribution would affect deployment of fuel production technologies.

A second related gap is the lack of adequate consideration of flexibility in fuel product slates from existing fossil liquids and alternative fuel production routes. For instance, petroleum refining assets in the United States and many other regions currently produce gasoline-heavy product slates [41], [42]. Although some operational changes can broaden the range of product slates served by existing refineries [41], substantial changes could require further investments to retrofit these assets. These considerations raise important questions regarding the extent to which new investment could align supply and demand for particular products [2], [3], particularly under deep decarbonization scenarios in which the fuel demand distribution could be different from today.

Similarly, for alternative fuel pathways using biomass and $CO_2$ feedstocks, TEA studies assume specific fuel product slates, but the impact of changing product slates on process-level cost and performance has not been fully characterized. Moreover, existing energy system studies often do not consider different rates of carbon conversion or $CO_2$ capture (CC) during biofuel production [2], [3], [43]. A related challenge for utilizing biofuel TEA data in energy system models is the lack of harmonization of the different feedstock and product composition assumptions used across studies.

Here we address the limitations identified above to enhance understanding of liquid fuel production pathways in cost-optimized, deeply decarbonized energy systems. Our approach utilizes a multi-sector CEM that includes the electricity, $H_2$, $CO_2$, biomass, liquid fuels, and natural gas (NG) supply chains. For liquid fuels, we represent 11 alternative fuel production processes with varying cost and performance assumptions, including different CC rates and fuel product distributions. This implementation involves: (a) harmonizing carbon and energy balances for different fuel technology pathways reported in the literature (details in Section S2.3 of the SI); and (b) developing a simple but general method to estimate the costs and energy requirements associated with varying CC rates when such data are not provided by the literature source (details in Section S2.2 of the SI). This approach enables more consistent cross-technology comparisons and shows how system-level assumptions affect the deployment of different liquid fuel pathways under emissions constrained scenarios. Additionally, we explicitly differentiate fuel product demands and impose constraints on fossil liquid product distributions based on historical refinery data, thereby allowing us to quantify how these constraints affect fossil liquid deployment across a range of possible future energy systems.

We apply the framework to a case study of the contiguous United States under a net-zero $CO_2$ emissions constraint, varying assumptions about biomass and $CO_2$ sequestration availability (which could reflect resource constraints as well as other factors), the flexibility of supply technologies to alter product mixes, and fuel demand distributions. Our results highlight a substantial role for biofuels with CCS in deeply decarbonized energy systems, with the extent of $CO_2$ capture within biofuel pathways primarily driven by biomass resource assumptions. Synthetic fuels deploy in several cases, particularly when $CO_2$ sequestration availability is limited, leading to greater utilization of biogenic $CO_2$, which effectively intensifies biofuel production. Additionally, while the flexibility of fossil liquids production and the liquid fuels product distribution do not strongly affect overall energy system results, these assumptions significantly affect the competition among individual biofuel processes.

The remainder of this paper is organized as follows. Section 2 outlines the modeling framework, input assumptions, and scenario design with details included in the supporting information (SI). Section 3 presents the key results from the core energy system scenarios and discusses the sensitivity to alternative demand assumptions, alternative fuel production flexibility assumptions, and alternative techno-economic assumptions. Section 4 discusses the implications for liquid fuel decarbonization strategies and directions for future research, and Section 5 concludes the paper.

## 2. Methods

### 2.1. Energy system modeling approach

This study builds upon and extends an open-source, multi-sector CEM, MACRO [44], to study liquid fuel decarbonization under emissions constraints, summarized in Figure 1. The model is formulated as a single year cost-optimization that minimizes the sum of system-wide annualized capital and operating costs, spanning the supply chains of electricity, $H_2$, $CO_2$, biomass, liquid fuels, and NG. Supplies and demands of these vectors are balanced at the hourly and regional level unless otherwise indicated.

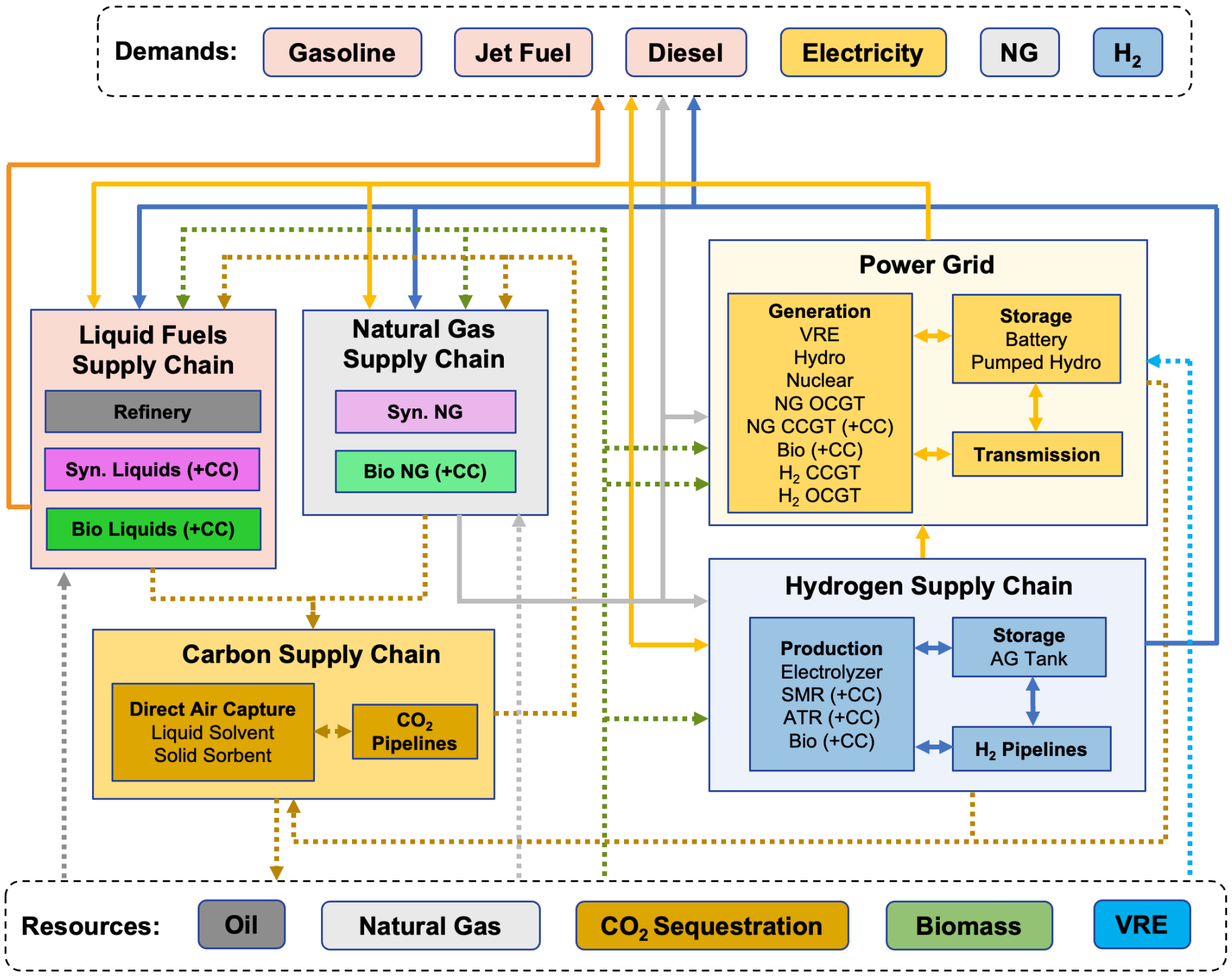

*Figure 1. A) Architecture of the MACRO model showing the various sectors represented (Refer to Section S5 of the SI for model formulation). Key exogeneous inputs to the model are highlighted in the dotted boxes labeled "Resources" and "Demands". CC = $CO_2$ capture NG = Natural gas, VRE = Variable renewable energy (solar and wind), OCGT = Open cycle gas turbines, CCGT = Combined cycle gas turbines, SMR = Steam methane reforming, ATR = Autothermal reforming, AG = Above ground storage, Syn. = Synthetic.*

In this study, we expanded MACRO to include representations of liquid and gaseous fuel supply chains and their integration with the rest of the energy system. This includes enforcing supply-demand balances for individual liquid fuels (gasoline, jet fuel, diesel) and NG, as well as introducing investment and operation decisions and constraints for respective production technologies. The model incorporates multiple fuel (both gas and liquids) production pathways, including: (1) biofuels, (2) synthetic fuels synthesized from captured $CO_2$ and $H_2$, and (3) continued use of fossil fuels (gas and liquids). For biofuel and synthetic fuel pathways, the model resolves the individual fuel outputs and tracks the disposition of unconverted $CO_2$, which may be either captured or vented depending on the process configuration. To reflect refinery product distribution constraints, the model imposes bounds on the relative outputs of fossil-derived jet fuel, gasoline, and diesel (Eq. S18-S19 in Section S5.1) [42]. In addition, the model tracks fuel use in other commodity supply chains, such as NG consumption for electricity generation, $H_2$ production, or direct air capture (DAC).

Resource availability for VRE, $CO_2$ sequestration, and biomass are defined via supply curves for each region as described further below. In addition, electricity, $H_2$, and captured $CO_2$ can be transported between regions via transmission networks and pipelines, subject to capacity limits. Transportation within a region is assumed to occur at zero cost. The model represents existing electricity transmission infrastructure and allows for endogenous investment in additional electricity lines and new pipelines for $H_2$ and $CO_2$ when cost-effective. NG transmission is not modeled explicitly; instead, we assume exogeneous NG prices at the regional level, which implicitly accounts for differences in transmission costs. Liquid fuel transmission costs are implicitly assumed to be small by enforcing the supply-demand balance at the national level rather than the regional level (Eq. S15-S17 in Section S5.1).

The MACRO model formulation also includes several additional operational and policy constraints: (1) Regional constraints requiring all captured $CO_2$ to be utilized, transported, or sequestered (Section

S5.4), (2) a resource adequacy constraint for the electricity sector that enforces a balance between supply and demand of firm capacity (Table S22); (3) ramping and minimum output constraints for technologies with limited flexibility, such as thermal power generators, fossil based $H_2$ production, DAC, and fuel production technologies (Table S23); (4) electricity and $H_2$ storage inventory balance constraints at hourly and weekly scales, described in [45], [46]; and (5) a system-wide $CO_2$ emissions policy limit (Section S5.5). The complete model implementation and documentation are available in a public GitHub repository [44], with detailed formulations for liquid and gaseous fuels modeling in Section S5 of the SI.

## 2.2. Case study description

We used MACRO to investigate future energy systems with net-zero emissions using a nine-zone representation of the contiguous U.S. energy system (Figure S17). All technologies except electricity generation and transmission, fossil liquids production, and NG production are modeled as greenfield developments, with investment decisions endogenously determined. Projected hourly demand for electricity, as well as annual demands for $H_2$, liquid fuels (gasoline, jet fuel, diesel, see Figure 2A) and NG for each region in 2050 were obtained from the low electrification (E-) demand scenario of the Net-Zero America Study [2] (Table S12). Although this scenario is labeled "low electrification" in the underlying literature source, it still represents a net-zero transition path that includes substantial electrification of end uses, leading to higher electricity demand and lower demand for liquid fuels and NG relative to today. Apart from electricity that is represented with hourly variations, the demand profile for each of the other commodities is represented as constant hourly demand that sums to the annual demand for the entire year.

The choice of 2050 as a model year is primarily motivated by the availability of outputs (e.g., final energy demands) from other studies that can be used as inputs for this study. However, the key findings from this paper related to the drivers of fuel process selection in deeply decarbonized systems are not likely to be sensitive to assumptions about when the net-zero year occurs. Put differently, because we solve for a single future state, it is possible to interpret these results as describing a future year in which these final demands and $CO_2$ emissions are attained, which need not be 2050.

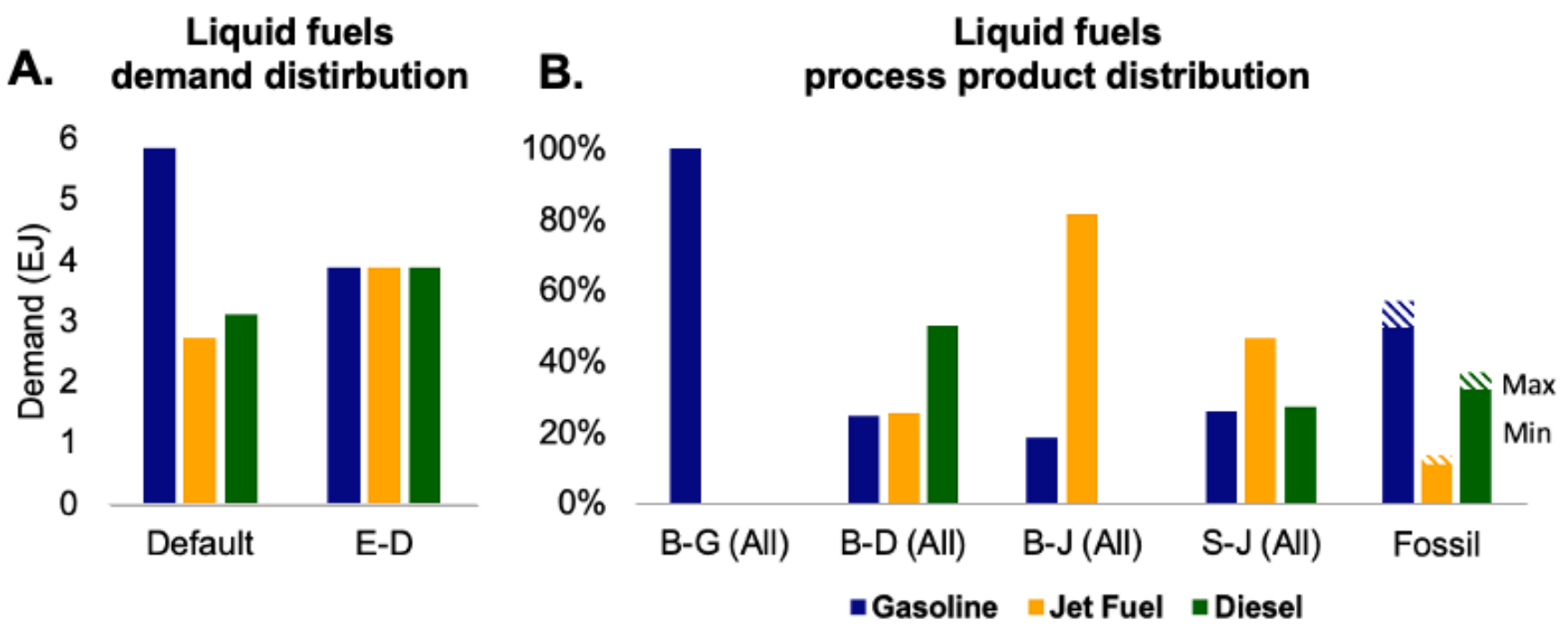

*Figure 2. A) Liquid fuels demand distributions used in this study for the future net-zero year. The default demand distribution is obtained from Princeton University's Net-Zero America study low electrification scenario as shown in Table S12 [2], and the "E-D" (equal demands of gasoline, jet fuel, and diesel) distribution is obtained by allocating the total default liquid fuels demand equally across the individual liquid fuel products. B) Fuel product (gasoline, jet fuel, and diesel) distribution of liquid fuel production pathways modeled in this study – see Table 1 for additional process information. For fossil liquids, refinery product slate constraints are highlighted by showing minimum (solid bars) and maximum (dashed bars) values for gasoline, jet fuel, and diesel. Labels refer to whether process is bio ("B") or synthetic ("S"), as well as the dominant fuel product ("G" – gasoline, "J" – jet fuel, "D" – diesel). "All" indicates that the product distribution is independent of CC level.*

To approximate year-round system operations, the model simulated operations over 26 representative weeks. Of these, 23 were derived using a time-domain reduction method based on k-means clustering applied to a single year of hourly energy demand and resource availability time series data for renewable resources (e.g. wind, solar, hydro) [47]. In addition, three additional "extreme" weeks corresponding to periods of peak electricity demand and minimum solar and wind availability were included to capture the impact of such operating variations on system capacity decisions. The resulting 26 weeks were sequenced chronologically and weighted to the number of actual weeks represented, allowing for a tractable and accurate approximation of full-year operations.

Cost and performance assumptions for technologies in the electricity, $H_2$, and $CO_2$ supply chains are summarized in Section S3.2 of the SI, and mainly follow the assumptions used in a previous study [22]. Annualized $CO_2$ sequestration rates and associated injection costs were sourced from the ReEDS model, with annualized rates estimated as the constant yearly injection rate that would fully utilize the available $CO_2$ storage over a 100-year period (Table S29 and Figure S18) [48]. Biomass supply assumptions were obtained from the 2023 Billion-Ton Report from which regional supply curves were constructed to represent the regional availability and costs of various biomass types – herbaceous, woody, and agricultural residues as shown in Figure S19-S24 and Table S38-S39 [49]. The average regional energy and carbon content of each biomass category was defined based on the 2023 Billion-Ton Report and primarily GREET 2023 data, respectively (Table S40) [50].

The supply of fossil liquid fuels is assumed to be unconstrained and priced using the national 2050 average fuel cost for the transportation sector projected by the U.S. Energy Information Administration's (EIA) Annual Energy Outlook (AEO) 2023 reference scenario [51] (Table S32). In addition, the product yields of fossil gasoline, jet fuel, and diesel from U.S. refineries from 2009 – 2023 were used to develop lower and upper bounds on the fossil liquid supply distribution [42]. The implications of refinery flexibility enabled by reconfigurations or new technology is considered in an additional sensitivity case discussed in Section 3.5.

Cost and performance assumptions for biofuel pathways were obtained from various TEAs in the literature (Table S34), each with unique product distribution mixes, as highlighted in Figure 2B. Modeled biofuels pathways include: (a) biomass gasification followed by methanol synthesis to produce gasoline exclusively (B-G) [20], (b) biomass gasification followed by FT [18] to primarily produce diesel (B-D), along with gasoline and jet fuel as co-products, (c) biomass gasification followed by FT [19] with different CC rates and an alternative product distribution tilted toward jet (B-J) but with some gasoline and electricity production. Including CC variations, a total of nine biofuel processes were represented. Cost and performance assumptions for synthetic liquid fuel production were derived from the TEA study of Zang et al. [21], which considers one process with and without CC, namely $H_2$ and $CO_2$ conversion to liquids using FT technology that produces a mix of gasoline, jet fuel, and diesel (S-J) described further in Table S33. In this study, capital and operating costs (other than feedstock costs) are assumed to be scenario-independent. While capital costs may decline with deployment in real markets, this interaction is challenging to represent explicitly, so as an alternative, we consider a sensitivity to capital costs in Section 3.6.

The supply of fossil NG is also assumed to be unconstrained and available at the average delivered price in EIA's AEO 2023 reference scenario. Synthetic and biomass-derived NG pathways are parametrized using available literature (see Table S36-S37 for details). Cost and performance assumptions for synthetic NG production are based on a TEA study of $CO_2$ methanation process [52], while bio-NG production assumptions are based on a TEA study of a process utilizing gasification to produce syngas followed by fermentation to produce NG [53]. All cost parameters in this study are standardized to 2022 dollars to account for differences in dollar-years between various references.

For biomass and synthetic fuel pathways not originally modeled with $CO_2$ capture, we developed alternative process variants with different capture rates using an approximation method discussed in Section S2.2, thereby ensuring that each fuel pathway includes configurations spanning a range of CC rates (Table 1). This method involves matching the $CO_2$ concentrations of the flue gas stream(s) in each pathway to analogous industrial point sources (e.g., flue gas in cement or ethanol production) and applying the corresponding capture costs and energy requirements reported in published techno-economic assessments (TEAs) of industrial point-source $CO_2$ capture [54], [55]. This approach omits pathway-specific energy integration opportunities and could be viewed as a conservative estimate of energy impacts of CC implementation. In addition, energy and carbon balances of the various pathways are harmonized as described in Section S2.3. This harmonization allows for the application of user-defined carbon content for biomass and fuels in the MACRO model, while preserving the energy conversion and product distribution as reported in the respective process-level TEAs.

*Table 1. Carbon conversion and $CO_2$ capture rates of alternative liquid fuels production pathways in this study: Carbon conversion = carbon content of fuel products divided by the carbon content of the feedstock. Reported biofuel carbon conversion values are based on weighted average value of 47.3% carbon content per biomass weight and 18.15 MMBtu/tonne biomass from the high mature market scenario of the 2023 Billion-Ton Report [49]. Carbon conversion is not a technology input parameter in the MACRO model. Instead, biofuel production technologies are modeled using energy conversion efficiency and product distribution inputs as described in Section S5.1 of the SI (see Eq. S8 – S10). As such, the actual carbon conversion might differ according to the type of biomass consumed as shown in Table S38-S39. Detailed input assumptions of the pathways are shown in Table S33-S34, with carbon balances shown in Figure S25-S28 of the SI. Process labels refer to whether process is bio ("B") or synthetic ("S"), the dominant fuel product ("G" – gasoline, "J" – jet fuel, "D" – diesel) as shown in Figure 2B, and the rate of $CO_2$ capture ("CCXX") if applicable.*

| Process | B-G | B-G-CC31 | B-G-CC99 | B-J-CC75 | B-J-CC84 | B-J-CC99 | B-D | B-D-CC53 | B-D-CC99 | S-J | S-J-CC99 |
|---|---|---|---|---|---|---|---|---|---|---|---|
| Carbon Conversion (%) | 28.3 | 28.3 | 28.3 | 24.3 | 32.0 | 24.3 | 38.2 | 38.2 | 38.2 | 47.2 | 47.2 |
| $CO_2$ Capture Rate (%) | - | 31.0 | 99.3 | 74.8 | 83.9 | 99.0 | - | 52.8 | 99.5 | - | 99.0 |

### 2.3. Scenarios

We evaluated systems with net-zero emissions, varying assumptions about biomass and $CO_2$ sequestration resource availabilities as shown in Figure 3A. Tighter constraints on these resources do not necessarily need to be interpreted as physical resource constraints but could represent preferences expressed through other forms of policy that restrict availability. The low (LB: 9.5 EJ/y) and high (HB: 15.9 EJ/y) biomass availability scenarios are based on the supply curves for low and high mature market, respectively, from the 2023 Billion-Ton Report [49]. The low (LS: 433 $MtCO_2$/year) and high (HS: 866 $MtCO_2$/y) availability scenarios for $CO_2$ sequestration assume that 5% and 10%, respectively, of the $CO_2$ sequestration capacity at each site included in the NREL ReEDS model is available [48]. The 10% limit is based on suggestions in the literature that around 10% of $CO_2$ sequestration sites in the U.S. may be economically viable [56].

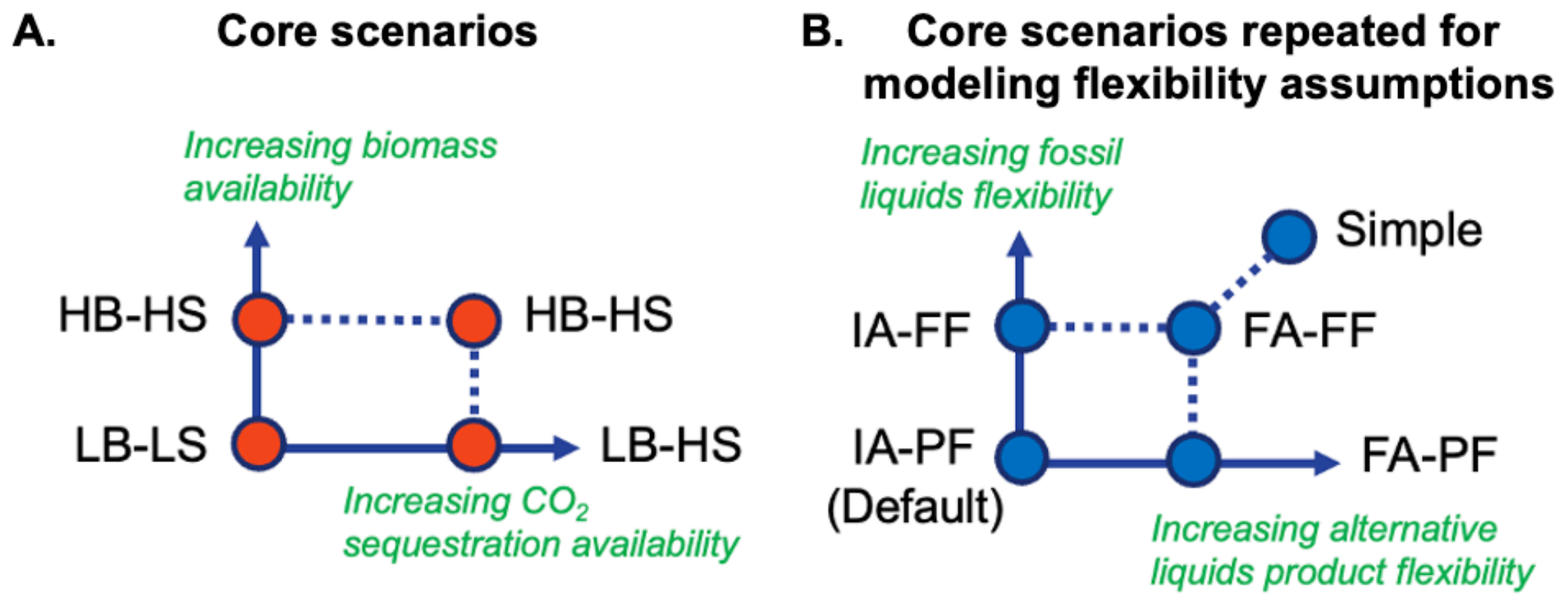

*Figure 3. A) Scenario matrix showing four core scenarios assessed in this study by varying assumptions about biomass and $CO_2$ sequestration resource availability. "HB" = High biomass of 15.9 EJ/y, "LB" = Low biomass of 9.5 EJ/y. "HS" = High $CO_2$ sequestration of 866 Mt $CO_2$/y, "LS" = Low $CO_2$ sequestration of 433 Mt $CO_2$/y. All scenarios assume a net-zero emissions constraint. B) Matrix of liquid fuels modeling flexibility in which the four core scenarios in Figure 3A are rerun with other assumptions about fossil liquid fuels and alternative liquid fuels supply flexibility. "IA" = Inflexible alternative liquid fuels where liquid fuel product slate strictly follows the original product distribution of their respective literature references. "FA" = Fully flexible alternative liquid fuels where the fuel product slate of each technology is allowed to vary while respecting total liquid fuel output (on an energy basis) reported by the literature reference. "PF" =*

*Partially flexible fossil liquid fuels where the ratio of purchased conventional fossil gasoline, jet fuel, and diesel lies within historical bounds of U.S. fossil petroleum refineries [42]. "FF" = Fully flexible fossil liquid fuels where conventional fossil gasoline, jet fuel, and diesel can be produced in any ratios. "Simple" = Simplified modeling assumptions of liquid fuels without any product differentiation of gasoline, jet fuel, and diesel, with aggregated liquid fuels demand and weighted average conventional fossil liquid fuels purchase price and emission factors. The "IA-PF" modeling assumption is used as the default in this study.*

For each of the above scenarios, we also evaluated cases with varying assumptions about fuel technology product flexibility as shown in Figure 3B. "FF" and "PF" refer to scenarios with full flexibility and partial flexibility in fuel product distribution from fossil liquids supply, respectively. Here partial flexibility refers to the product distributions that are constrained by historical refinery product distributions. Similarly, "FA" and "IA" refer to scenarios with full flexibility and no flexibility, respectively in product distribution from each alternative fuel production technology. In effect, under "FA", we assume that the processes can produce the product distribution with greatest system value without any additional costs (see Section S5.2 for formulation), while under "IA", the product distribution of each process is restricted to the distributions reported in the original literature sources (seen in Figure 2B). We also evaluate an additional case, termed "simple model", without any fuel product differentiation in demand or supply, to quantify the impacts of simplifying liquid fuels modeling assumptions, given the prevalence of this approach in coarser-resolution models. The "IA-PF" scenario is used as the base case representation of fuels supply product flexibility. We also undertake additional sensitivity analysis to consider alternative assumptions about liquid fuel demand distributions and capital costs of alternative fuel production technologies.

## 3. Results

### 3.1. Energy system impacts of biomass and $CO_2$ storage availability

Figure 4 summarizes the energy system impact of varying biomass and $CO_2$ sequestration availability in the base case ("IA-PF" in Figure 3B), which assumes a constrained fuel product distribution from each fuel production pathway. In all scenarios, biomass is fully utilized to produce liquid fuels, indicating that biomass used for fuels outcompetes the use of biomass in other sectors as explained further below. Biofuels account for 59% of total liquid fuels in high biomass (HB) scenarios and 34-35% in low biomass (LB) scenarios. As the product slates of both alternative and fossil fuel pathways are constrained, multiple biofuel pathways are deployed, providing the fuel specific products needed to exactly match the specified fuel demand distribution.

Biofuel deployment also enables the use of fossil liquid fuels when paired with CCS, as the sequestered biogenic $CO_2$ provides removals that offset fossil emissions (Figure 4C and 4D). Accordingly, for the same biomass resource availability, the HS scenario enables more fossil fuel use compared to LS scenario (41% vs. 16% of total demand in HB, and 49% vs. 30% in LB scenarios). Conversely, when sequestration is more constrained, synthetic fuel production increases (from 0% to 25% of fuel demand in HB, and from 16% to 38% in LB scenarios). Constraining biomass resource availability limits the amount of biofuel that can be produced, with the remaining fuel demand supplied by fossil fuels, if the assumed $CO_2$ sequestration availability allows for it, and synthetic fuel production. These results indicate a tendency for liquids production in highly decarbonized systems to rely on both fossil fuels (when removals can offset remaining emissions) and biofuels, with synthetic fuels deploying only when biomass availability is constrained (which limits biofuel production) or when $CO_2$ sequestration availability is constrained (which limits fossil liquid production by limiting removal potential). It is worth noting that synthetic fuels can be produced from biogenic $CO_2$ (e.g., $CO_2$ captured from biofuel production), which could be interpreted as biofuel produced from an intensified process in which more of the carbon from primary biomass is converted to fuel.

The optimal carbon capture (CC) rate across the biofuel pathways is primarily affected by the biomass resource assumption, as seen by the reliance on processes with higher CC rates in the LB scenarios than in the HB scenarios. Higher CC rates are consistent with higher marginal $CO_2$ abatement costs in LB ($549-664/tonne) compared to HB ($119-299/tonne) scenarios (see Table S1). These carbon prices are also consistent with the observation that DAC only deploys in the LB scenarios (see Figure 4D) in which biomass cannot provide the amount of atmospheric carbon required for CDR or utilization.

Furthermore, the availability of $CO_2$ sequestration impacts the choice of DAC technologies – only the lower cost natural gas-driven solvent-based DAC pathway gets deployed under the HS scenario, while under the LS scenario with higher $CO_2$ abatement costs, we also see deployment of the more expensive sorbent-based, electricity-powered DAC pathway, which requires less $CO_2$ sequestration per tonne $CO_2$ removed.

In the NG production sector (Figure 4E), there is no deployment of biomass-based NG, with fossil NG dominating supply and synthetic NG production sensitive to $CO_2$ sequestration availability (0% in HS vs. 11-45% in LS scenarios). Greater deployment of biofuels in the liquid fuels sector relative to the NG sector is driven in part by differences in abatement costs between the biomass technologies in the two sectors. Compared to fossil liquids, fossil NG is less expensive ($4.89-8.85/MMBtu vs. $22.18-28.53/MMBtu) and has a lower $CO_2$ emission factor (5.4 vs. 7.2-7.3 kg $CO_2$/MMBtu) (Tables S30, S31, S35). Hence, biomass use in the liquid fuels sector delivers greater emissions reduction for a given increase in cost relative to the incumbent fossil technology.

Across the scenarios, electrolytic $H_2$ dominates production. Prior work suggests that the choice between electrolytic and natural gas-based hydrogen production depends in part on infrastructure and flexibility assumptions [22]. In addition, constraints on $CO_2$ storage increase the cost of options that require carbon capture and sequestration. Regardless of the type of hydrogen produced, the magnitude of $H_2$ production is substantially affected by deployment of synthetic fuels. For instance, $H_2$ production in the LB-LS scenario, with the most synthetic fuel production, is about five times exogeneous $H_2$ demand. The strong power-$H_2$ sector coupling due to electrolyzer deployment in these cases also causes electricity generation to increase with increasing synthetic fuel production. Within the power sector, VRE accounts for 86-95% of total electricity generation, with the balance provided by a combination of other resources, including nuclear, hydro and unabated natural gas capacity dispatched infrequently (2-6% utilization). Overall, the energy system results in Figure 4 illustrate how biofuel, fossil fuel, and synthetic fuel shares are affected by biomass and $CO_2$ sequestration assumptions, as well as how the overall energy system responds.

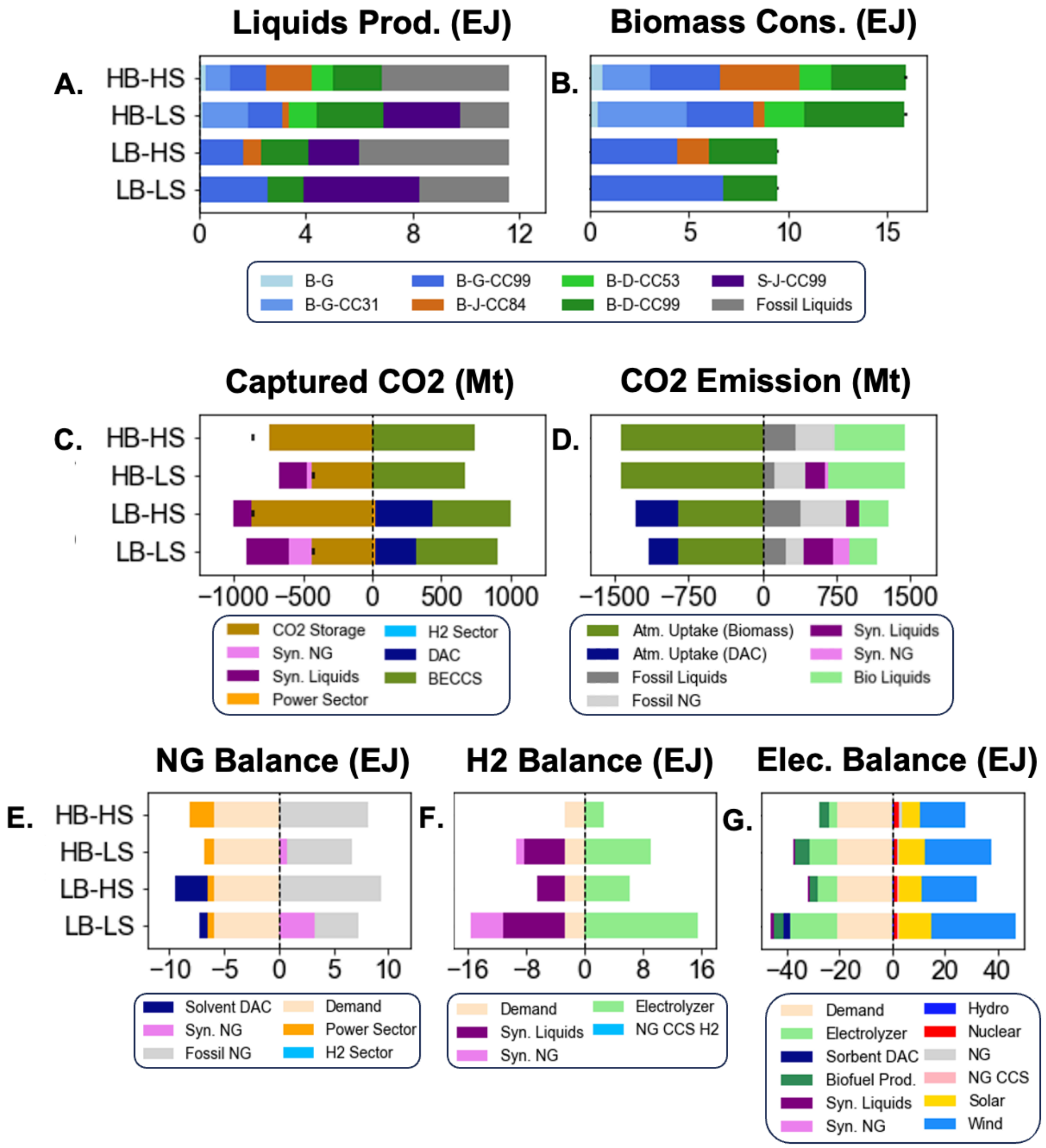

*Figure 4. Energy system impact of varying biomass and $CO_2$ sequestration availability in the net-zero year. These cases use default "IA-PF" modeling assumptions described in Figure 3B and default demand distribution described in Figure 2A; A) Total liquid fuels production by technologies, B) Biomass consumption by technologies, C) Captured $CO_2$ balance, D) $CO_2$ emission balance, E) Natural gas balance, F) $H_2$ balance, G) Electricity balance. Liquid fuels production technology labels follow Table 1. "DAC" = direct air capture, "BECCS" = bioenergy with $CO_2$ capture and sequestration, "NG" = natural gas, "CCS" = $CO_2$ capture and sequestration. The resource limits for biomass and $CO_2$ sequestration are indicated by the bars in Figure 4B and Figure 4C respectively. In each panel for marketable commodities, positive values indicate production, and negative values indicate consumption. "Demand" indicates exogenously assumed final demand. For $CO_2$ emissions, positive values indicate combustion emissions, while negative values indicate atmospheric uptake. "Atm. uptake (biomass)" and "Atm. uptake (DAC)" refers to the carbon content of the biomass resource and $CO_2$ removal by DAC processes, respectively. The contributions from different liquid production technologies to each liquid fuel product demand are shown in Figure S1.*

### 3.2. Disposition of atmospheric carbon

Two metrics summarize the main differences in fuels technology deployment across the various scenarios in Figure 4: (a) the disposition of carbon emissions from biofuel production – either emitted or captured (Figure 5A) and (b) the disposition of captured atmospheric carbon, which can be either converted to fuel, i.e., utilized, or sequestered (Figure 5B).

Across scenarios, biomass availability emerges as the primary factor affecting the share of atmospheric carbon that is re-emitted, which is effectively determined by endogenously choosing the $CO_2$ capture rates for the different biofuel technologies. Under constrained biomass availability in the LB scenarios, system-level biofuel CC rates are 97-99% (Figure 5A), whereas in the HB scenarios, the system level biofuel CC rates are lower at 70-76%. For a given amount of remaining fossil liquids, lower biomass resource availability implies that more of the biogenic carbon must be captured and sequestered to offset fossil emissions. This incentive to increase biogenic $CO_2$ capture further increases in scenarios with lower biomass availability because these scenarios tend to also have more remaining fossil liquids than scenarios with higher biomass availability, all else equal. The tendency toward higher carbon capture rates when biomass is limited is consistent with the higher observed carbon prices in these runs, which would support the additional cost of higher capture rates (Table S1).

Whereas biomass availability drives the biofuel process CC rates, the availability of $CO_2$ sequestration resources largely determines the disposition of captured atmospheric $CO_2$ between sequestration or utilization. It is worth noting that atmospheric carbon can come from combusting or processing biomass or from DAC, with the latter only deployed under LB scenarios, contributing 26-33% of total atmospheric $CO_2$ uptake. In LS scenarios, 35-53% of captured $CO_2$ is utilized for fuel production, whereas in the HS scenarios, 0-14% of captured $CO_2$ is used for fuel production (Figure 5B). This increased coupling between biofuel and synthetic fuel production in the LS scenarios effectively creates a combined bio-synthetic fuel pathway with atmospheric carbon (including biogenic $CO_2$) conversion to fuel of 49-63% under the LS scenario as compared to 32-33% in the HS scenario (Table S2). When biogenic $CO_2$ from biofuel production is captured and used to produce more synthetic fuel, this combination could alternatively be viewed as a single intensified biofuel production pathway.

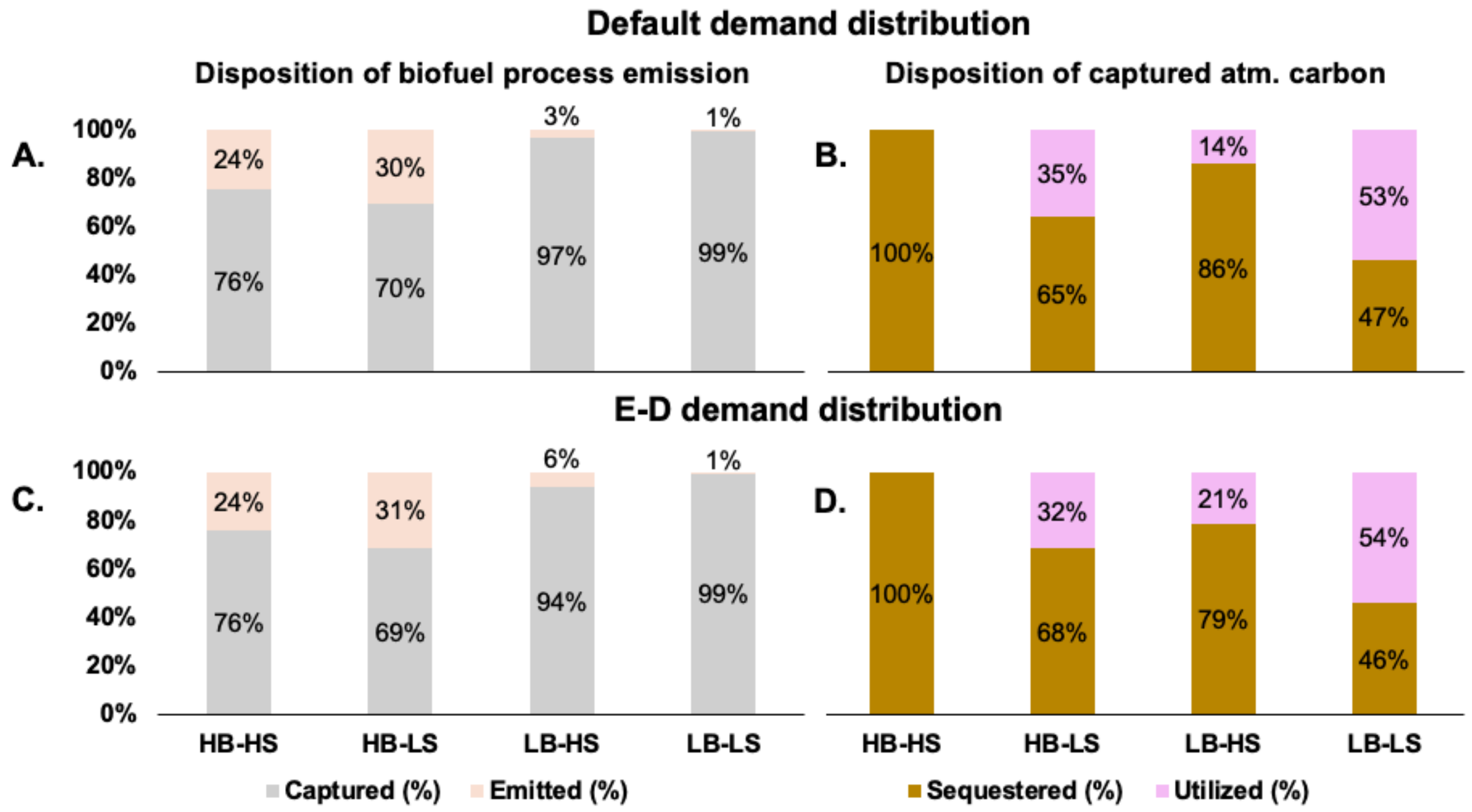

*Figure 5. A,C) Disposition of biofuel production carbon emissions (excludes the carbon content in biofuel products), B, D) Disposition of captured atmospheric carbon from biofuel processes and direct air capture across the core scenarios described in Figure 3A for default "IA-PF" modeling assumptions as described in Figure 3B, under default (panel A,B) and "E-D" (panel C,D) demand distributions as described in Figure 2A. Emitted = carbon vented (i.e., not captured) in biofuel production processes, sequestered = captured atmospheric carbon that is sequestered in geological storage, utilized = captured atmospheric carbon that is utilized for synthetic fuels production (both liquid fuels and NG).*

### 3.3. Economic drivers of fuel pathway selection

To assess differences in the deployment of individual liquid fuel technologies across scenarios, we evaluate their levelized production costs ($/GJ) and compare them with the system-level average prices of fuels in each scenario (Figure 6). The levelized production cost includes: (a) scenario-dependent costs, such as the cost of energy and other inputs for each technology, which are estimated based on their scenario-dependent system-average shadow prices (i.e., dual variables of commodity supply-demand balances averaged over regions and time) (see Eq. S2 – S4 in Section S2.1), and (b) scenario-independent costs such as capital costs (CAPEX) as well as fixed and variable operation and maintenance costs (FOM and VOM) that are also used to parameterize technologies in the CEM.

Figure 6 highlights how the absolute levelized production cost of alternative fuel technologies is impacted more by scenario-dependent costs, reflecting resource availability assumptions, rather than scenario-independent, technology specific factors such as capital cost and non-energy, non-feed operating costs, which are generally the focus of most TEA studies. For example, the cost of biofuel technologies producing gasoline via MeOH with 99% CC varies by more than 2x ($36.5/GJ vs. $76.3/GJ), across the scenarios, while its scenario-independent costs only amount to $24.1/GJ. In the case of synthetic fuels, although $H_2$ is a major cost component, as noted by various TEA studies, we find that its contribution is comparable to the cost of captured $CO_2$ in two of three scenarios with synthetic fuels deployment (LB-LS, LB-HS). We also note that correlations between commodity shadow prices, which are generally not considered in TEA studies, play an important role in setting the levelized production cost of individual technologies. For instance, even though biomass prices are 3-4x greater in LB than in HB scenarios (Figure S4), these increased costs are partly offset by increased revenue from $CO_2$ capture due to the higher carbon prices (i.e., $CO_2$ abatement costs – see Table S1) associated with these scenarios.

Besides showing absolute levelized costs, Figure 6 also highlights relative differences in costs among different technologies. Notably, when carbon prices are relatively low, such as in the HB-HS scenario, differences among system-average levelized costs of biofuel pathways with different $CO_2$ capture rates are relatively small. In this case, regional differences in costs play a more prominent role in driving technology adoption. For example, in regions with limited $CO_2$ storage capacity such as in the Northeast and Mid-Atlantic, the system adopts lower CC pathways to avoid investment in $CO_2$ pipelines to transport captured $CO_2$ to neighboring regions. On the other hand, where $CO_2$ storage is available locally such as in the North-Central U.S., higher CC pathways are adopted.

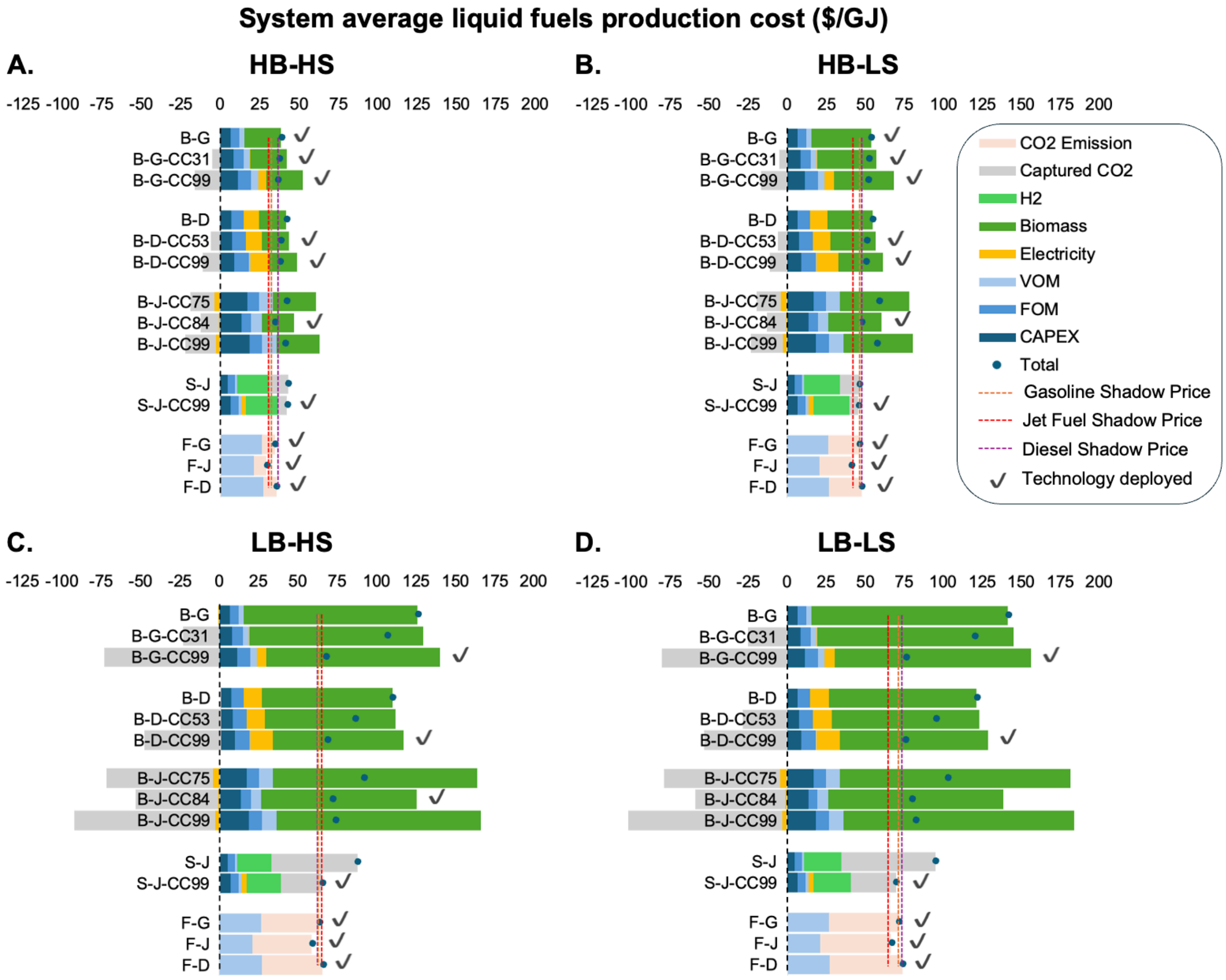

*Figure 6. System-level average production costs of liquid fuel pathways across core scenarios described in Figure 3A for default "IA-PF" (Inflexible alternative liquid fuels with partially flexible fossil liquid fuels) modeling assumptions described in Figure 3B. The stacked bars show the average cost per unit of fuel ($/GJ) for each technology in Table 1, with product distribution shown in Figure 2B. Scenario-dependent costs for biomass, electricity, hydrogen, and captured $CO_2$ are calculated using system-averaged time-weighted values according to Eq. S2 – S4 in Section 2.1 of the SI. These values are based on the shadow prices for each commodity supply-demand balance for a given scenario. Additional costs, such as capital costs (CAPEX), fixed operating and maintenance (FOM) costs, and variable operating and maintenance (VOM) costs, are model input assumptions. The vertical dashed lines represent the system-average shadow prices for gasoline, jet fuel, and diesel. Since deployment decisions occur at the regional level, and the figure uses system-averaged prices to report scenario-dependent costs, the estimated cost of each deployed technology may not exactly equal fuel revenue.*

### 3.4. Sensitivity to liquid fuel demand distribution

Given uncertainty in demands for different fuel products, we examine the robustness of the system outcomes to alternative assumptions regarding the liquid fuels demand distribution. Changing the fuel demand distribution could impact the fuel production mix, particularly under the scenario in which the fuel product distribution of each technology is constrained, as is true in our base case ("IA-PF"). We tested this hypothesis by considering an alternative equal fuel demand distribution ("E-D") case in which demand for gasoline, jet, and diesel is equal rather than dominated by gasoline as in the base case (Figure 2A). This alternative demand scenario is motivated by economy-wide trends related to growing aviation and freight demand (increasing jet and diesel fuel demands, respectively) and electrification of light-duty vehicles (declining gasoline demand).

Figure 7A-D presents the resulting liquid fuel production mix, along with broader system outcomes including electricity, $H_2$, and NG production. We find that key system outcomes identified previously such as the deployment of high CC biofuel pathways in LB scenarios, and the increased reliance on synthetic fuels under constrained $CO_2$ sequestration resources in LS scenarios remain largely unchanged. Likewise, the two summary metrics identified earlier – disposition of biofuel process emissions and disposition of captured atmospheric carbon – follow the same patterns observed in the base case fuel demand scenario (Figure 5C, D). This consistency underscores the dominance of resource constraints (which may reflect physical resource constraints or policy preferences) rather than fuel demand distributions in determining the optimal system deployment of biofuels, synthetic fuels, and fossil fuels.

While fuel demand distributions have a modest impact on the relative deployment of biofuels, fossil fuels, and synthetic fuels, they have a greater impact on the competition among biofuel production technologies. In the default demand scenario, gasoline-dominated (B-G) biofuel pathways contribute a significant share of liquid fuels production, with gasoline accounting for 50-74% of biofuels (Figure S1). On the other hand, the share of gasoline in biofuels decreases to 22-32% in the E-D scenarios (Figure S2), as the system shifts toward diesel and jet-dominated (B-D and B-J) biofuel processes. This reallocation reflects the flexibility of the biofuel portfolio to adjust output mixes to align with changes in demand distributions. In contrast, the flexibility of fossil liquid supply is restricted by historical refinery yield constraints that favor a gasoline-heavy product distribution (see Figure 2B). The historical refinery product mix aligns well with the base case fuel demand distribution, especially under LB scenarios with increased reliance on fossil and synthetic fuels. However, when demands are assumed to shift, as in the E-D scenarios, fossil refinery yields no longer match the demand distribution, and the share of fossil liquid supply decreases by 4-18% across the core scenarios.

When biomass resource availability is high, the model deploys a mix of biofuel processes that produces an optimal distribution of gasoline, jet, and diesel such that the remaining liquid fuels demand can be fulfilled with the least-cost mix of fossil and synthetic fuels. Therefore, the proportion of total bio, synthetic, and fossil liquid fuels under the base and E-D demand distributions are similar in the HB scenarios. However, under constrained biomass availability, the ability to optimize product distribution through biofuel process selection is more limited. Combined with the limited flexibility of fossil liquid fuels, more synthetic liquid fuels fulfill the demand distributions that are less gasoline-heavy (Figure 7I and M).

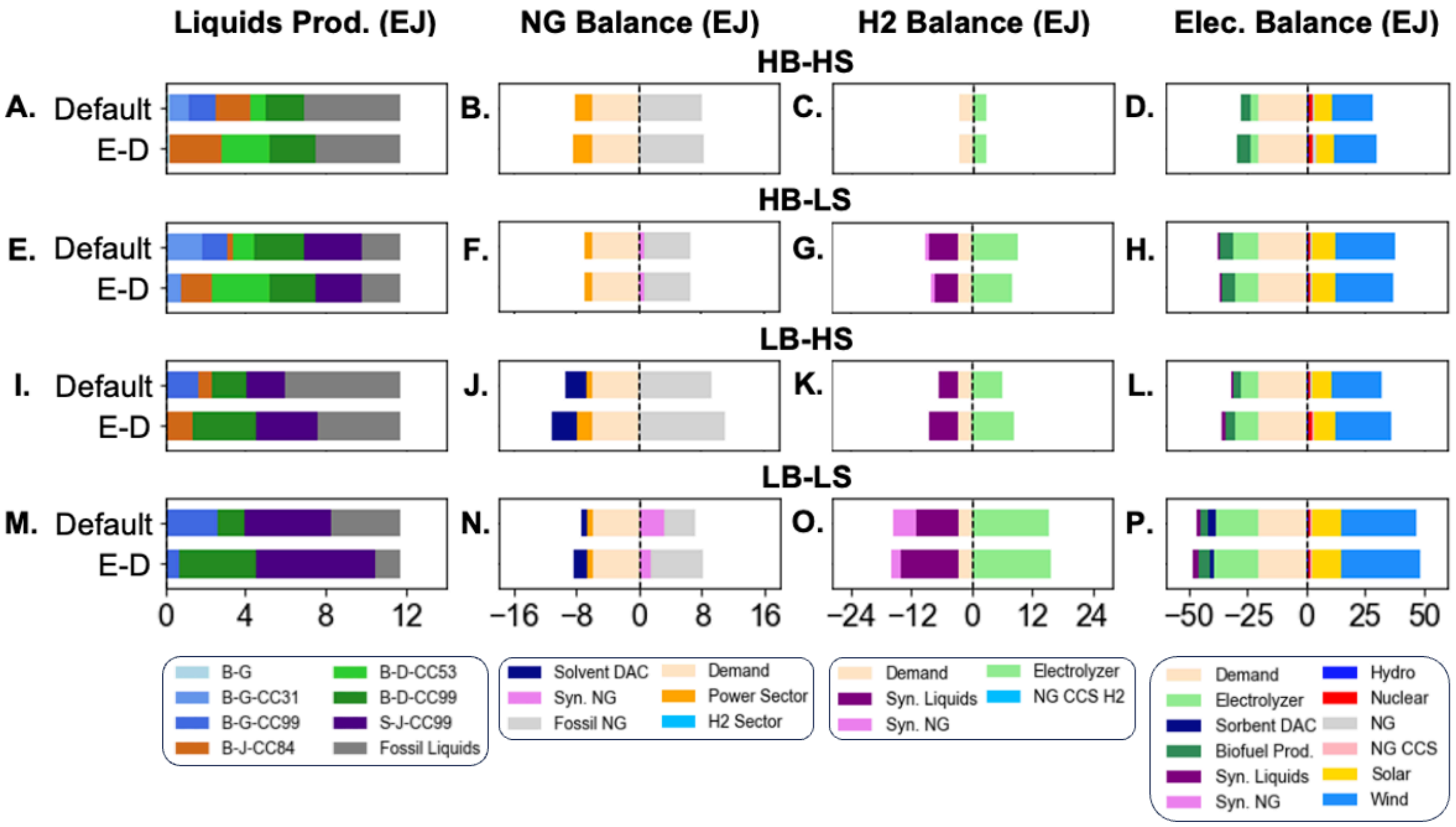

*Figure 7. Impacts of fuel demand distribution on energy system outcomes: Liquid fuels production by pathway (Panels A,E,I,M), NG balance (Panels B,F,J,N), $H_2$ balance (Panels C,G,K,O), and electricity balance (Panels D,H,L,P) across core scenarios described in Figure 3A for "IA-PF" modeling assumptions described in Figure 3B. The scenarios are evaluated under the default and "E-D" demand distributions described in Figure 2A. Liquid fuels production technology labels follow Table 1. The contributions from different liquid production technologies to each liquid fuel product demand are shown in Figure S2.*

### 3.5. Sensitivity to liquid fuels supply flexibility

To assess the robustness of energy system outcomes to modeling assumptions about liquid fuel technology product supply flexibility, we simulated the core matrix under four alternative sets of assumptions about production flexibility, as described in Section 2.3 and shown in Figure 3B. These cases span a range of possible ways to represent liquid fuels in energy system models and enable us to assess how product distribution constraints affect process selection and system-level decarbonization outcomes. In particular, the FF cases allow refineries to vary the fuel production mix without restriction, and the FA cases allow alternative liquid production technologies to do the same thing.

Figure 8 illustrates the resulting liquid fuel production mix and key system-wide energy balances across these flexibility assumptions for the core scenarios of biomass and $CO_2$ sequestration assumptions (Figure 3A). Overall, we find that the total mix of bio, synthetic, and fossil liquid fuels as well as broader energy system outcomes are similar across the flexibility cases. In other words, key findings driven by biomass availability and $CO_2$ sequestration limits are mostly unaffected by modeling choices regarding flexibility of liquid fuels processes. Nonetheless, the extent of flexibility can affect process-level deployment decisions. Under fully flexible alternative fuels configurations, process selection is guided primarily by raw production costs as the processes can freely adjust product outputs to match the required demand distribution. In these cases, biofuel technologies are deployed for their cost competitiveness and not limited by their native product distributions.

However, when alternative fuels production flexibility is restricted in IA-FF and IA-PF cases, liquid fuels deployment reflects both process economics and compatibility with the assumed demand profile. For example, in IA-FF, biofuels are allocated mainly to gasoline to satisfy the high gasoline demand, while fossil liquid fuels are mainly in the form of jet fuel due to lower assumed costs. In contrast, the base case (IA-PF) requires all technologies to adhere to their predefined product yields, resulting in a diversified mix of bio, synthetic, and fossil pathways to balance demand across liquid fuel types. Across flexibility variants, we find that outcomes related to the disposition of carbon (see Figure S7) such as increased biogenic $CO_2$ venting in HB cases and increased utilization of captured $CO_2$ in LS scenarios are preserved. Similarly, repeating the alternative modeling flexibility cases with the E-D liquid fuels demand distribution produced comparable outcomes in system-wide energy system results and carbon disposition trends (see Figure S10 – S12). This consistency suggests that key system-level results are driven by biomass and $CO_2$ sequestration constraints rather than assumptions about liquid fuels production flexibility.

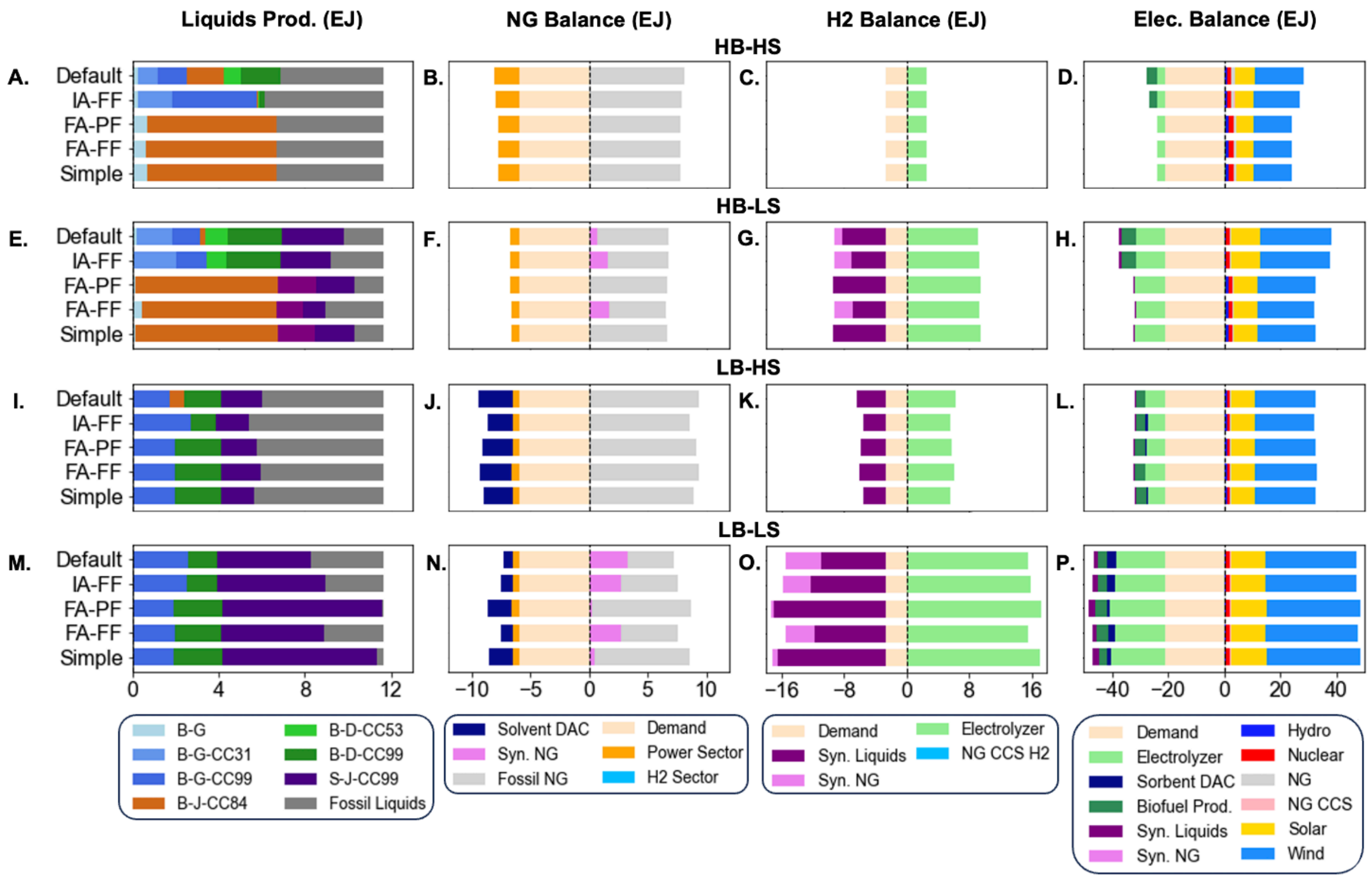

*Figure 8. Effect of liquid fuel flexibility modeling assumptions on energy system outcomes: Liquid fuels production by pathway (Panels A,E,I,M), NG balance (Panels B,F,J,N), $H_2$ balance (Panels C,G,K,O), and electricity balance (Panels D,H,L,P) across the core scenarios described in Figure 3A , under the default demand distribution described in Figure 2A. Each row compares the five flexibility modeling assumptions described in Figure 3B. Default: "IA-PF": inflexible alternative and partially flexible fossil fuels, which is the default assumption used in this study, "IA-FF": inflexible alternative fuels with fully flexible fossil fuels, "FA-PF": fully flexible alternative fuels with partially flexible fossil fuels, "FA-FF": fully flexible alternative and fossil fuels, Simple: aggregated liquid fuels with no product differentiation. Liquid fuels production technology labels follow Table 1.*

### 3.6. Sensitivity to alternative liquid fuel capital cost assumptions

To evaluate the robustness of system outcomes to economic assumptions for alternative liquid fuel production pathways, we simulated two alternative capital costs scenarios in addition to the base case: (1) 2x-CAPEX: Doubled capital costs for alternative liquid fuel technologies, and (2) 3x-CAPEX: Tripled capital costs for alternative liquid fuel technologies. Many literature estimates are developed for Nth-of-a-kind projects, with initial costs often substantially higher. Therefore, higher capex assumptions could reflect a world in which costs stay higher for longer despite deployment, whereas lower capex assumptions could reflect a world in which technologies reach the assumed costs of mature plants by the net-zero year examined here.

System-level findings remain largely consistent across scenarios for various capital cost assumptions (see Figure 9), with deployment of processes with higher $CO_2$ capture rates in LB scenarios, and increased synthetic fuel production in LS scenarios. However, there are some shifts in the disposition of biomass and utilized $CO_2$ as capital costs are varied. For example, synthetic fuel production shifts from liquid fuels towards NG (Figure 9E, F) under higher liquids capex assumptions, reflecting the lower abatement costs associated with displacing NG when the cost of alternative liquids increases. Nonetheless, the share of captured carbon utilized for synthetic pathways (both liquid fuels and NG) remains similar to the default case (see Figure S14). Similarly, as the capital cost of liquid fuel production increases, biomass use shifts from liquid fuels to hydrogen production in the HB-HS scenario. Since BECCS-$H_2$ requires a larger biomass input per unit of energy produced compared to biofuels, we do not observe its deployment in scenarios with constrained biomass and $CO_2$ availability, in which biomass prices are higher than in the HB-HS scenario. Overall, the carbon disposition trends (see Figure S14) remain similar across scenarios, indicating that assumptions regarding biomass and $CO_2$ sequestration continue to drive system decisions.

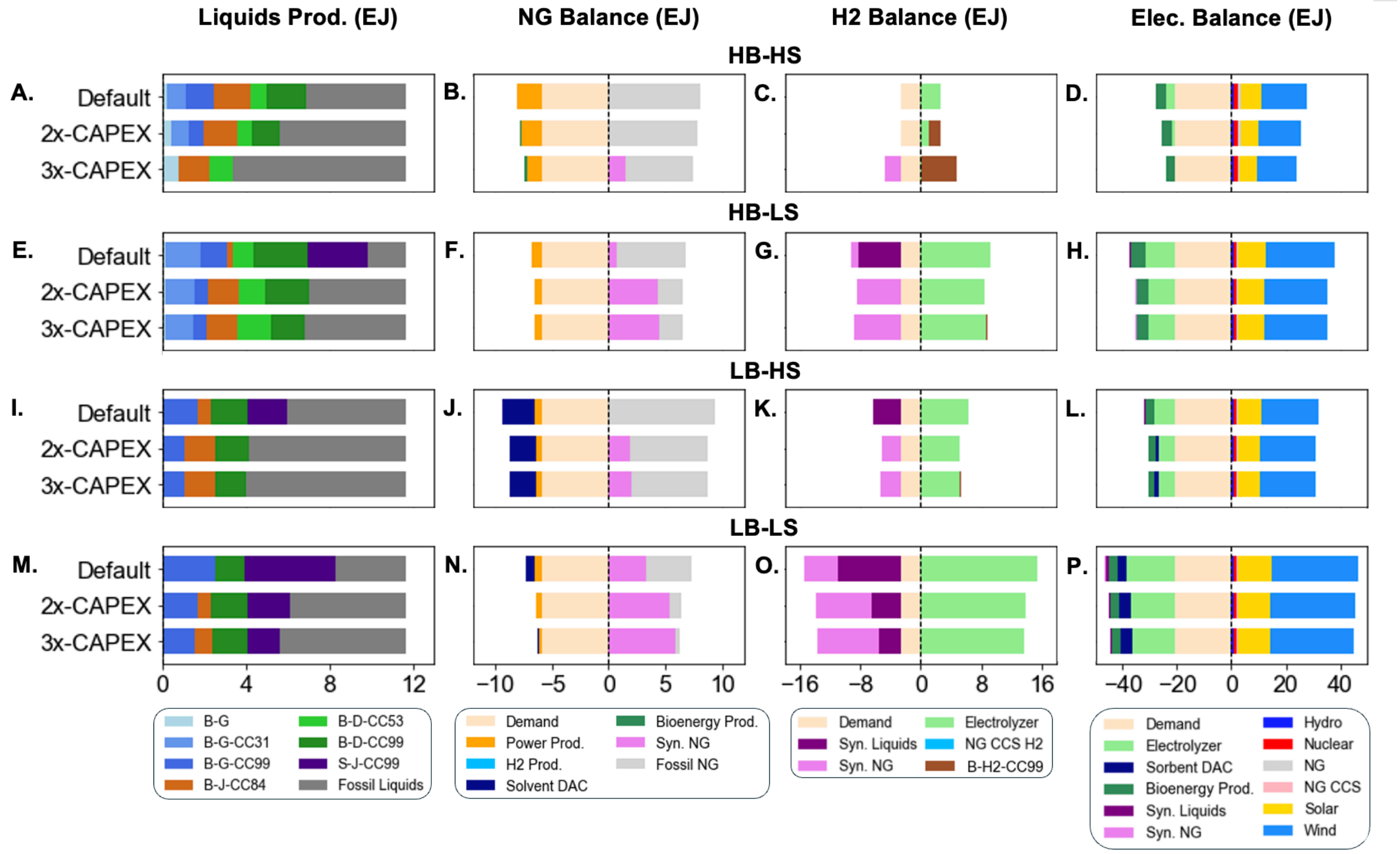

*Figure 9. Energy system outcomes: Liquid fuels production by pathway (Panels A,E,I,M), NG balance (Panels B,F,J,N), $H_2$ balance (Panels C,G,K,O), and electricity balance (Panels D,H,L,P) across various capital cost assumptions across core scenarios as described in Figure 3A for "IA-PF" modeling assumptions as described in Figure 3B under default demand distribution as described in Figure 2A. Liquid fuels production technology labels follow Table 1. "2x-CAPEX" = Doubled capital costs for bio and synthetic liquid fuels production technologies, "3x-CAPEX" = Tripled capital costs for bio and synthetic liquid fuels production technologies.*

# 4. Discussion

## 4.1 Insights relevant to energy system decarbonization

Our analysis highlights that resource constraints, namely biomass availability and $CO_2$ sequestration, are key drivers of the competition between biofuel, fossil liquid fuel, and synthetic liquid fuel pathways in deeply decarbonized energy systems. While biomass availability generally determines the optimal level of $CO_2$ capture for biofuel production, the availability of $CO_2$ sequestration primarily affects the choice between fossil and alternative liquid fuels. We also find that biomass tends to be allocated to liquid fuel production, as its use in other sectors (e.g., hydrogen) emerges only when biofuel process capital costs are substantially higher than default assumptions, and even then, only under the most resource-abundant scenario (HB-HS). This result suggests that the abatement cost of BECCS in liquid fuels is lower than the abatement cost of BECCS in other sectors under the assumptions used in this study. In addition, technologies that capture the most $CO_2$ per unit energy, such as BECCS in power or hydrogen production, become less competitive when $CO_2$ sequestration resources are limited, which reinforces the preference for allocating biomass to liquid fuels in our results. These system-level insights are mostly insensitive to changes in assumptions regarding fossil and alternative fuel supply flexibility as well as to the assumed fuel demand distribution.

Across the evaluated net-zero energy systems, low-carbon fuels play a major role in meeting liquid fuel demand, but their contribution to gaseous fuel supply remains limited, except when the cost of producing alternative liquids is substantially increased and resources are more constrained. This result reflects the lower abatement cost of alternative liquid fuel pathways relative to gaseous fuel pathways, driven by the lower carbon intensity and cost of fossil natural gas compared to fossil liquids assumed in this study.

The modeling results also reveal large variations in system-wide hydrogen consumption, with the most demand – nearly eight times the demand in the case with the least hydrogen consumption – occurring in scenarios with the greatest synthetic fuel production. As hydrogen use increases under limited $CO_2$ sequestration, this additional demand is satisfied using electrolytic hydrogen production, which increases overall electricity demand and low-carbon power generation. In this way, constraints on biomass availability and $CO_2$ sequestration are managed through an expansion of low-carbon electricity supply, largely VRE generation, which itself may be limited by policy or societal constraints not considered in this analysis [40], [57].

All net-zero scenarios exhibit substantial increases in biomass primary energy, $CO_2$ sequestration, electricity generation, and in some cases, hydrogen production. $CO_2$ sequestration and hydrogen increase the most relative to today, although hydrogen production varies considerably by scenario. On the other hand, among the different types of energy that expand, electricity increases the most in absolute terms.

Finally, the range of energy system outcomes resulting from varying constraints highlights tradeoffs that will need to be examined and managed as part of an energy transition. For example, scenarios differ in terms of system costs (scenarios with more stringent resource constraints are more costly) but also in terms of the amount and type of land use (e.g., scenarios that use less land for bioenergy may require more land for electricity infrastructure). In addition, scenarios with limited bioenergy tend to use more $CO_2$ sequestration, illustrating that constraining one resource may put more pressure on another [40].

### 4.2 Comparison with prior system-level studies on fuel decarbonization

A systematic comparison of our model results to those from other energy system studies on liquid fuel decarbonization is challenging, given differences in model structure (e.g., temporal resolution of grid operations, sectoral scope and interactions), technology parameterization, and demand assumptions. Nevertheless, documenting similarities and differences remains valuable, as it highlights the sensitivity, or insensitivity of certain outcomes to modeling and data assumptions and underscores the need for further analysis to resolve discrepancies across models and scenarios. In this context, our finding that synthetic fuel deployment is driven primarily by constraints on $CO_2$ sequestration (and less so on biomass availability) agrees with the findings of several other studies focused on fuel decarbonization pathways that consider sector-coupled energy systems frameworks [34], [40], [58], [59].

The continued role of fossil liquids and their sensitivity to available $CO_2$ sequestration is consistent with findings from other studies that explicitly examine competition among biofuels, synthetic fuels, and fossil fuels [3], [34], [40], [60] . Fewer studies, however, examine how product distribution constraints affect the competitiveness of fossil liquids. We find that these constraints only modestly affect the relative competition between fossil fuels, biofuels, and synthetic fuels (Figure 8), but they play a larger role in altering the competition among alternative fuel production technologies.

Even though there is broad agreement among energy system modeling studies about the importance of biomass as a means to provide atmospheric carbon for CDR to achieve deep decarbonization, there is less consensus on technological pathways for biomass utilization across electricity, $H_2$, gaseous fuel, and liquid fuel production sectors [3], [34], [40], [60], [61], [62], [63]. As pointed out by other studies [62], it is difficult to glean insights about the disposition of biomass when opportunities to deploy BECCS in different sectors are not consistently represented in models. For instance, in contrast to our finding that biomass is used entirely for biofuel production, Millinger et al., who examine the role of biomass in a high-electrification European decarbonization context using a multi-sector CEM framework [34], report a relatively small role for biofuels, with biomass primarily allocated to heat and power generation. However, biomass will tend to be utilized in sectors in which it can be coupled to CCS, so this result may reflect the lack of assumed availability of CCS on biofuels production in that study than anything more fundamental.

The existence of several near-optimal solutions in CEMs with several binding constraints, which is common under deep decarbonization scenarios [64], also creates the possibility that small deviations in assumptions could lead to differences in biomass utilization. This point is evident in our cost sensitivity analysis (Figure 9), in which BECCS shifts from liquid fuels to $H_2$ production under high capital cost assumptions for liquids, but only when biomass and $CO_2$ sequestration resources are abundant. This sensitivity illustrates that the sectoral allocation of biomass depends not only on resource availabilities, but also on technology cost assumptions of both biofuels and competing uses of biomass.

### 4.3. Insights regarding fuel production technologies and their representation in system models

Our analysis highlights several key takeaways for alternative fuel technology development and associated process-level assessment. First, the optimal liquid fuel decarbonization pathways across the evaluated biomass and $CO_2$ sequestration assumptions rely on high CC rates (> 70%) associated with biofuel production. This suggests a need for closer process-level examination of the incremental energy and non-energy cost requirements (e.g., capital costs) associated with increasing CC rates in biofuel processes.

Second, the system-level integration of biofuel and synthetic fuel production increases the fraction of atmospheric carbon converted to fuel under resource-constrained scenarios. This increase in carbon conversion efficiency highlights the potential value of novel processing strategies that enhance biofuel carbon conversion. Specifically, rather than using $CO_2$ as a feed, using biomass as a feed while making process changes to recover and convert process carbon as part of the same biofuel production process could improve overall efficiency [65], [66]. Such approaches would have the added advantage of increasing biomass conversion efficiency, which could be valuable if biomass feedstocks are limited or costly and could help to diminish concerns about the land use implications of bioenergy production. On the other hand, increasing the conversion efficiency of biogenic carbon would increase the demand for hydrogen and electricity for fuels production.

Third, the sensitivity of fuel process selection to assumptions about product demand distribution underscores the need to characterize the incremental costs and energy requirements associated with varying product slates at the process level. There is a similar need to understand the flexibility of incumbent refineries and the incremental costs associated with changing product slates. In general, consistency in assumptions across different fuel production processes with different capture rates, product slates, and feeds would be very useful for evaluating the competitiveness of these technologies in scenarios generated by system models.

Finally, unlike sectors such as electricity and $H_2$, in which technology alternatives are well-characterized and can often be represented in energy system models with relatively few parameters, analyzing liquid fuel decarbonization pathways requires extensive data integration between process-level engineering (including TEA) models and energy system models. In this context, greater transparency and rigor in documenting process-level studies – including the ability to track carbon and energy flows within the process – would facilitate consistent parametrization of competing fuel technologies in energy system models, even when sourced from multiple, heterogeneous process-level studies. For example, reporting the cost factors used to scale equipment costs to total overnight costs could help harmonize these scaling factors across similar technologies before they are applied in energy system models.

### 4.4. Study limitations and future directions

We identify several study limitations that point towards areas for future work. First, we did not consider the time path of fuel technology deployment between the present and the net-zero year (nominally 2050), thereby omitting possible path dependencies driven by near-term policy or market dynamics. For example, existing corn ethanol blending mandates for gasoline are excluded in this analysis. If CCS were applied to existing corn ethanol plants, it is possible that these would deploy and would continue to be economic under deep decarbonization. Second, although our analysis incorporates supply curves for various renewable energy and carbon storage resources, we adopt simplified assumptions for fossil liquids and natural gas. This reflects the expectation that fossil fuel utilization will decline under deep decarbonization, reducing the likelihood that supply constraints will materially affect outcomes.

In addition, while our analysis included multiple technological pathways for biofuels, we represented only a single technological pathway for synthetic liquid fuels with a fixed product distribution, reflecting the focus on jet-dominant processes in the literature. This representation limits our ability to evaluate liquid fuel demand distributions that diverge substantially from current patterns, particularly under constrained biomass and $CO_2$ sequestration availability. Expanding the representation of synthetic liquid fuels options is therefore an important priority for future work.

We also note that an increasing share of synthetic fuels leads to higher average liquid fuel prices [34], which could incentivize reductions in liquid fuel demand in favor of direct electricity use or deployment of other energy carriers such as $H_2$. Although not explicitly modeled here, such final energy demand substitution would likely result in lower total electricity and $H_2$ generation due to the increased energy efficiency of direct electrification versus its indirect use to produce synthetic fuels. Therefore, improving the characterization of fuel demand elasticity to fuel prices in energy system models could further clarify the role for synthetic fuels in energy system decarbonization pathways.

Similarly, more information about variation in future final energy demands for both liquid fuels and other energy commodities would enable exploration of a wider scenario space. Lower aggregate fuel demand under the same biomass and $CO_2$ resource assumptions could reduce the sensitivity of fuel technology deployment to resource constraints, while higher demand driven by growth in fuel-intensive services (e.g., aviation) or a slower rate of electrification could amplify the importance of these resource constraints. In addition, there are substantial opportunities to more explicitly integrate fuel production with carbon-intensive industrial sectors (e.g., chemicals production) to capture competition for feeds as well as co-production of fuels and chemicals that could alter process selection across sectors.

## 5. Conclusion

This study examines liquid fuel decarbonization pathways using a multi-sector capacity expansion modeling framework applied to a net-zero case study of the contiguous United States. We find that two factors – biomass and $CO_2$ sequestration resource constraints – shape competition among biofuels, synthetic fuels, and fossil liquids, as well as the overall disposition of atmospheric carbon in the energy system. Biomass availability determines the extent of biofuel deployment and $CO_2$ capture rates in liquid fuel production pathways, whereas $CO_2$ sequestration availability governs the balance between carbon storage and utilization for synthetic fuels production.

These system-level findings are robust to assumptions regarding liquid fuel demand distributions, production flexibility, and variations in capital cost. At the same time, these assumptions shape competition among biofuel processes, as process selection is the primary means by which fuel supply is aligned with the product demand distribution. Moreover, the utilization of captured biogenic $CO_2$ for synthetic fuel production under resource-constrained scenarios enhances the overall conversion of atmospheric carbon to liquid fuels, highlighting the potential value of coordinated process development and system-level integration.

Finally, the implications of this analysis extend beyond the liquid fuels sector. Increased reliance on synthetic fuels would inherently link liquid fuel decarbonization to large-scale deployment of low-carbon hydrogen and electricity, highlighting the potential value of coordinated cross-sector infrastructure planning. In addition, future research could examine the temporal evolution of these pathways, broaden the representation of synthetic fuel technologies with diverse product distributions, incorporate endogenous fuel demand responses, consider other assumptions about final energy demands, and improve the representation of industrial sector coupling. Collectively, these extensions would enable a more comprehensive understanding of alternative transition pathways toward net-zero energy systems.

## 6. Acknowledgements

This research was supported by the ExxonMobil Technology and Engineering Company. The authors gratefully acknowledge Youssef Shaker for contributing towards the development related to $CO_2$ and liquid fuels infrastructure representation, and Nicole Shi for preparing the power system dataset used in the study. The authors acknowledge the MIT SuperCloud and Lincoln Laboratory Supercomputing Center for providing high-performance computing, database, and consultation resources that contributed to the results reported in this work. The views expressed herein are solely those of the authors.

**Supporting Information**

# Decarbonization pathways for liquid fuels: A multi-sector energy system perspective

Jun Wen Law[1], Bryan K. Mignone[2], Dharik S. Mallapragada[3*]

1. MIT Energy Initiative, Massachusetts Institute of Technology, Cambridge, MA 02139

2. ExxonMobil Technology and Engineering Company, Annandale, NJ 08801

3. Chemical and Biomolecular Engineering Department, Tandon School of Engineering, New York University, Brooklyn, NY 11201

*Correspondence: Dharik S. Mallapragada (dharik.mallapragada@nyu.edu)

## Table of Contents

## S1. Additional results

### S1.1 Default case scenario sets

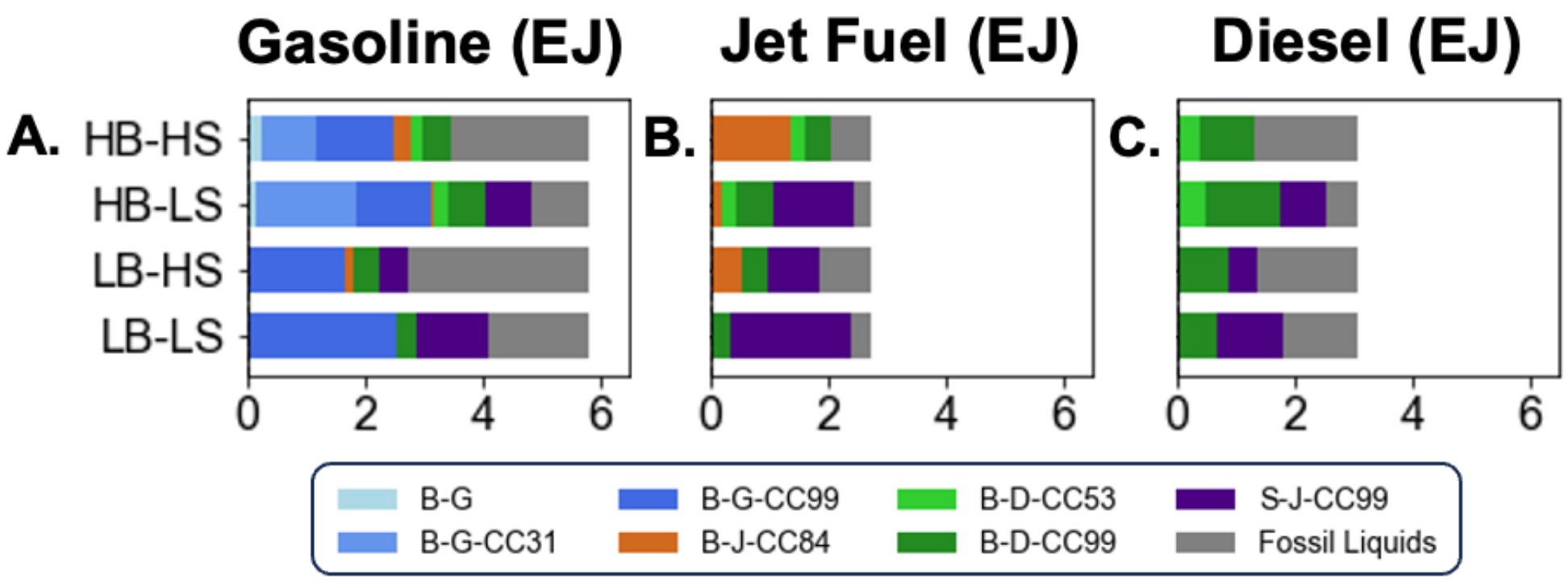

Figure S1. A) Gasoline production pathways, B) Jet fuel production pathways, and C) Diesel production pathways across the core scenarios as described in Figure 3A under default "IA-PF" modeling assumptions as described in Figure 3B, and default demand distribution in Figure 2A. Liquid fuels production technology labels follow Table 1.

Table S1. $CO_2$ marginal abatement cost (defined as the dual value of the $CO_2$ emission policy constraint as described in Eq. S51 in Section S5.5), and total levelized system cost (defined as total system cost divided by total exogeneous energy demand of 11385 TWh) of core scenarios described in Figure 3A of the main text, under default "IA-PF" modeling assumptions as described in Figure 3B, and default demand distribution as described in in Figure 2A.

| Scenarios | HB-HS | HB-LS | LB-HS | LB-LS |
|---|---|---|---|---|
| $CO_2$ marginal abatement cost ($/tonne) | 119.23 | 298.56 | 548.48 | 663.55 |
| Total levelized system cost ($/MWh) | 57.96 | 62.14 | 67.89 | 75.29 |

Table S2. Disposition of atmospheric $CO_2$ uptake, which includes biomass $CO_2$ content and $CO_2$ captured from DAC across core scenarios described in Figure 3A, under default "IA-PF" modeling assumptions as described in Figure 3B, and default demand distribution as described in Figure 2A. Converted to fuels = carbon content in biofuels and synthetic fuels, sequestered = percentage of atmospheric carbon that is captured and sequestered in geological storage, emitted = carbon vented in biofuel processes as a result of lower $CO_2$ capture (CC) rates.

| Scenarios | HB-HS | HB-LS | LB-HS | LB-LS |
|---|---|---|---|---|
| Converted to fuels (%) | 33% | 49% | 32% | 63% |
| Sequestered (%) | 51% | 30% | 66% | 36% |
| Emitted (%) | 16% | 21% | 2% | 1% |

Table S3. System average fuel production costs calculated using commodity shadow prices of energy system results, (detailed method described in Section S2.1) across the core scenarios as described in Figure 3A, under default "IA-PF" modeling assumptions as described in Figure 3B, and default demand distribution as described in Figure 2A. Liquid fuels production technology labels follow Table 1.

| Fuel production cost ($/GJ) | B-G | B-G-CC31 | B-G-CC99 | B-D | B-D-CC53 | B-D-CC99 | B-J-CC75 | B-J-CC84 | B-J-CC99 | S-J | S-J-CC99 |
|---|---|---|---|---|---|---|---|---|---|---|---|
| HB-HS | 38.76 | 37.34 | 36.48 | 42.29 | 38.53 | 38.29 | 42.21 | 34.54 | 40.93 | 43.15 | 42.44 |
| HB-LS | 53.75 | 52.22 | 51.63 | 54.88 | 50.97 | 50.86 | 58.81 | 47.41 | 57.55 | 46.29 | 45.70 |
| LB-HS | 126.06 | 106.94 | 67.66 | 109.25 | 85.55 | 67.94 | 91.56 | 71.44 | 73.34 | 87.47 | 64.94 |
| LB-LS | 141.54 | 120.18 | 76.25 | 121.69 | 95.46 | 75.76 | 102.62 | 79.92 | 82.30 | 95.03 | 69.74 |

## S1.2 Equal demand distribution case (E-D)

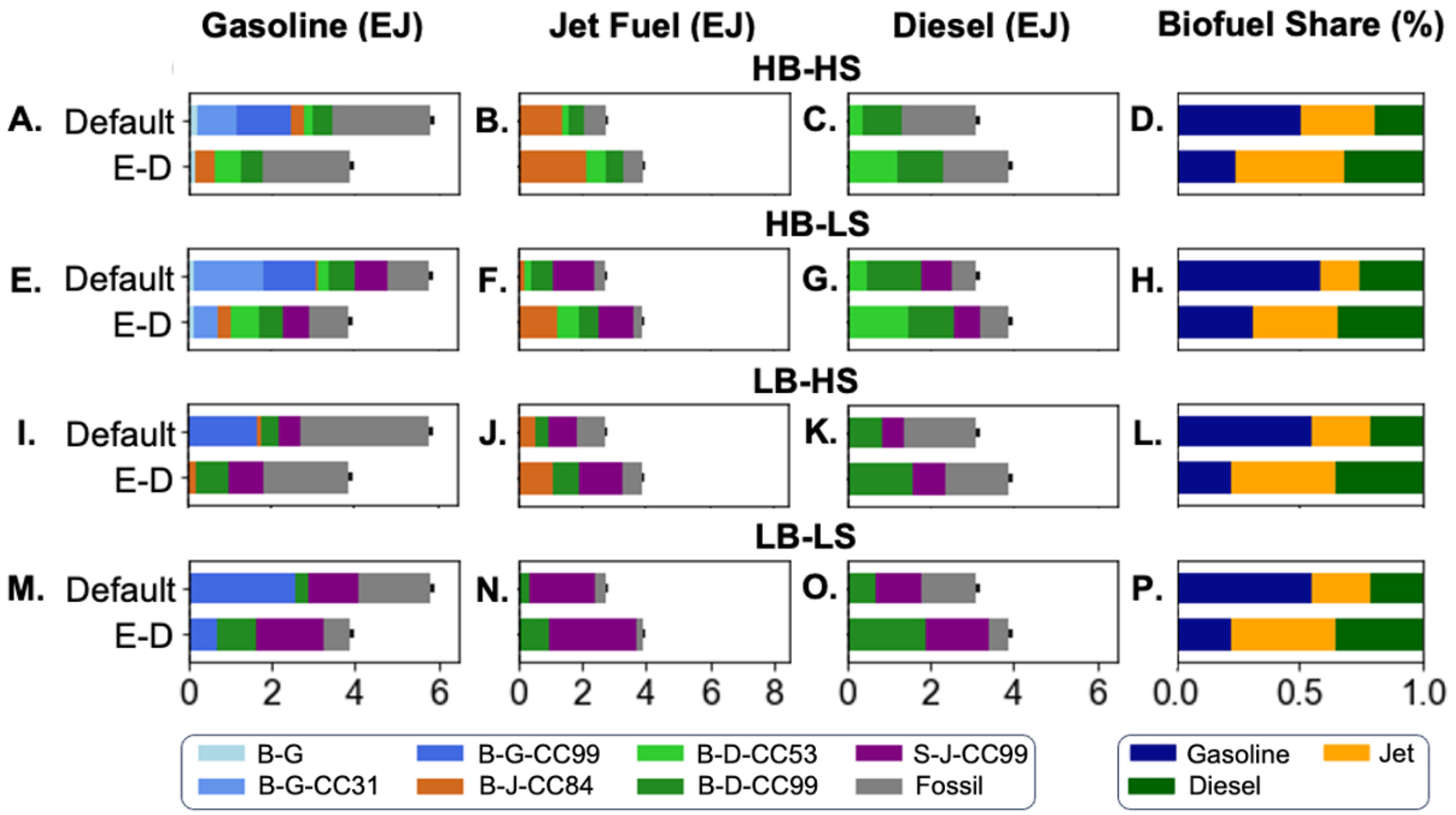

Figure S2. Gasoline production pathways (Panels A, E, I, M), jet fuel production pathways (Panels B, F, J, N), diesel production pathways (Panels C, G, K, O), as well as biofuel shares of gasoline, jet fuel, and diesel (Panels D, H, I, P) across the core scenarios as described in Figure 3A under default "IA-PF" modeling assumptions as described in Figure 3B, and both default and "E-D" demand distribution as described in Figure 2A. Liquid fuels production technology labels follow Table 1.

Table S4. Disposition of atmospheric carbon uptake from biofuel processes and direct air capture across the core scenarios as described in Figure 3A under default "IA-PF" modeling assumptions as described in Figure 3B, and "E-D" demand distribution as described in Figure 2A.

| E-D scenarios | HB-HS | HB-LS | LB-HS | LB-LS |
|---|---|---|---|---|
| Converted to fuels (%) | 36% | 49% | 39% | 65% |
| Sequestered (%) | 49% | 30% | 58% | 34% |
| Emitted (%) | 15% | 20% | 3% | 1% |

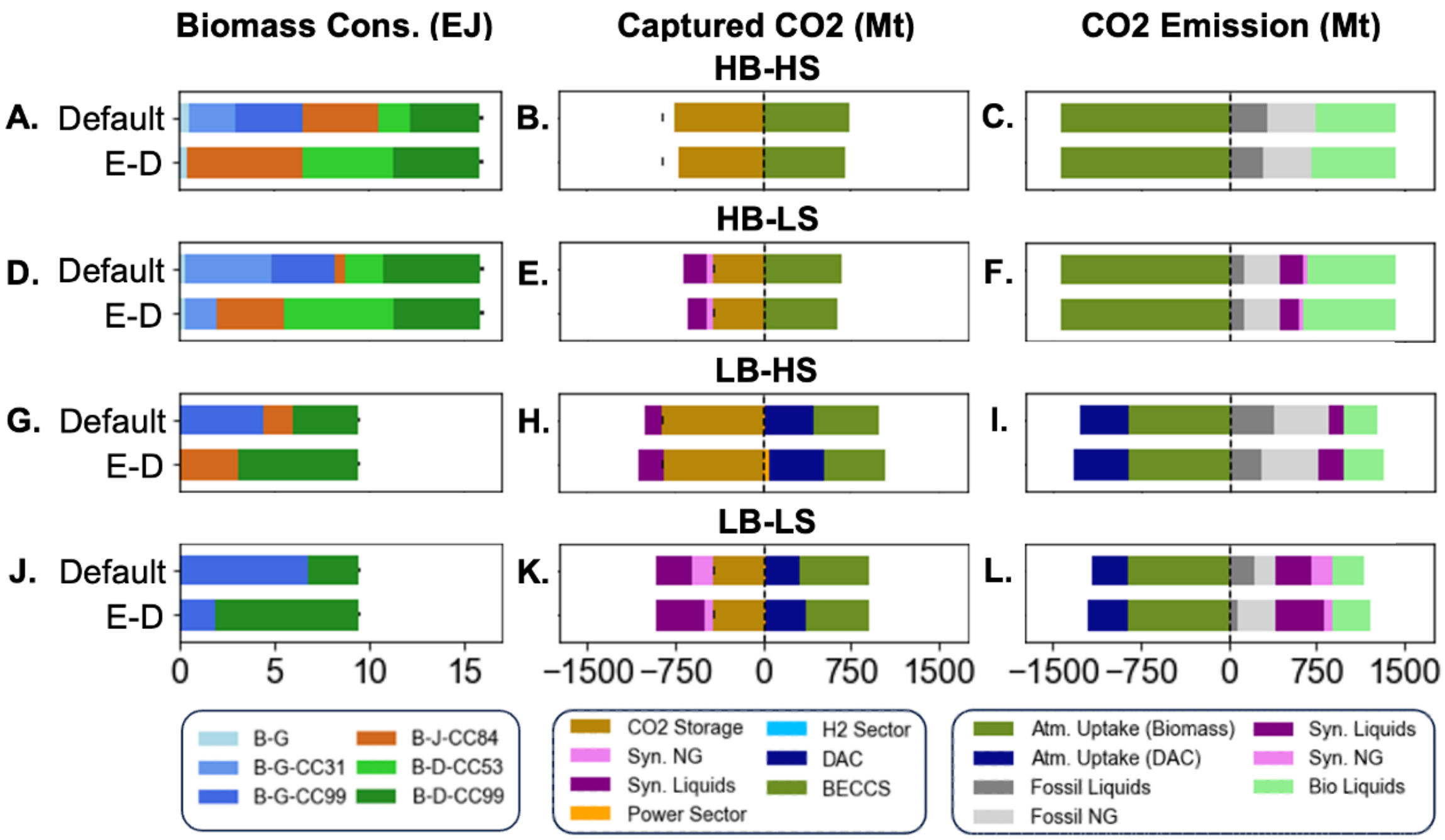

Figure S3. Biomass consumption by technologies (Panels A, D, G, J), captured $CO_2$ balance (Panels B, E, H, K), and $CO_2$ emission balance (Panels C, F, I, L) results across the core scenarios as described in Figure 3A under default “IA-PF” modeling assumptions as described in Figure 3B, and both default and “E-D” demand distribution as described in Figure 2A. Liquid fuels production technology labels follow Table 1. The resource limit of biomass and $CO_2$ sequestration are indicated by the bars in their respective plots.

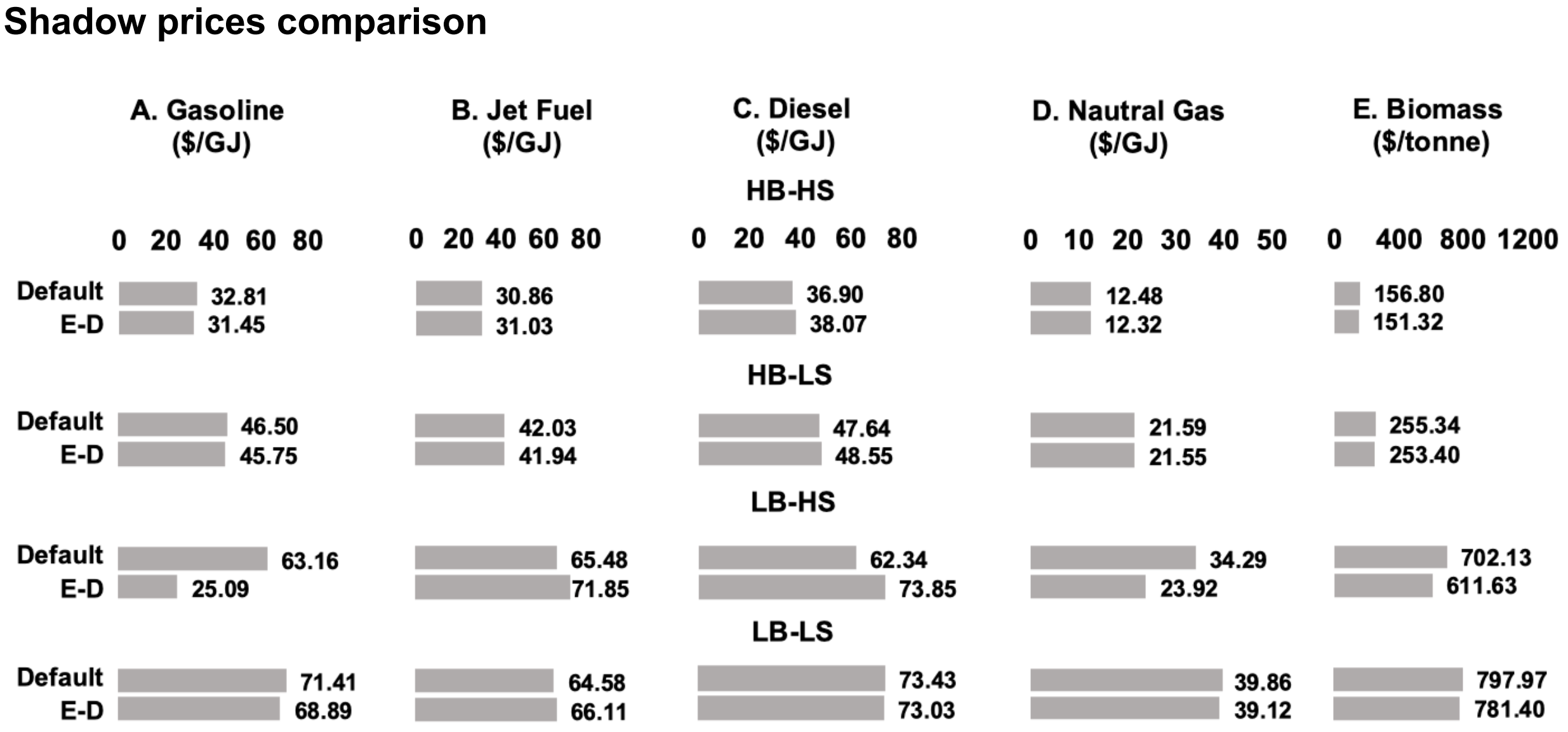

Figure S4. System average shadow prices for A) gasoline, B) jet fuel, C) diesel, D) natural gas, and E) biomass across the core scenarios as described in Figure 3A under default “IA-PF modeling assumptions as described in Figure 3B, and both default and “E-D” demand distribution as described in Figure 2A. Shadow prices of commodities represent the dual value of the supply-balance constraints of the respective commodities in MACRO, and system average shadow prices are calculated according to Eq. S1 in Section S2.1.

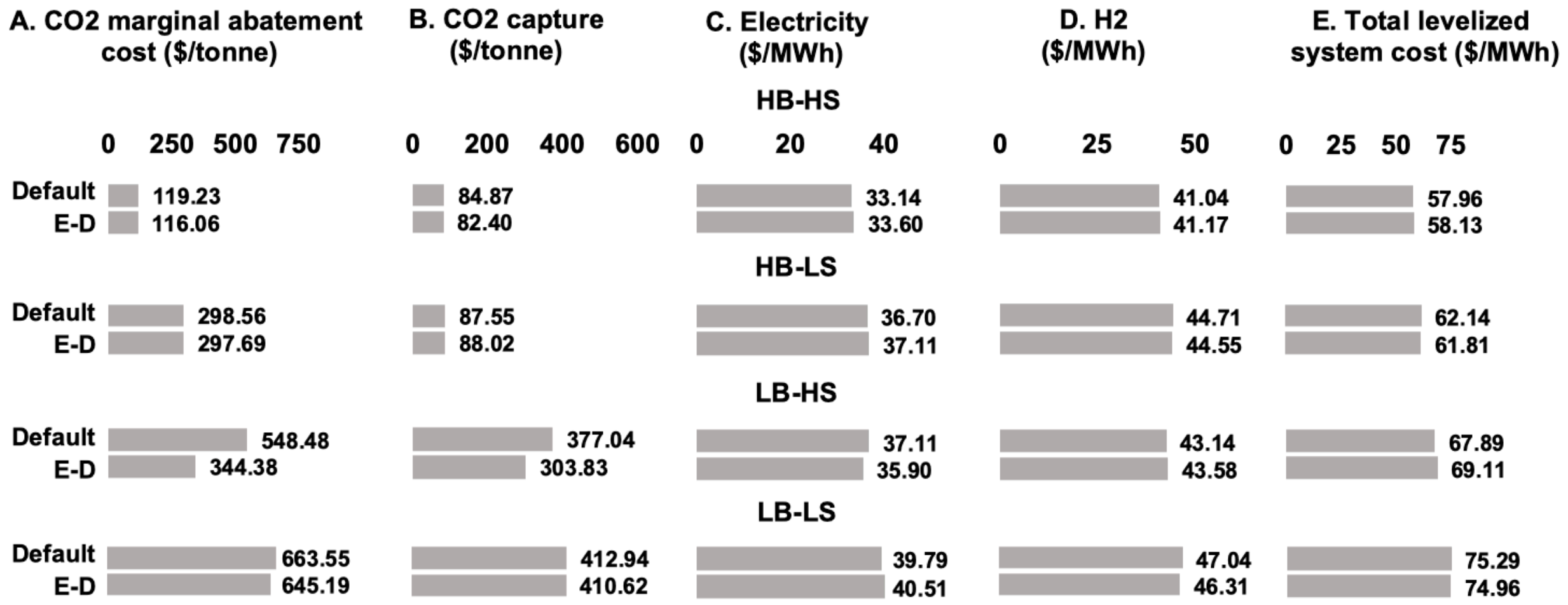

Figure S5. A) $CO_2$ marginal abatement cost, B) system average $CO_2$ capture price, C) system average electricity shadow price, D) system average H2 shadow price, E) Total levelized system cost across the core scenarios as described in Figure 3A under default "IA-PF modeling assumptions as described in Figure 3B, and both default and "E-D" demand distribution as described in Figure 2A. Shadow prices of commodities represent the dual value of the supply-balance constraints of the respective commodities in MACRO, and system average shadow prices are calculated according to Eq. S1 in Section S2.1. $CO_2$ marginal abatement cost = dual value of the $CO_2$ emission policy constraint as described in Eq. S51 in Section S5.5. System average $CO_2$ capture price = $CO_2$ marginal abatement cost – system average dual value of the captured $CO_2$ balance constraint. This represents the value of avoided $CO_2$ emission – cost of utilization and sequestration of captured $CO_2$. Total levelized system cost = total system cost divided by total exogeneous energy demand of 11385 TWh.

## S1.3 Other flexibility assumption cases (Default demand distribution)

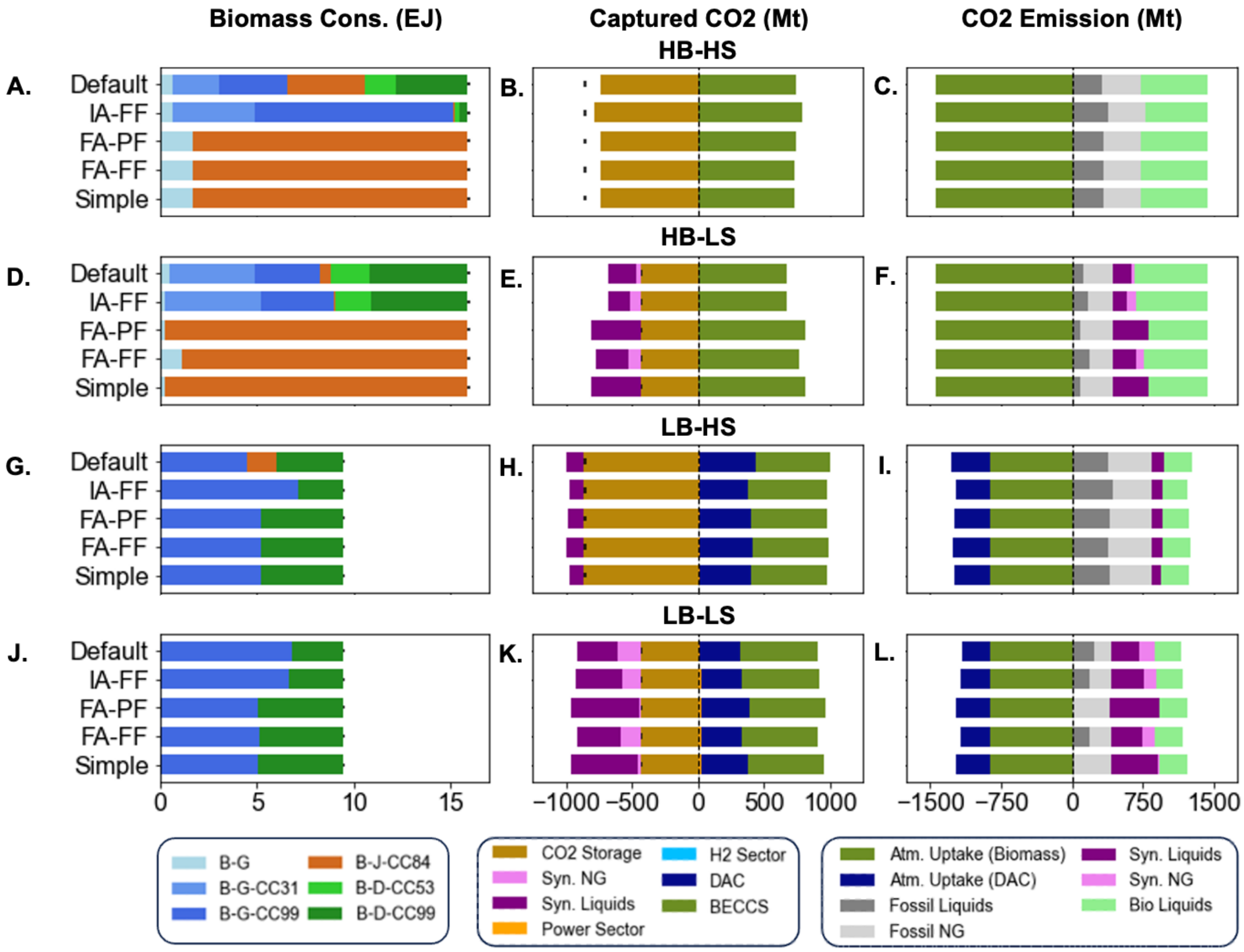

Figure S6. Effect of liquid fuel flexibility modeling assumptions on biomass consumption by technologies (Panels A, D, G, J), captured $CO_2$ balance (Panels B, E, H, K), and $CO_2$ emission balance (Panels C, F, I, L) across the core scenarios as described in Figure 3A, under default demand distribution as described in Figure 2A. Each row compares the five flexibility modeling assumptions as described in Figure 3B. Liquid fuels production technology labels follow Table 1, and the resource limit of biomass and $CO_2$ sequestration are indicated by the bars in their respective plots.

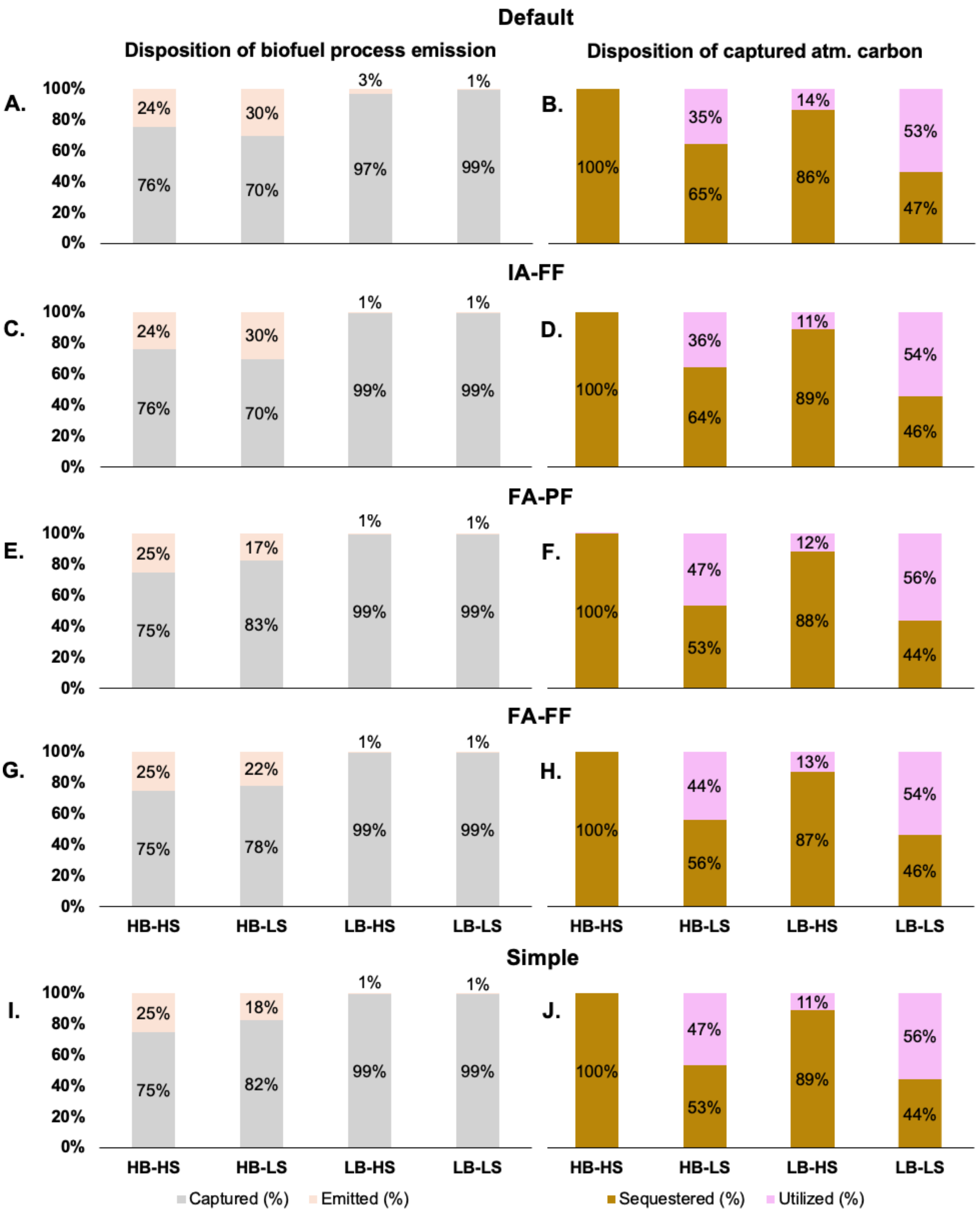

Figure S7. A,C,E,G,I) Disposition of biofuel production carbon emissions (excludes the carbon content in biofuel products), B,D,F,H,J) Disposition of captured atmospheric carbon from biofuel processes and direct air capture across the core scenarios described in Figure 3A under default demand distribution as described in Figure 2A, for the five flexibility modeling assumptions as described in Figure 3B: “Default” (Panels A,B), “IA-FF” (Panels C,D), “FA-PF” (Panels E,F), “FA-FF” (Panels G,H), “Simple” (Panels I,J).

Table S5. Disposition of atmospheric carbon uptake from biofuel processes and direct air capture across the core scenarios as described in Figure 3A under “IA-FF” modeling assumptions as described in Figure 3B, and default demand distribution as described in Figure 2A.

| IA-FF scenarios | HB-HS | HB-LS | LB-HS | LB-LS |
|---|---|---|---|---|
| Converted to fuels (%) | 29% | 49% | 30% | 64% |
| Sequestered (%) | 54% | 30% | 70% | 35% |
| Emitted (%) | 17% | 21% | 0% | 1% |

Table S6. Disposition of atmospheric carbon uptake from biofuel processes and direct air capture across the core scenarios as described in Figure 3A under “FA-PF” modeling assumptions as described in Figure 3B, and default demand distribution as described in Figure 2A.

| FA-PF scenarios | HB-HS | HB-LS | LB-HS | LB-LS |
|---|---|---|---|---|
| Converted to fuels (%) | 32% | 49% | 32% | 66% |
| Sequestered (%) | 51% | 30% | 68% | 34% |
| Emitted (%) | 17% | 21% | 0% | 1% |

Table S7. Disposition of atmospheric carbon uptake from biofuel processes and direct air capture across the core scenarios as described in Figure 3A under “FA-FF” modeling assumptions as described in Figure 3B, and default demand distribution as described in Figure 2A.

| FA-FF scenarios | HB-HS | HB-LS | LB-HS | LB-LS |
|---|---|---|---|---|
| Converted to fuels (%) | 32% | 48% | 32% | 64% |
| Sequestered (%) | 51% | 30% | 67% | 35% |
| Emitted (%) | 17% | 22% | 0% | 1% |

Table S8. Disposition of atmospheric carbon uptake from biofuel processes and direct air capture across the core scenarios as described in Figure 3A under “Simple” modeling assumptions as described in Figure 3B, and default demand distribution as described in Figure 2A.

| Simple model scenarios | HB-HS | HB-LS | LB-HS | LB-LS |
|---|---|---|---|---|
| Converted to fuels (%) | 32% | 49% | 31% | 65% |
| Sequestered (%) | 51% | 30% | 68% | 34% |
| Emitted (%) | 17% | 21% | 0% | 1% |

## Shadow prices comparison

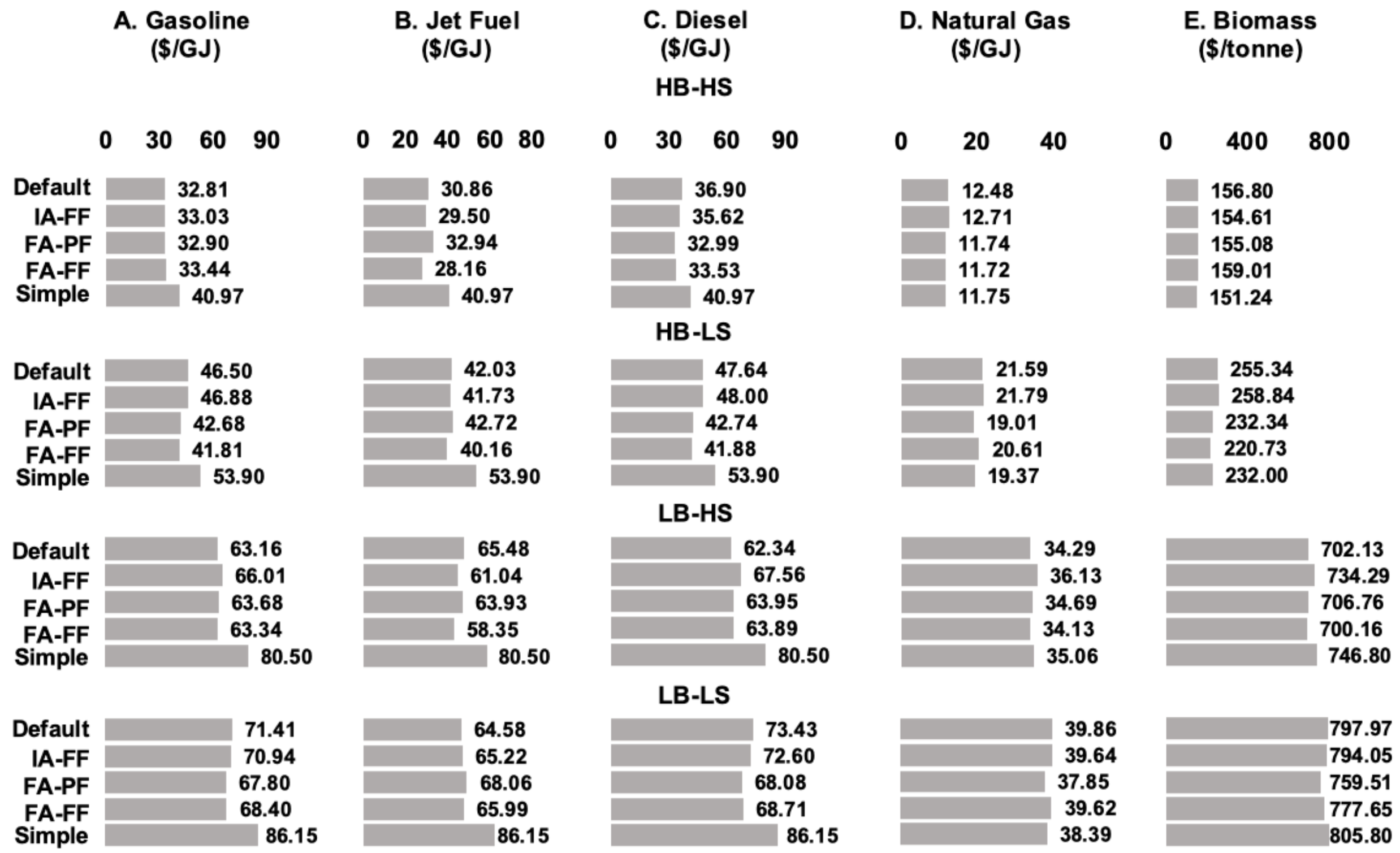

Figure S8. System average shadow prices for A) gasoline, B) jet fuel, C) diesel, D) natural gas, and E) biomass across the core scenarios as described in Figure 3A. Each row compares five flexibility modeling assumptions as described in Figure 3B under default demand distribution as described in Figure 2A.

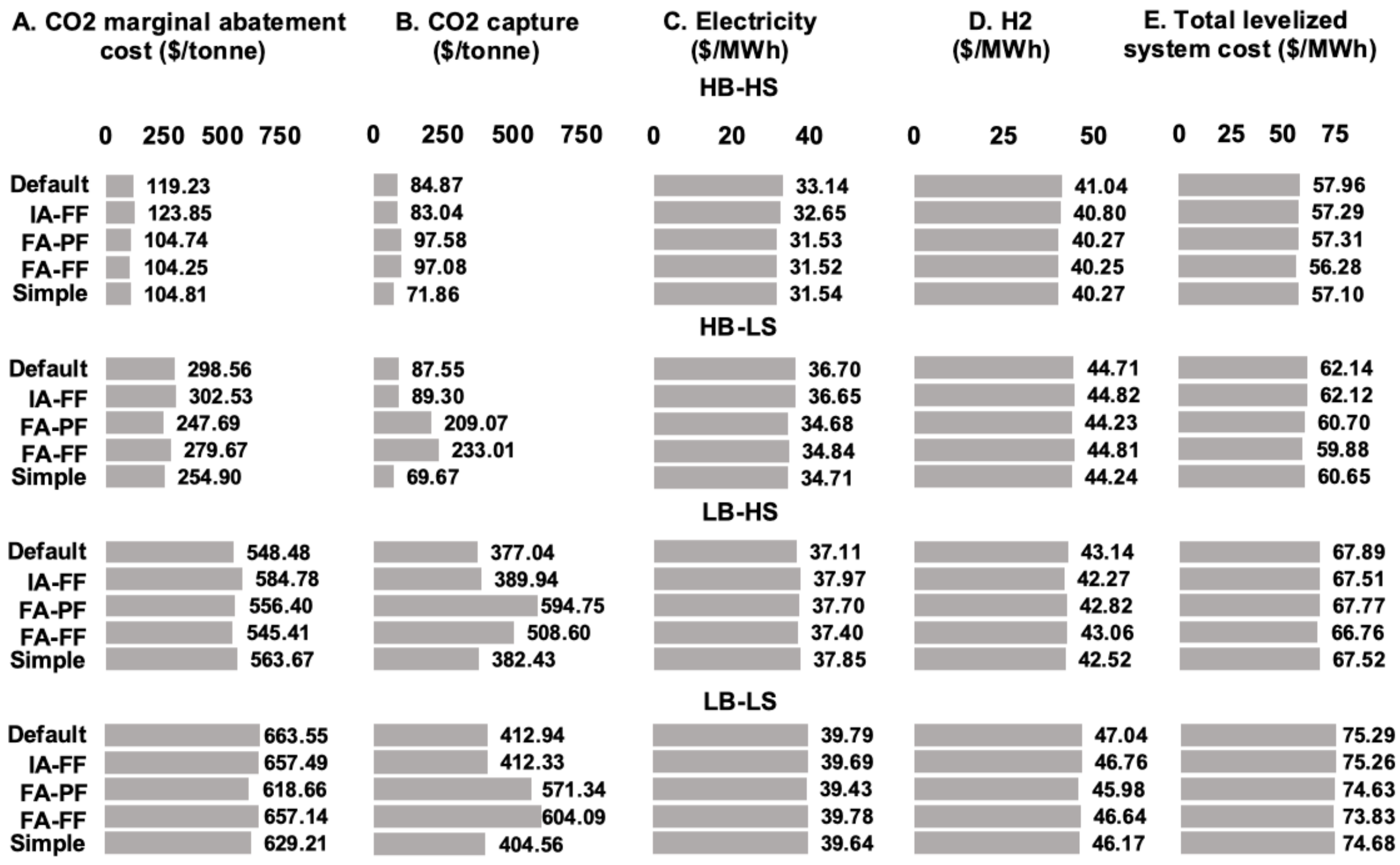

Figure S9. A) $CO_2$ marginal abatement cost, B) system average $CO_2$ capture price, C) system average electricity shadow price, D) system average $H_2$ shadow price, E) Total levelized system cost across the core scenarios as described Figure 3A. Each row compares five flexibility modeling assumptions as described in Figure 3B under default demand distribution as described in Figure 2A.

## S1.4 Other flexibility assumption cases (E-D demand distribution)

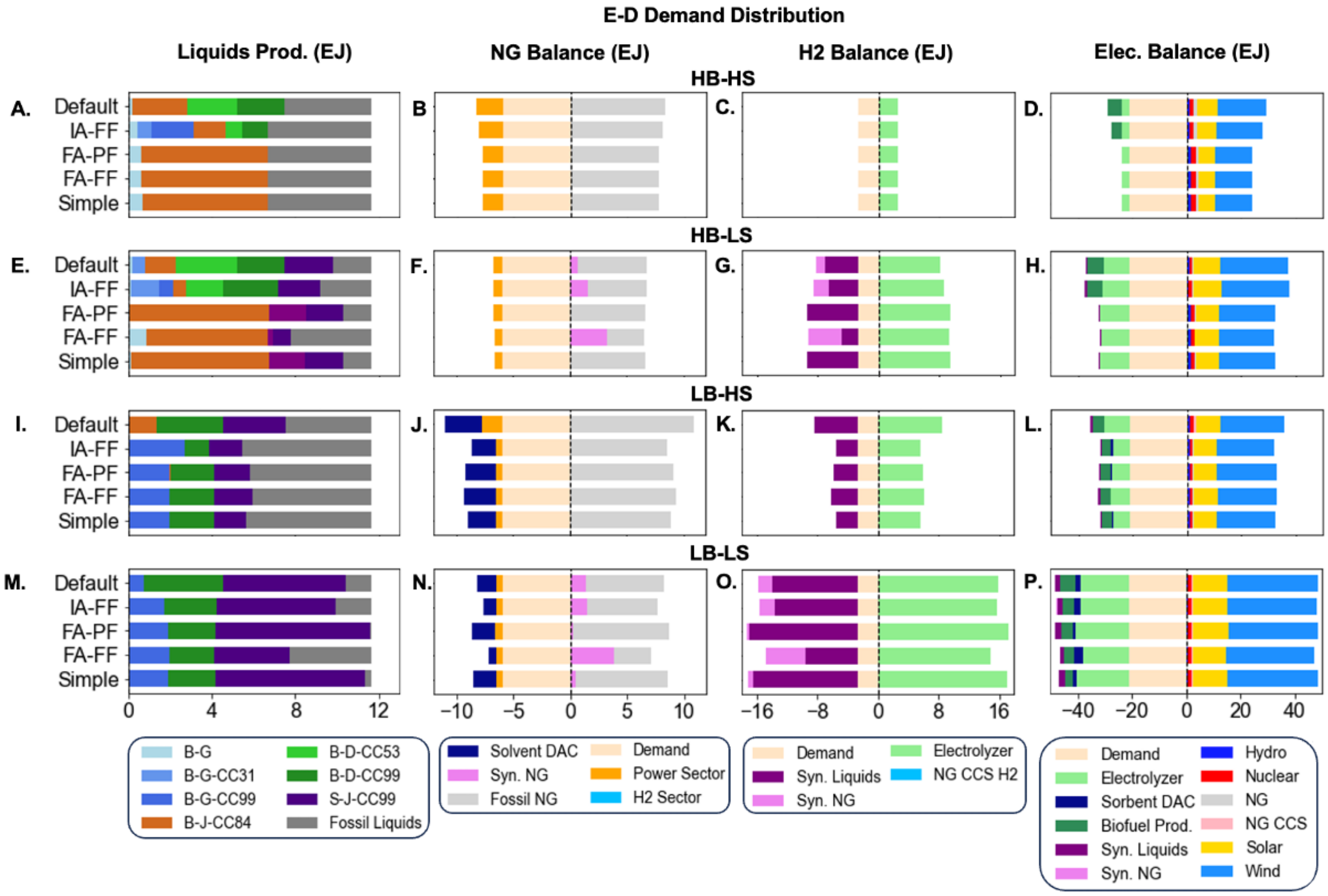

Figure S10. Effect of liquid fuel flexibility modeling assumptions on energy system outcomes: Liquid fuels production by pathway (Panels A,E,I,M), NG balance (Panels B,F,J,N), $H_2$ balance (Panels C,G,K,O), and electricity balance (Panels D,H,L,P) across the core scenarios as described in Figure 3A, under “E-D” demand distribution as described in Figure 2A. Each row compares the five flexibility modeling assumptions as described in Figure 3B. Liquid fuels production technology labels follow Table 1.

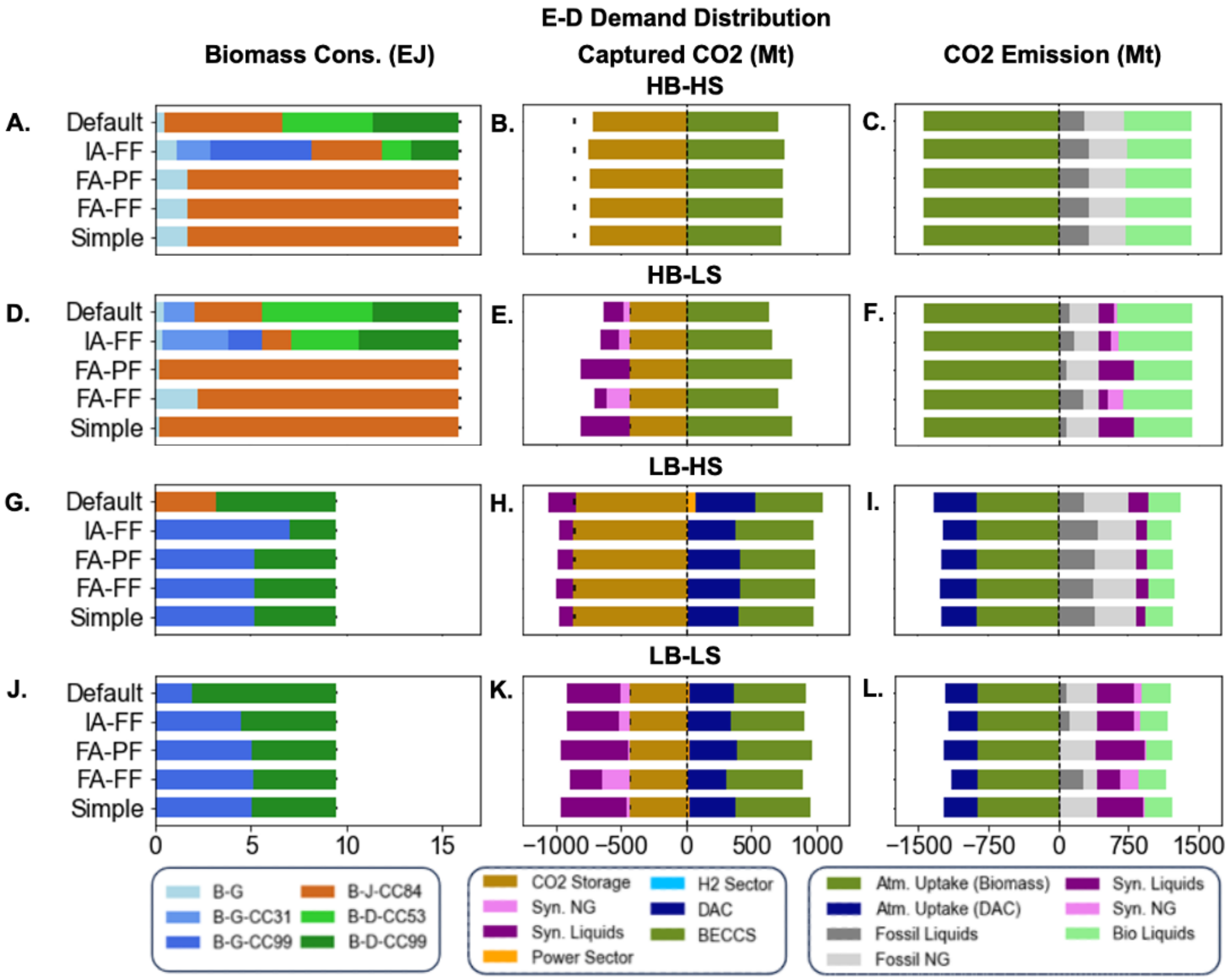

Figure S11. Effect of liquid fuel flexibility modeling assumptions on biomass consumption by technologies (Panels A, D, G, J), captured $CO_2$ balance (Panels B, E, H, K), and $CO_2$ emission balance (Panels C, F, I, L) results across the core scenarios as described in Figure 3A, under "E-D" demand distribution as described in Figure 2A. Each row compares the five flexibility modeling assumptions as described in Figure 3B. Liquid fuels production technology labels follow Table 1, and the resource limit of biomass and $CO_2$ sequestration are indicated by the bars in their respective plots.

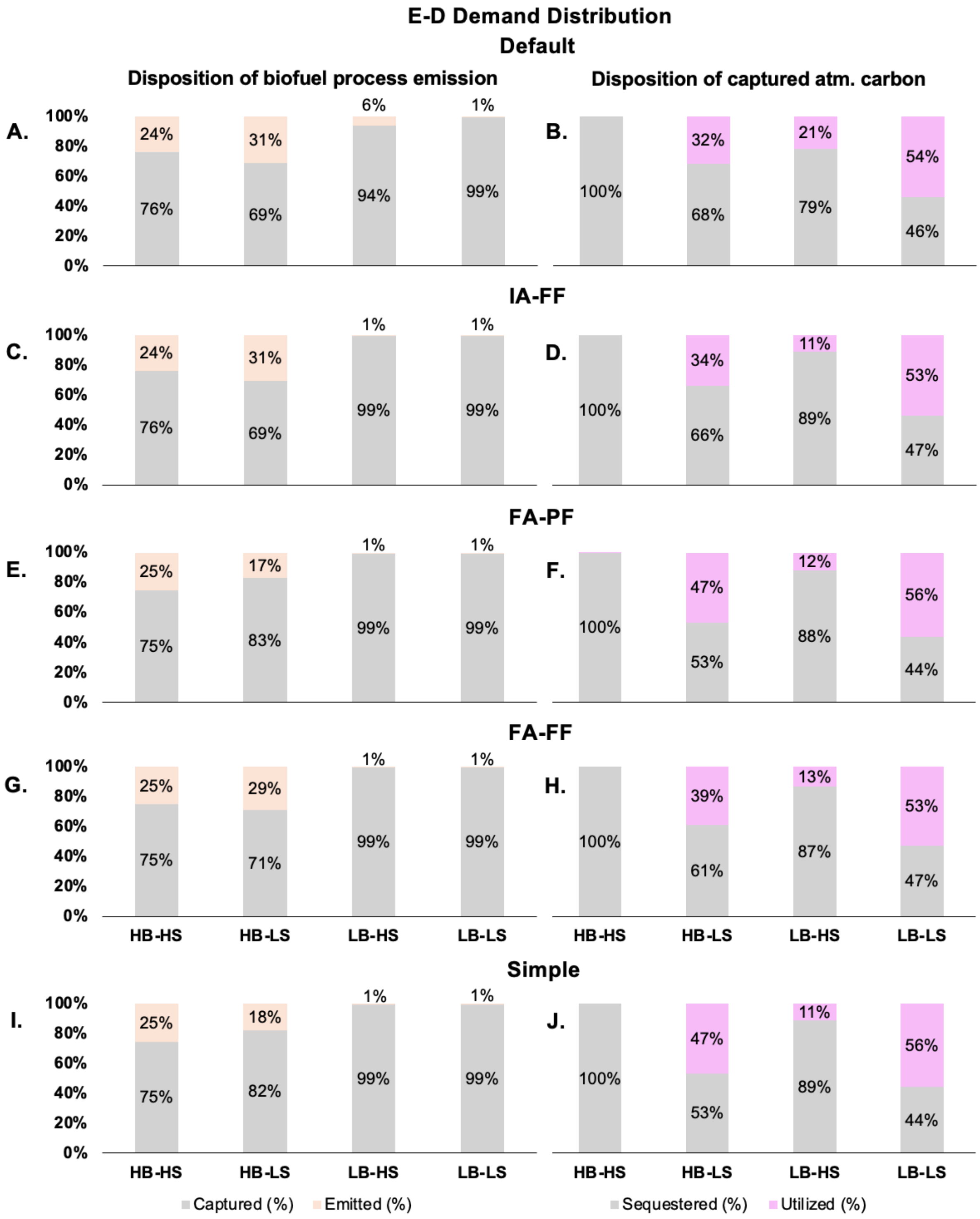

Figure S12. Panels A,C,E,G,I) Disposition of biofuel production carbon emissions (excludes the carbon content in biofuel panels products), panels B,D,F,H,J) Disposition of captured atmospheric carbon from biofuel processes and direct air capture across the core scenarios described in Figure 3A under "E-D" demand distribution as described in Figure 2A, for the five flexibility modeling assumptions as described in Figure 3B: "Default" (Panels A,B), "IA-FF" (Panels C,D), "FA-PF" (Panels E,F), "FA-FF" (Panels G,H), "Simple" (Panels I,J).

## S1.5 Cost sensitivity cases

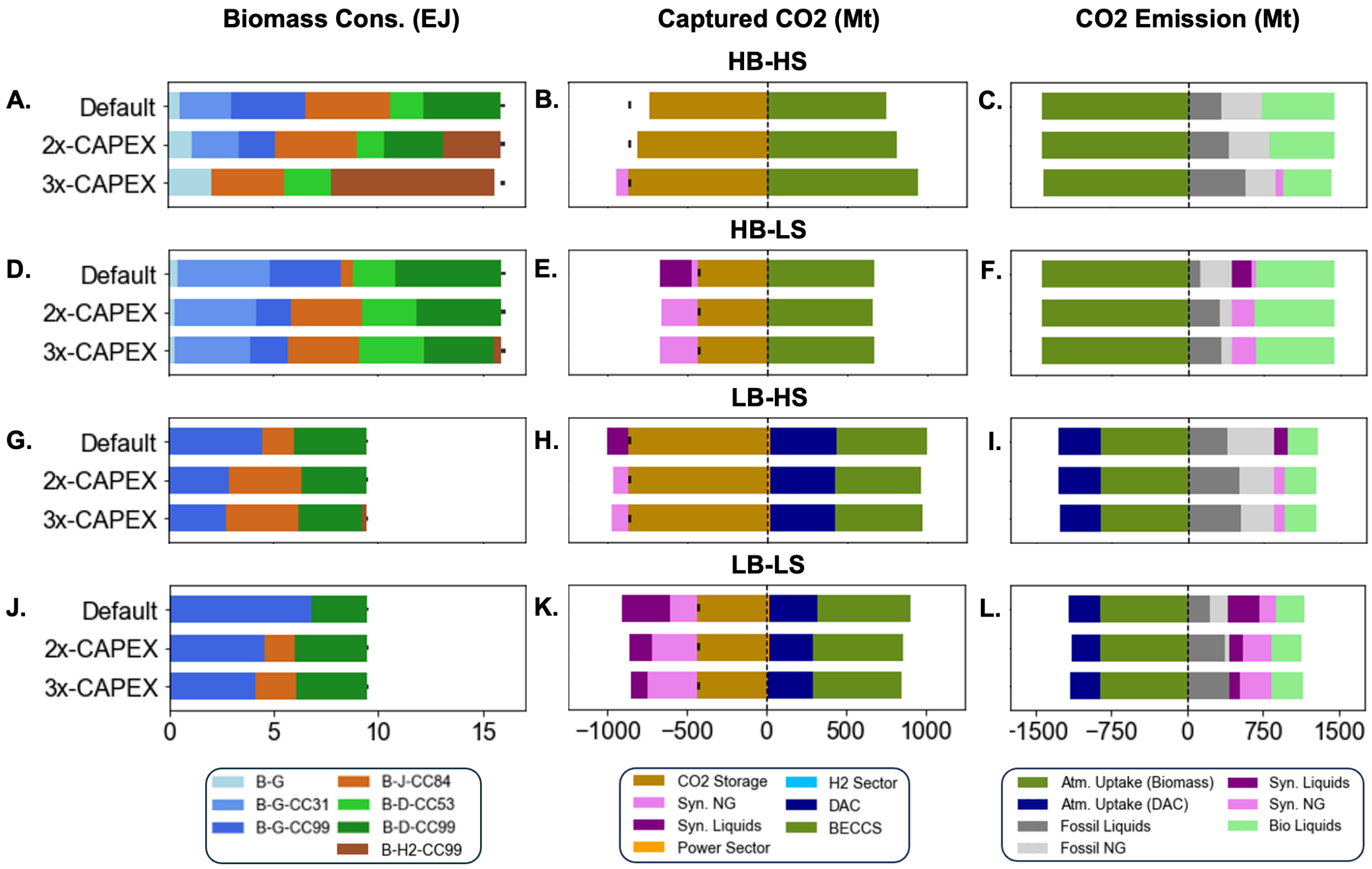

Figure S13. Additional energy system outcomes: Biomass consumption by technologies (Panels A, D, G, J), captured $CO_2$ balance (Panels B, E, H, K), and $CO_2$ emission balance (Panels C, F, I, L) across various capital cost assumptions across core scenarios as described in Figure 3A for “IA-PF” modeling assumptions as described in Figure 3B under default demand distribution as described in Figure 2A. Liquid fuels production technology labels follow Table 1 and the resource limit of biomass and $CO_2$ sequestration are indicated by the bars in their respective plots. “2x-CAPEX” = Doubled capital costs for bio and synthetic liquid fuels production technologies, “3x-CAPEX” = Tripled capital costs for bio and synthetic liquid fuels production technologies

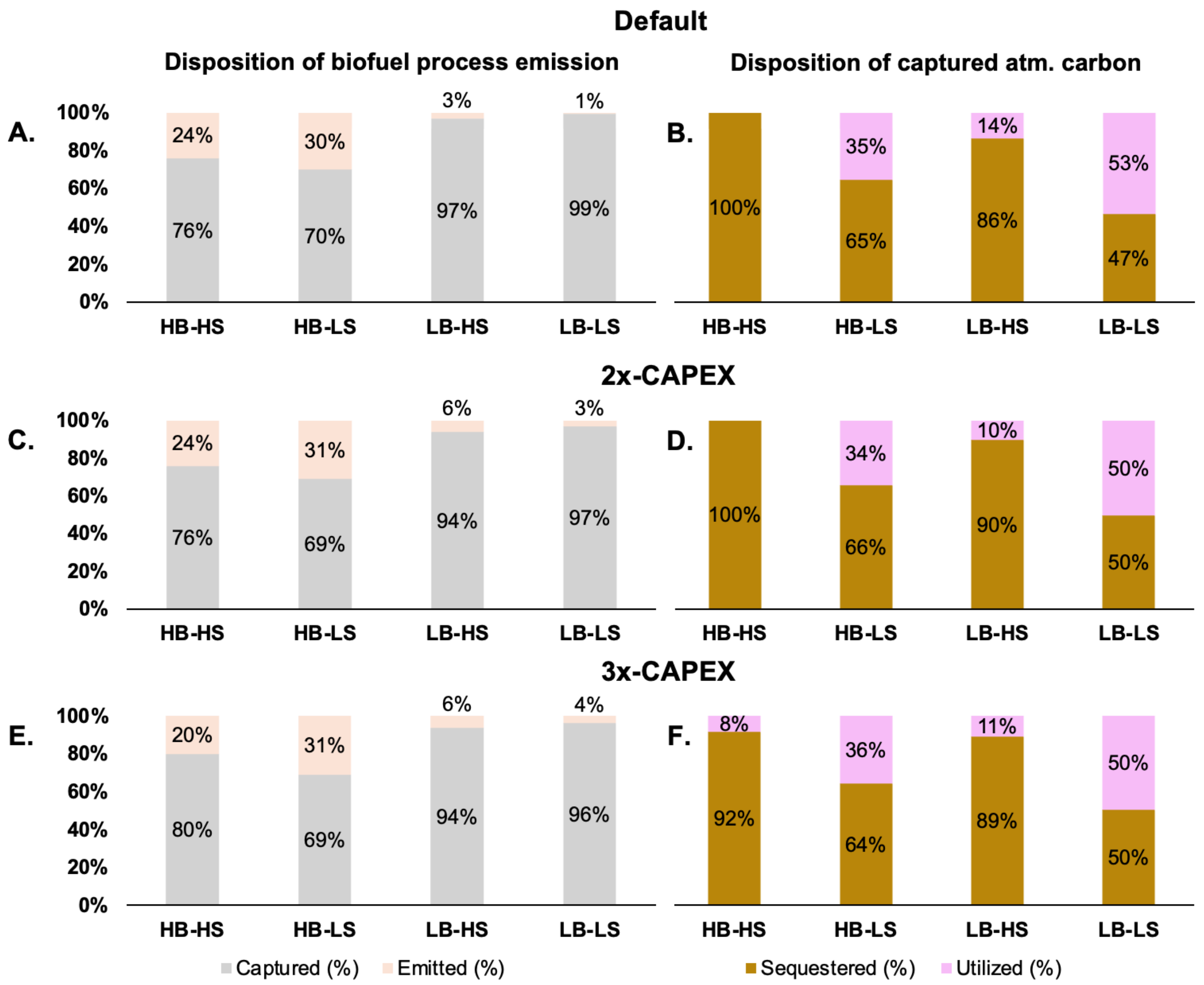

Figure S14. A,C,E,G) Disposition of biofuel production carbon emissions (excludes the carbon content in biofuel products), B,D,F,H) Disposition of captured atmospheric carbon from biofuel processes and direct air capture for various cost assumptions: Default (panel A,B), "2x-CAPEX" (panel C.D), and "3x-CAPEX" (panel E,F) as described in Section 3.6 across the core scenarios as described in Figure 3A, under default "IA-PF" modeling assumption as described in Figure 3B and default demand distribution as described in Figure 2A.

Table S9. Disposition of atmospheric carbon uptake from biofuel processes and direct air capture across the core scenarios as described Figure 3A for with "2x-CAPEX" cost assumptions, under default "IA-PF" modeling assumptions as described in Figure 3B, and default demand distribution as described in Figure 2A.

| 2x-CAPEX scenarios | HB-HS | HB-LS | LB-HS | LB-LS |
|---|---|---|---|---|
| Converted to fuels (%) | 27% | 48% | 30% | 61% |
| Sequestered (%) | 55% | 30% | 67% | 37% |
| Emitted (%) | 18% | 21% | 3% | 3% |

Table S10. Disposition of atmospheric carbon uptake from biofuel processes and direct air capture across the core scenarios as described Figure 3A for with "3x-CAPEX" cost assumptions, under default "IA-PF" modeling assumptions as described in Figure 3B, and default demand distribution as described in Figure 2A.

| 3x-CAPEX scenarios | HB-HS | HB-LS | LB-HS | LB-LS |
|---|---|---|---|---|
| Converted to fuels (%) | 22% | 49% | 30% | 60% |
| Sequestered (%) | 61% | 30% | 67% | 37% |
| Emitted (%) | 17% | 22% | 3% | 3% |

## Shadow prices comparison

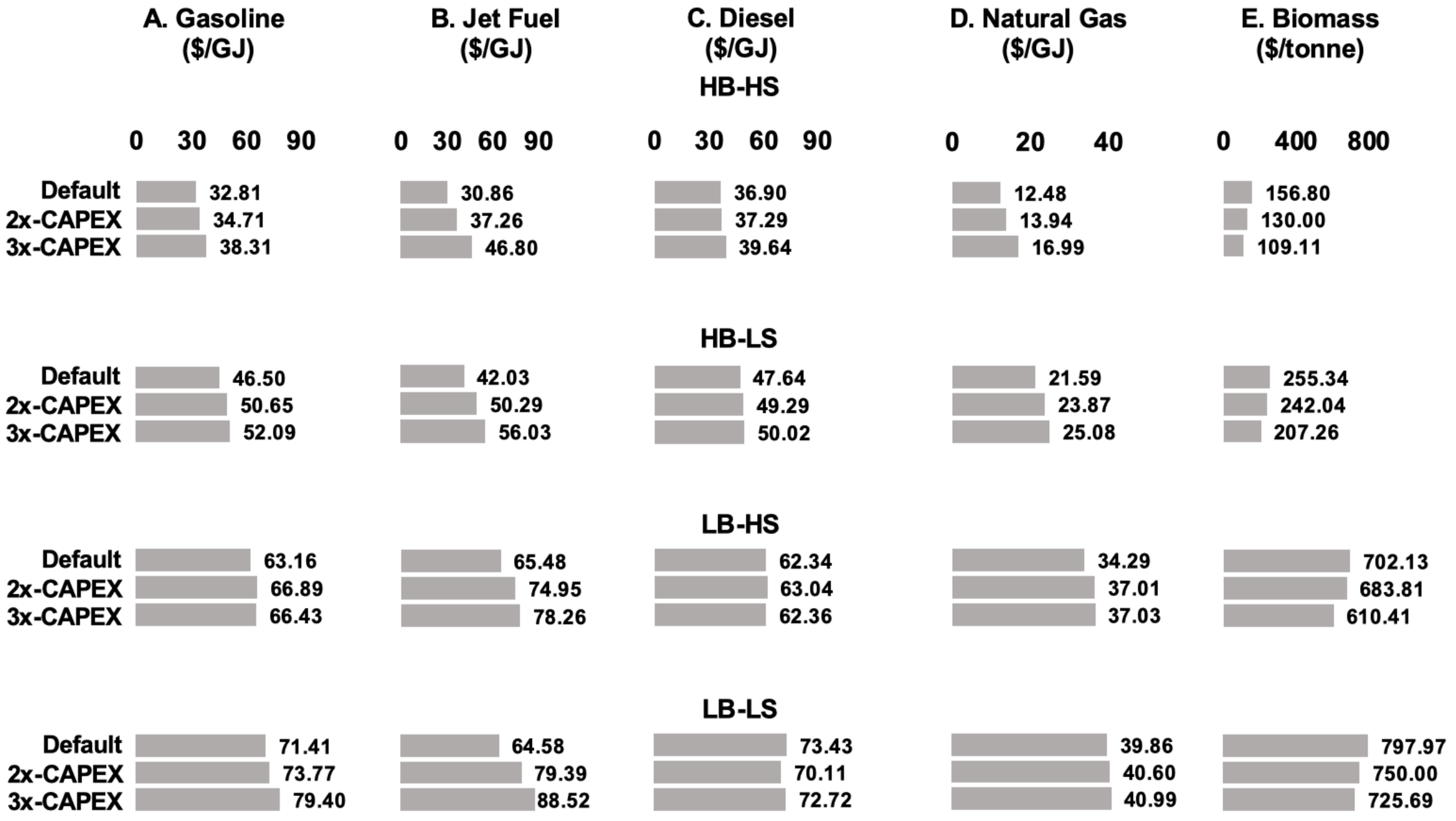

Figure S15. System average shadow prices for A) gasoline, B) jet fuel, C) diesel, D) natural gas, and E) biomass calculated according to Eq. S1 across the core scenarios as described in Figure 3A for various cost assumptions described in Section 3.6, under default "IA-PF" modeling assumptions as described in Figure 3B, and default demand distribution as described in Figure 2A.

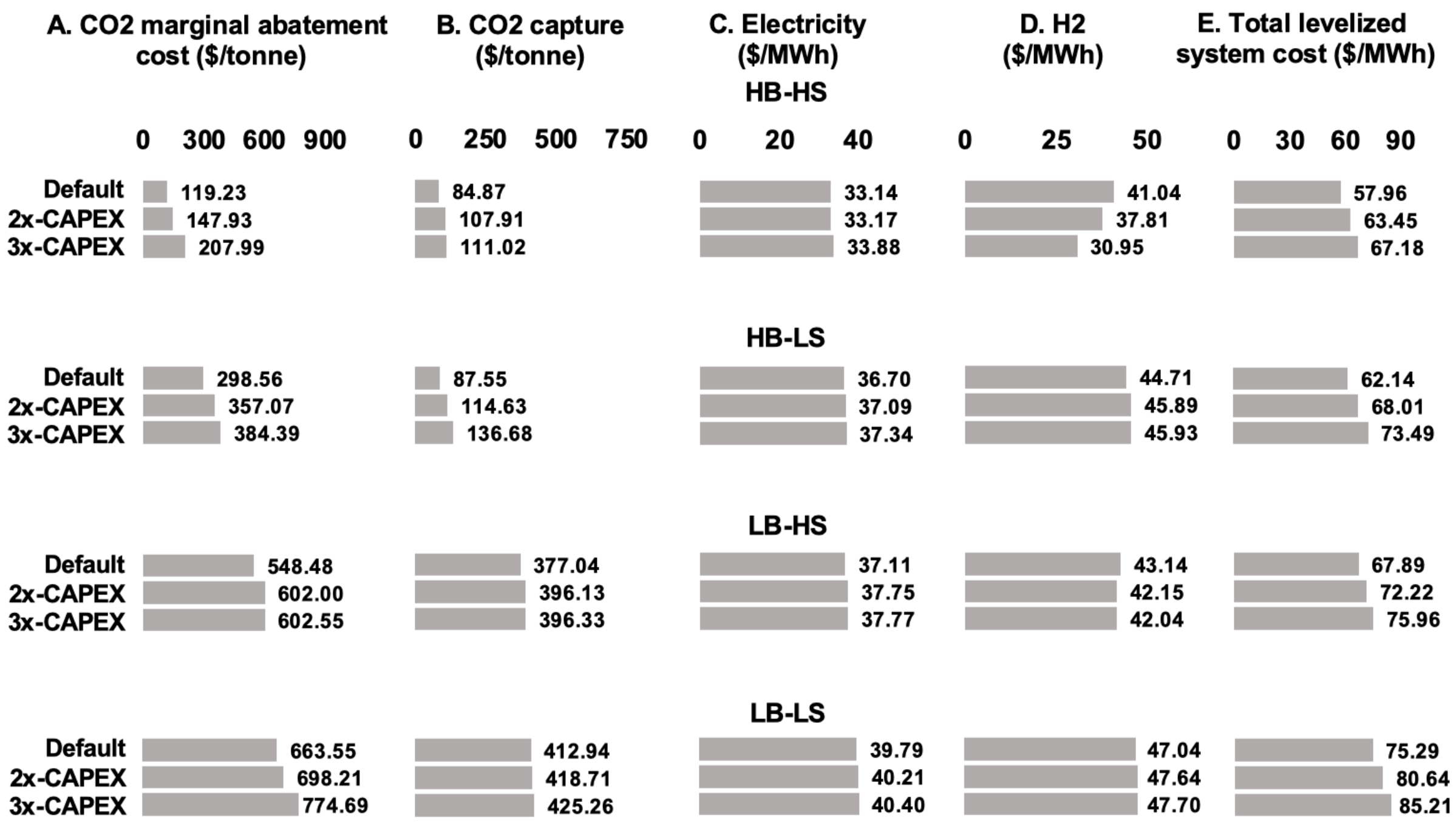

Figure S16. A) $CO_2$ marginal abatement cost, B) system average $CO_2$ capture price, C) system average electricity shadow price, D) system average $H_2$ shadow price, E) Total levelized system cost across the core scenarios as described Figure 3A for various cost assumptions described in Section 3.6, under default “IA-PF” modeling assumptions as described in Figure 3B, and default demand distribution as described in Figure 2A.

## S2. Methods

### S2.1. Energy system shadow prices

#### Method of calculating system average shadow prices

Liquid fuels shadow prices represent system-level prices as the supply demand balance constraints are enforced at system-level (Eq. S15– S17). System average shadow prices of other resource $\overline{\lambda_r}$ with supply-demand balance constraints enforced at the regional level such as electricity, hydrogen, and natural gas are calculated using Eq. S1:

$$\overline{\lambda_r} = \frac{\sum_{t=1}^{T} \sum_{z=1}^{Z} \omega_t \cdot \lambda_{t,r,z}}{\sum_{t=1}^{T} \omega_t} \qquad \text{(Eq. S1)}$$

Where $\lambda_{t,r,z}$ is the shadow price of each resource in region z in time t, and $\omega_t$ is the time weight of each time t.

#### Method of calculating system average liquid fuels production cost

The system average liquid fuels production cost for each production resource i ($\overline{C_{prod,i}}$) presented in Section 3.3 is calculated according to Eq. S2:

$$\overline{C_{prod,i}} = \frac{\sum_{z=1}^{Z} C_{prod,i,z} \cdot LF_{prod,z}}{\sum_{z=1}^{Z} LF_{prod,z}} \qquad \text{(Eq. S2)}$$

Where,

$C_{prod,i,z}$ = Liquid fuels production cost for resource i in region z calculated according to Eq. S3 – S4

$LF_{prod,z}$ = Total liquid fuels produced in region z

For biofuel production resource i:

$$C_{prod,i,z} = \text{Fixed Cost}_i + \text{VOM}_i + \overline{\lambda_{Bio}} \cdot \theta_{Bio,i} + \overline{\lambda_{electricity,z}} \cdot \xi_{elec,i} - \overline{\lambda_{CO2\ Capt,z}} \cdot \theta_{CO2\ Capt,i} \qquad \text{(Eq. S3)}$$

For synthetic fuels production resource i:

$$C_{prod,i,z} = \text{Fixed Cost}_i + \text{VOM}_i + \overline{\lambda_{electricity,z}} \cdot \xi_{elec,i} + \overline{\lambda_{H2,z}} \cdot \xi_{H2,i} + \overline{\lambda_{CO2\ Capt,z}} \cdot \xi_{,CO2\ in,i} \qquad \text{(Eq. S4)}$$

Where,

$\text{Fixed Cost}_i$ = CAPEX + Fixed maintenance and operation (FOM) costs ($/GJ)

$\text{VOM}_i$ = Variable maintenance and operation costs ($/GJ)

$\theta_{CO2\ Capt,i}$ = $CO_2$ captured by the biofuel process (tonne $CO_2$ captured/GJ)

$\theta_{Bio,i}$ = biomass input (tonne biomass/GJ) for biofuel production

$\xi_{elec,i}$ = electricity consumption (MWh/GJ)

$\xi_{H2,i}$ = hydrogen consumption (MWh/GJ)

$\xi_{CO2\ in,i}$ = captured $CO_2$ input (tonne $CO_2$/GJ) for synthetic fuel production

$\overline{\lambda_{CO2\ Capt,z}}$ = average regional $CO_2$ capture price (\$/tonne $CO_2$ captured) = $CO_2$ marginal abatement cost – dual value of the captured $CO_2$ balance constraint. This represents the value avoided $CO_2$ emission – cost of utilization and sequestration of captured $CO_2$.

For electricity, hydrogen, and biomass, $\overline{\lambda_{r,z}} = \frac{\sum_{t=1}^{T} \omega_t \cdot \lambda_{t,r,z}}{\sum_{t=1}^{T} \omega_t}$, where $\lambda_{tz}$ is the shadow price of each resource in region z in time t, and $\omega_t$ is the time weight of each time t.

## S2.2 Standardized approximation method for estimating $CO_2$ capture costs and energy requirements

This section presents an approximation methodology for estimating the costs and energy requirements of retrofitting $CO_2$ capture units onto fuel production processes that were originally designed without $CO_2$ capture in their respective reference TEAs. This approach enables the representation of a wide variety of $CO_2$ capture rates for biofuels and synthetic fuels processes in this study.

The method described in this section involves systematically identifying major $CO_2$ emission streams from published process flow diagrams and mass balances in reference TEAs of fuel production processes, then matching those streams to analogous industrial processes with established post-combustion capture technologies. Cost and energy penalty estimates are derived from the U.S. National Energy Technology Laboratory (NETL) TEAs of $CO_2$ capture in industrial point sources and power plants [1], [2] as shown in Table S11. By mapping emission streams to comparable industrial sources based on flue gas composition, this framework enables consistent estimation of $CO_2$ capture system costs and energy requirements across process types. It preserves the integrity of the original process design and energy conversion, while enabling evaluation of $CO_2$ capture configurations not modeled in existing literature.

For each identified $CO_2$ emission stream from published process flow diagrams in reference TEAs, the selection of appropriate $CO_2$ capture technology is based on the concentration of $CO_2$ in each emission stream. Assumptions were made to the concentration in each emission stream when the data is unavailable. For example, industrial process flue gas emissions were assumed to be 10-20% wt. $CO_2$, while natural gas related flue gas emissions were assumed to be <4% wt. $CO_2$. The framework employs a systematic approach to match emission streams based on their $CO_2$ concentrations to representative $CO_2$ capture processes in NETL TEAs:

- High-Concentration $CO_2$ Streams (~100% wt. $CO_2$): For streams with nearly pure $CO_2$, such as those from syngas cleaning in gasification processes, ethanol plant $CO_2$ compression parameters are assumed with 100% CC rate [1].

- Medium-Concentration $CO_2$ Streams (10-20% wt. $CO_2$): Emission streams with moderate $CO_2$ concentrations are matched to cement plant $CO_2$ capture parameters with 99% CC rate [1].

- Low-Concentration $CO_2$ Streams (<4% wt. $CO_2$): Dilute $CO_2$ streams, are mapped to $CO_2$ capture parameters of natural gas combined cycle (NGCC) Power Plant with 95% CC rate [3].

The NETL TEAs provide breakdowns of capital expenditures (CAPEX), operating expenditures (OPEX), and energy penalties associated with different $CO_2$ capture technologies and configurations (see Table S11). We assumed that all thermal energy input requirements for the capture process are met through electricity. Assuming both the cost and energy requirements of $CO_2$ capture systems scale linearly with the input $CO_2$ mass flow rate, we adapt NETL's reference $CO_2$ capture data to the specific scale and $CO_2$ emission characteristics of the various emission streams of fuel production technologies in this study using the following steps:

1. The costs and energy requirements of the $CO_2$ capture processes from NETL TEAs are normalized to a per-tonne $CO_2$ mass inflow basis.

2. These normalized costs and energy requirements are scaled to match the targeted $CO_2$ emission streams of the fuel production technologies based on their respective reference TEAs, and then scaled to a per-unit biomass input basis for biofuel production processes, and per-unit $CO_2$ feedstock input basis for synthetic fuels processes.

3. The scaled $CO_2$ capture costs and energy requirements are added to the original plant input parameters and used as inputs for MACRO.

4. Finally, the overall CC rate of the fuel production technology is calculated as the ratio of the sum of captured $CO_2$ from all streams to the sum of original $CO_2$ emission streams and used as inputs for MACRO.

The pathways characterized using this method are included in Table S33, Table S34, and Table S37, and Section S4 shows carbon balance diagrams of various variants of biofuels and synthetic fuels production technologies including those parameterized using the above-mentioned approximation method.

Table S11. Costs and energy requirements of $CO_2$ capture technology used in the approximation method to develop variants of biofuel and synthetic fuels technologies with various $CO_2$ capture rates. Cost assumptions are scaled according to per tonne CO2 input basis and converted to 2022 dollars, a technology lifetime of 30 years and discount rate of 4.5% is used to annualize investment costs.

| Industrial $CO_2$ capture | Input $CO_2$ concentration range (% wt.) | $CO_2$ capture rate (%) | CAPEX ($/(tonne $CO_2$ input/h)) | Annualized CAPEX ($/(tonne $CO_2$ input/h)/y) | FOM ($/(tonne $CO_2$ input/h)/y) | VOM ($/tonne $CO_2$ input) | Electricity input (MWh/tonne $CO_2$ input) |
|---|---|---|---|---|---|---|---|
| Ethanol plant $CO_2$ capture | ~ 100% | 100.0 | 1,384,619 | 85,004 | 76,359 | 1.66 | 0.11 |
| Cement plant 99% $CO_2$ capture | 10 - 20 % | 99.0 | 2,746,462 | 168,610 | 87,115 | 5.54 | 1.14 |
| NGCC 95% $CO_2$ capture | <4 % | 95.0 | 2,233,695 | 137,130 | 74,912 | 6.36 | 0.91 |

## S2.3. Harmonization of carbon and energy balances for fuels production technologies

This section details the methodology used to harmonize the carbon and energy balances of various fuels production processes in this study obtained from techno-economic analyses (TEAs) in the literature, including the additional variants with various $CO_2$ capture rates described in Section S2.2. This harmonization allows for the application of user-defined carbon content for biomass and fuels in the model while preserving the original process's energy conversion and product distribution as reported in their respective TEAs, and ensuring the balance of carbon and energy balances in the inputs and outputs of the processes.

### Harmonization of energy balance for biofuel production in MACRO

The following steps are involved to obtain harmonized biofuel energy conversion parameters for MACRO across the biofuel technologies modeled in this study.

1. Determining biomass energy converted to biofuels for MACRO biofuel modeling (Eq. S8 – S10)

   As we modeled energy conversion of biofuel processes based on higher heating value (HHV) efficiency, defined as fraction of input biomass energy converted into fuels in MACRO, these values are directly utilized if reported in the technology's reference TEA, or calculated from using available data within the reference. Any assumptions to infer energy content of fuels were obtained from the GREET 2023 database. For example, if the reference TEA only reported the volume of fuels produced, GREET 2023 database is used to estimate the energy content of the fuels product, and the process energy efficiency is calculated by dividing the energy content of the products by the energy content of the original input biomass [4]. Any necessary conversions from lower heating value (LHV) to HHV are performed using appropriate assumptions. This allows MACRO to model biofuel technology for biomass types with unique energy contents, reflecting differences in fuel output yields per unit mass of biomass across different biomass types. For example, a biomass type with higher energy content will produce more fuel products per tonne of purchased biomass.

2. Determining the fraction of individual fuel products

   For biofuel processes yielding multiple fuel products, the distribution is required in MACRO (Eq. S8 – S10). The energy distribution ratios of individual fuels according to HHV energy content are directly utilized if the reference provides them. However, if only mass ratios are available in the reference, the energy ratios are calculated using GREET 2023 energy content data [4].

3. Assumptions used for harmonizing biofuel energy modeling in MACRO

   1) Process energy efficiency and fuel product distribution remain constant regardless of the biomass type.
   2) Costs are assumed to be based on a per-unit biomass mass input and remain consistent irrespective of the biomass type.
   3) Any electricity inputs are also assumed to be on a per-unit biomass mass input basis and remain constant regardless of the biomass type (see Eq. S22)
   4) The cost and energy requirements of the plant are assumed to scale linearly with the input biomass input.

## Harmonization of carbon balance for biofuel production in MACRO

The carbon balance modeling for biofuel processes in MACRO is described in Eq. S13 – S14, where the amount of vented $CO_2$ emissions and captured $CO_2$ are calculated according to user defined 1) biomass carbon content and 2) $CO_2$ capture (CC) rates for various processes. Eq. S13 and S14 ensures that any carbon content not in biofuel products are either emitted into the atmosphere (see Eq. S51), or captured for utilization or sequestration (see Eq. S50). As such, this ensures that no carbon is unaccounted for in the energy system.

The carbon content in the biofuels is calculated by applying emission factors obtained from the GREET 2023 database shown in Table S31 to the fuel product outputs of the biofuel production processes (see Eq. S8 – S10). This allows the carbon balance to be complete in MACRO with the user providing input parameters on 1) biomass carbon content and 2) CC rates for various processes.

## Harmonization of carbon and energy balance for synthetic fuels production in MACRO

As compared to biofuels where different feedstocks have unique carbon and energy content, synthetic fuels production is scaled to per unit $CO_2$ feedstock inputs, and we can directly utilize the production yield of various fuels product per unit $CO_2$ feedstock input from the reference in the literature, as modeled in Eq. S5 – S7 in MACRO. Similarly, the electricity and $H_2$ input requirements are scaled to per unit $CO_2$ feedstock input and can be directly utilized from the original reference, as modeled in Eq. S22 – S23 in MACRO. Nonetheless, the amount of vented $CO_2$ emissions and captured $CO_2$ of the synthetic fuels process are calculated in MACRO according to user defined $CO_2$ capture (CC) rates as shown in Eq. S11 – S12. This ensures that any carbon content not in the synthetic fuel products are either emitted into the atmosphere (see Eq. S51), or captured for utilization or sequestration (see Eq. S50), with all carbon accounted for in the energy system.

## S3. Modeling input assumptions

### S3.1. Energy demand across sectors

Table S12. Annual energy demand for various commodities in this study: Power, $H_2$, natural gas, and liquid fuels (gasoline, jet fuel, diesel). Hourly power demand profiles were obtained from projected 2050 state-level data from Princeton University's Net-Zero America low electrification scenario [5]. Annual $H_2$, natural gas, and liquid fuels demands were obtained from projected 2050 state-level end-uses likewise from Net-Zero America low electrification scenario, with the assumption of constant hourly demand in creating hourly demand profiles. State-level demands were then aggregated into respective regions based on the states within each region. Electricity, $H_2$, and natural gas supply-balance constraints are modeled based on hourly regional resolution in MACRO, while liquid fuels are modeled as a systemwide supply-demand constraint as shown in Eq. S15 – S17 to reflect the low cost of liquid fuels transportation, using the "Total" column of this table. "CA" = California, "NW" = Northwest, "SW" = Southwest, "TX" = Texas, "NCEN" = North Central, "CEN" = Central, "SE" = Southeast, "MIDAT" = Mid-Atlantic, "NE" = Northeast.

| Energy Demand | CA | NW | SW | TX | NCEN | CEN | SE | MIDAT | NE | Total |
|---|---|---|---|---|---|---|---|---|---|---|
| Power (EJ/year) | 1.78 | 0.99 | 1.30 | 2.19 | 3.32 | 1.96 | 4.53 | 3.08 | 1.65 | 20.79 |
| $H_2$ (EJ/year) | 0.13 | 0.12 | 0.10 | 0.34 | 0.45 | 0.74 | 0.47 | 0.23 | 0.06 | 2.64 |
| Natural Gas (EJ/year) | 0.43 | 0.33 | 0.35 | 0.52 | 1.36 | 0.57 | 0.75 | 0.92 | 0.66 | 5.89 |
| Gasoline (EJ/year) | 0.63 | 0.26 | 0.38 | 0.49 | 0.84 | 0.50 | 1.37 | 0.86 | 0.49 | 5.84 |
| Jet Fuel (EJ/year) | 0.50 | 0.15 | 0.27 | 0.26 | 0.28 | 0.11 | 0.55 | 0.30 | 0.31 | 2.73 |
| Diesel (EJ/year) | 0.25 | 0.15 | 0.19 | 0.34 | 0.47 | 0.38 | 0.64 | 0.41 | 0.26 | 3.10 |

### S3.2. Power and $H_2$ sector modeling assumptions

Data assumptions for power and $H_2$ sectors are mainly based on those from a previous work on net-zero power-$H_2$ coupled system in similar geographical regions [6], with costs updated to 2022 USD and greenfield power generation technology cost and performance data updated to NREL Annual Technology Baseline (ATB) 2024 [7].

Table S13. Existing capacity and cost parameters in 2022 dollars for network expansion for power network transmission lines obtained from the U.S. Environmental Protection Agency (EPA) version of the Integrated Planning Model (IPM) [8].

| Network Lines | Transmission Path | Distance (Miles) | Existing Capacity (MW) | Line Reinforcement Cost ($/MW-mile) | Annualized Line Reinforcement Cost ($/MW-mile/y) |
|---|---|---|---|---|---|
| 1 | CA to NW | 622 | 6,533 | 2,089 | 128 |
| 2 | CA to SW | 490 | 11,964 | 2,090 | 128 |
| 3 | NW to SW | 577 | 4,530 | 1,286 | 79 |
| 4 | SW to CEN | 732 | 610 | 1,476 | 91 |
| 5 | TX to CEN | 494 | 2,525 | 1,476 | 91 |
| 6 | NCEN to CEN | 485 | 9,851 | 1,262 | 77 |
| 7 | NCEN to SE | 890 | 3,745 | 928 | 57 |
| 8 | NCEN to MIDAT | 751 | 9,083 | 1,262 | 77 |
| 9 | CEN to SE | 780 | 4,872 | 1,333 | 82 |
| 10 | SE to MIDAT | 491 | 5,552 | 1,333 | 82 |
| 11 | MIDAT to NE | 474 | 1,915 | 2,143 | 132 |

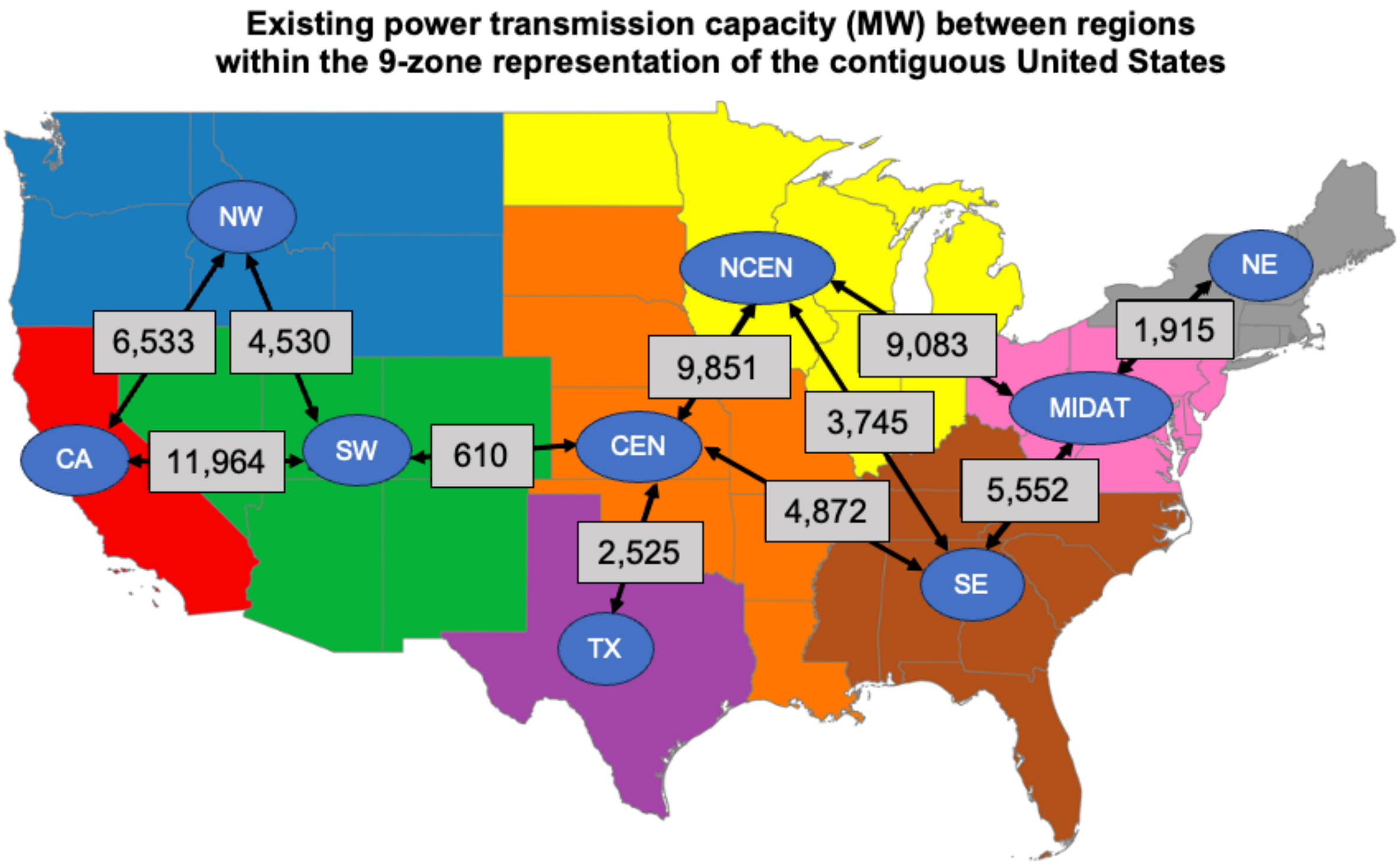

Figure S17. Existing power transmission capacity (MW) within the nine representative regions of the contiguous U.S. based on the Open Energy Outlook 2022 [9], with aggregated initial power transmission capacity between regions obtained from the U.S. Environmental Protection Agency (EPA) version of the Integrated Planning Model (IPM) [8]. "CA" = California, "NW" = Northwest, "SW" = Southwest, "TX" = Texas, "NCEN" = North Central, "CEN" = Central, "SE" = Southeast, "MIDAT" = Mid-Atlantic, "NE" = Northeast.

Table S14. Greenfield power generation technology cost and performance parameters. Data corresponds to 2045 costs (moderate) reported by the NREL Annual Technology Baseline 2024 in 2022 dollars[7], with the assumption that 2050 energy system would consist of technologies built five years earlier. A discount rate of 4.5% is used to annualize investment costs according to assumed technology lifetime and costs are converted to 2022 dollars. CC = Combined cycle, CT = Combustion turbine, CC = $CO_2$ capture and storage.

| Technology | Lifetime (year) | Investment cost | | Annualized CAPEX | | Fixed operation and maintenance (FOM) cost | | Variable operation and maintenance (VOM) cost ($/MWh) | Heat Rate (MMBtu/MWh) |
|---|---|---|---|---|---|---|---|---|---|
| | | Power ($/kW) | Energy ($/kWh) | Power ($/kW/year) | Energy ($/kWh/year) | Power ($/kW/year) | Energy ($/kWh/year) | | |
| Natural Gas (CC) | 30 | 1014 | - | 62 | - | 28 | - | 1.85 | 6.16 |
| Natural Gas (CC-CCS) | 30 | 1,792 | - | 110 | - | 48 | - | 3.76 | 6.88 |
| Natural Gas (CT) | 30 | 908 | - | 56 | - | 23 | - | 6.94 | 9.72 |
| Nuclear | 40 | 4,250 | - | 231 | - | 175 | - | 2.80 | 10.50 |
| Solar | 30 | 628 | - | 39 | - | 14 | - | - | - |
| Land Based Wind | 30 | 1,021 | - | 63 | - | 26 | - | - | - |
| Offshore Wind | 30 | 2,823 | - | 173 | - | 57 | - | - | - |
| Battery | 15 | 283 | 179 | 26 | 17 | 7 | 4 | - | - |

Table S15. $H_2$ production technology cost and performance parameters. Cost and performance parameters for electrolyzer obtained from IEA ("Long term" scenario) [10], data for natural gas reforming technologies obtained from NETL techno-economic analysis study [11], and data for tank storage obtained from Papadias & Aluwalia [12]. Feedwater cost for electrolyzers assumed to be negligible compared to the cost of electricity. The capital cost for $H_2$ storage in terms of rate of production and electricity input are associated with charging components. SMR = Steam methane reforming, ATR = Autothermal reforming, CCS = $CO_2$ capture and storage. Units of $/MWH$_2$ are converted based on higher heating value of $H_2$. A discount rate of 4.5% is used to annualize investment costs and costs are converted to 2022 dollars.

| Technology | Lifetime (year) | Investment cost | | Annualized CAPEX | | Fixed operation and maintenance (FOM) cost | | Variable operation and maintenance (VOM) cost ($/MWh $H_2$) | Electricity input (MWh/ MWh $H_2$) | Natural gas input (MMBtu/ MWh $H_2$) |
|---|---|---|---|---|---|---|---|---|---|---|
| | | $H_2$ production ($/MW $H_2$) | Energy ($/MWh $H_2$) | $H_2$ production ($/MW $H_2$/year) | Energy ($/MWh $H_2$/year) | $H_2$ production ($/MW $H_2$/year) | Energy ($/MWh $H_2$/year) | | | |
| Electrolyzer | 20 | 596,881 | - | 45,886 | - | 1,175 | - | - | 1.14 | - |
| $H_2$ Storage (tank) | 30 | 59,240 | 16,065 | 3,637 | 986 | - | 32 | - | 0.01 | - |
| SMR-CCS | 25 | 1,194,764 | - | 80,574 | - | 33,260 | - | 2.90 | 0.05 | 4.72 |
| ATR-CCS | 25 | 944,296 | - | 63,682 | - | 25,798 | - | 2.13 | 0.10 | 4.44 |

Table S16. Power generation $CO_2$ emissions and $CO_2$ captured based on a natural gas fuel emission factor of 0.054 t$CO_2$ /MMBtu [4].

| Technology | Fuel combustion emissions (t$CO_2$ /MWh) | $CO_2$ capture rate (%) | $CO_2$ captured (t$CO_2$ /MWh) |
|---|---|---|---|
| Natural Gas (CC) | 0.33 | - | - |
| Natural Gas (CC CCS) | 0.01 | 97.0 | 0.36 |
| Natural Gas (CT) | 0.52 | - | - |

Table S17. Hydrogen production $CO_2$ emissions and $CO_2$ captured based on a natural gas fuel emission factor of 0.054 $tCO_2$ /MMBtu [4].

| Technology | Fuel combustion emissions ($tCO_2/tH_2$) | $CO_2$ capture rate (%) | $CO_2$ captured ($tCO_2/tH_2$) |
|---|---|---|---|
| SMR-CCS | 0.38 | 96.2 | 9.59 |
| ATR-CCS | 0.55 | 94.5 | 9.42 |

Table S18. Cost and input assumptions for biomass to electricity and biomass to $H_2$ production technologies, from Net-Zero America study assumptions and NREL techno-economic study respectively [5], [13]. "B" – Bio, "E" – electricity, "$H_2$" – hydrogen. "$CO_2$ capture rate (CCX)" = percentage of biomass carbon content captured, "biomass energy conversion" = percentage of biomass energy converted into electricity or $H_2$ based on higher heating value basis. Amount of carbon captured and bioenergy produced would differ across biomass types and regions according to their respective carbon and energy content as shown in Table S38 and Table S39. B-$H_2$ was not modeled with $CO_2$ capture in its original reference [13], and the approximation method in Section S2.2 was used to parameterize the B-$H_2$-CC99 variant used in this study. A minimum and maximum operation output of 40% - 90%, and 85% - 90% of built capacity is enforced for B-E and B-$H_2$ respectively. Cost assumptions are scaled according to per tonne biomass input basis and converted to 2022 dollars, a technology lifetime of 30 years and discount rate of 4.5% is used to annualize investment costs.

| Technology | Lifetime (year) | Investment cost ($/(tonne biomass/h)) | Annualized Investment cost ($/(tonne biomass/h)/y) | Fixed operation and maintenance (FOM) cost ($/(tonne biomass/h)/y) | Variable operation and maintenance (VOM) cost ($/tonne biomass) | $CO_2$ capture rate (%) | Biomass energy conversion (% HHV) | Electricity input (MWh/tonne biomass) | Natural gas input (MMBtu/tonne biomass) |
|---|---|---|---|---|---|---|---|---|---|
| B-E-CC93 | 30 | 12,465,242 | 765,260 | 212,409 | 47.20 | 93.2 | 30.1 | - | - |
| B-$H_2$ | 30 | 2,814,068 | 172,760 | 190,982 | 13.35 | 0.0 | 58.6 | 0.13 | 0.46 |
| B-$H_2$-CC99 | 30 | 7,937,862 | 487,318 | 353,503 | 23.68 | 99 | 58.6 | 1.26 | 0.46 |

Table S19. Existing power generator capacity by resource in each region in 2050. The existing generation capacity in 2050 is estimated from capacity in 2021 after removing natural gas capacity that has exceeded their lifetime (40-50 years depending on natural gas plant type). Data obtained from EIA860 and Public Utility Data Liberation using PowerGenome [14]. Plant lifetimes of natural gas generators are based on NREL ReEDS model input assumption [15]. Based on the Inflation Reduction Act (IRA), second lifetime extensions were assumed for existing nuclear power generators [16]. CC = Combined Cycle. CT = Combustion Turbine.

| Technology | CA | NW | SW | TX | NCEN | CEN | SE | MIDAT | NE |
|---|---|---|---|---|---|---|---|---|---|
| Conventional Hydroelectric (GW) | 8.77 | 33.92 | 4.44 | 0.47 | 0.46 | 5.74 | 11.12 | 2.81 | 4.73 |
| Hydroelectric Pumped Storage (GW) | 3.94 | 0.31 | 0.78 | - | 2.00 | 0.94 | 6.26 | 5.24 | 3.21 |
| Natural Gas CC (GW) | 19.96 | 6.83 | 21.61 | 38.18 | 21.53 | 29.89 | 79.62 | 57.82 | 6.12 |
| Natural Gas CT (GW) | 8.32 | 1.82 | 6.53 | 4.92 | 16.80 | 13.46 | 25.62 | 15.21 | 1.41 |
| Nuclear (GW) | - | 1.16 | 4.00 | 5.02 | 14.35 | 7.26 | 35.86 | 20.21 | 5.50 |
| Onshore Wind (GW) | 6.06 | 10.85 | 5.28 | 24.93 | 22.12 | 27.69 | - | 3.23 | 3.65 |
| Small Hydroelectric (GW) | 0.34 | 0.73 | 0.15 | 0.02 | 0.48 | 0.14 | 0.25 | 0.26 | 0.83 |
| Solar Photovoltaic (GW) | 12.31 | 0.54 | 7.72 | 0.60 | 1.26 | 1.94 | 8.50 | 2.97 | 1.63 |

Table S20. Regional cost multipliers for investment cost of new power generation technologies according to region as obtained from the electricity market module in EIA's Annual Energy Outlook (AEO) 2021 [17]. CC = Combined Cycle. CT = Combustion Turbine. CCS = $CO_2$ capture and storage. The regional cost multipliers are applied to the baseline investment costs of each greenfield technology in Table S14 as model inputs for each region to account for regional variations of costs.

| Technology | CA | NW | SW | TX | NCEN | CEN | SE | MIDAT | NE |
|---|---|---|---|---|---|---|---|---|---|
| Natural Gas (CC) | 1.30 | 0.98 | 0.88 | 0.91 | 1.07 | 0.96 | 0.93 | 1.09 | 1.33 |
| Natural Gas (CC-CCS) | 1.07 | 0.95 | 0.86 | 0.95 | 1.02 | 0.96 | 0.96 | 1.01 | 1.10 |
| Natural Gas (CT) | 1.17 | 0.98 | 0.86 | 0.91 | 1.07 | 0.96 | 0.93 | 1.03 | 1.22 |
| Nuclear | 1.22 | 1.09 | 1.02 | 0.97 | 1.08 | 1.02 | 1.00 | 1.03 | 1.19 |
| Solar | 1.07 | 0.99 | 0.98 | 0.96 | 1.01 | 0.97 | 0.97 | 1.00 | 1.07 |
| Land Based Wind | 1.89 | 1.11 | 0.96 | 0.94 | 1.12 | 0.96 | 1.10 | 1.18 | 1.39 |
| Offshore Wind | 1.15 | 1.00 | - | - | - | - | - | - | 1.00 |
| Battery | 1.04 | 1.03 | 1.01 | 1.01 | 1.00 | 1.01 | 1.02 | 1.00 | 1.02 |

Table S21. Uranium costs in 2022 dollars for nuclear power resources from EIA's AEO 2023 [5]. Costs for fossil natural gas in Table S35.

| Fuel Cost | CA | NW | SW | TX | NCEN | CEN | SE | MIDAT | NE |
|---|---|---|---|---|---|---|---|---|---|
| Uranium ($/MMBtu) | 0.71 | 0.71 | 0.71 | 0.71 | 0.71 | 0.71 | 0.71 | 0.71 | 0.71 |

Table S22. Capacity reserve margin considered for each region based on the planning reserve margin constraint. The data sourced from IPM allocations based on electric reliability reports from NERC [8]. The planning reserve margin constraint enforces the need to procure "firm" generation capacity in excess of demand by the specified amount (i.e. reserve margin), where firm capacity contribution of each resource is calculated by either derating its installed capacity (in case of thermal plants) or available generation (in case of non-dispatchable resources like renewables and energy storage and flexible demand).

| | CA | NW | SW | TX | NCEN | CEN | SE | MIDAT | NE |
|---|---|---|---|---|---|---|---|---|---|
| Capacity Reserve Margin | 0.16 | 0.16 | 0.15 | 0.14 | 0.16 | 0.12 | 0.15 | 0.17 | 0.16 |

Table S23. Unit commitment parameters for resources. Parameters from nuclear power generators obtained from [19], [20], existing gas power generators from [14], new gas power generators from[21]. For NG-based H2 technologies, start cost and minimum up and down time not modeled, and ramping rates are assumed. Nameplate capacity contributes to the capacity reserve margin (CRM) constraint based on the respective derating factor of each technology, except for electrolyzers where its electricity consumption is the contributing factor. More information on the CRM can be found in the MACRO model documentations [22].

| Technology | Start cost ($/MW) | Min up time (h) | Min down time (h) | Max hourly ramping rate (% of nameplate capacity) | Min output (% of nameplate capacity) | Max output (% of nameplate capacity) | Derating factor for capacity reserve margin |
|---|---|---|---|---|---|---|---|
| Existing Natural Gas (CC) Power Generator | 111 | 6 | 6 | 64% | 10% - 60% | 100% | 0.93 |
| Existing Natural Gas (CT) Power Generator | 131 | 1 | 1 | 64% | 12% - 52% | 100% | 0.93 |
| Existing Nuclear Power Generator | 1,130 | 36 | 36 | 25% | 50% | 100% | 0.93 |
| New Natural Gas (CC) Power Generator | 69 | 4 | 4 | 100% | 30% | 100% | 0.93 |
| New Natural Gas (CC-CCS) Power Generator | 110 | 4 | 4 | 100% | 50% | 100% | 0.93 |
| New Natural Gas (CT) Power Generator | 158 | - | - | 100% | 25% | 100% | 0.93 |
| New Nuclear Power Generator | 1,130 | 36 | 36 | 25% | 30% | 100% | 0.93 |
| SMR-CCS H2 Resource | - | - | - | 50% | 85% | 90% | - |
| ATR-CCS Resource | - | - | - | 50% | 85% | 90% | - |
| Electrolyzer | - | - | - | 100% | 10% | 90% | 0.95 |
| Direct air capture technologies | | | | 100% | 85% | 90% | - |
| All fuels production technologies | - | - | - | 100% | 85% | 90% | - |

Table S24. $H_2$ pipeline data obtained from Hydrogen Delivery Scenario Analysis Model (HDSAM) v2 from Argonne National Laboratory for 100 km pipelines with 20 tonne $H_2$/h capacity [23]. HDSAM also provided the pipeline loss fraction, as well as the investment and operational costs, and energy requirements of 2 required compressors per 100 km that were included in the model inputs. This translates into an investment cost of $2228/MW-mile. Discount rate for annualization is 4.5%. Unless otherwise reported, all costs are in 2022 dollars. The total number of $H_2$ pipes (modeled as a continuous variable) along each line will be determined by the model based on the capacity requirement of $H_2$ transportation respectively. Candidate pipelines paths (i.e. regional source-sink pairs) and pipeline lengths are the same as power transmission network shown in Figure S17.

| Technology | Lifetime | Pipeline Investment Cost ($/MW $H_2$-mile) | Pipeline Annualized Investment Cost ($/MW $H_2$-mile/y) |
|---|---|---|---|
| $H_2$ Pipeline | 30 | 2,228 | 137 |

Table S25. Weighted regional average annual capacity factor and maximum available capacity expansion of greenfield variable renewable energy (VRE). VRE profiles were quantified using ZEPHYR (Zero-emissions Electricity system Planning with HourlY operational Resolution), a public repository where solar and wind site capacities and variability profiles were generated using historical weather data from 2011 from NREL NSRDB (National Solar Radiation Database) and WTK (WIND Toolkit) [24], [25], [26], to be consistent with the weather year in the Net-Zero America power demand profiles [5]. Together with the input of existing transmission line capacity data from NREL ReEDS (Regional Energy Development Model), the ZEPHYR model generates the available land area for solar and wind power development and determines the interconnection cost of each site (spur-line to substation and trunk-line reinforcement to urban areas) [15]. For each state, three bins (supply curves) containing hourly capacity factor, maximum capacity expansion, and interconnection transmission costs each for wind and solar resources were obtained. In ZEPHYR, state-level data was obtained by utilizing 4x4 km grids and translating hourly wind speed and irradiance into capacity factors using data from NSRDB and WTK respectively. National parks, urban areas, mountains, water bodies, and native land were excluded from the available area for greenfield VRE development, The maximum capacity expansion were obtained by multiplying the available area for VRE development by power generation density of 28 MW/km2 in wind and 1.6 MW/km2 in solar sites respectively [27]. For regions with a single state such as CA and TX, all 3 bins were utilized to represent a diverse choice of VRE availability. On the other hand, in regions with multiple states, the first bin of each state within that region was utilized (For example, NE region with 7 states modeled will have 7 bins of solar and wind resources each). The interconnection transmission cost of each bin was added onto the investment cost of greenfield wind and solar technology from NREL ATB in Table S14 as the model inputs. The weighted average capacity factor presented here are based on the 26 weeks model inputs after time domain reduction was performed on the VRE capacity factor profiles of 2011. The capacity factor in each hour of the 26 representative weeks was multiplied by their respective time weight, and weighted average across the total maximum capacity expansion of the bins utilized in each zone.

| Region | CA | NW | SW | TX | NCEN | CEN | SE | MIDAT | NE |
|---|---|---|---|---|---|---|---|---|---|
| Weighted Average Capacity Factor | | | | | | | | | |
| Solar | 0.31 | 0.25 | 0.31 | 0.30 | 0.22 | 0.26 | 0.26 | 0.22 | 0.20 |
| Land Based Wind | 0.26 | 0.39 | 0.39 | 0.46 | 0.46 | 0.47 | 0.36 | 0.40 | 0.43 |
| Maximum Capacity Expansion (GW) | | | | | | | | | |
| Solar | 5,953 | 8,913 | 9,404 | 18,172 | 10,908 | 9,021 | 8,503 | 4,328 | 2,738 |
| Land Based Wind | 340 | 945 | 736 | 10,38 | 811 | 864 | 671 | 338 | 206 |

### S3.3. $CO_2$ sector

Table S26. Cost and assumptions for DAC technologies obtained from NETL techno-economic study [28], [29]. NGCC = Natural gas combined cycle power plant, CCS = $CO_2$ capture and storage. DAC technologies with NGCC CCS are self-sufficient by producing electricity in a built-in NGCC unit and capturing natural gas combustion emissions using CCS. In the case of solvent DAC (with NGCC CCS), excess electricity is exported to grid. Investment and fixed operation and maintenance (FOM) costs are based on DAC input $CO_2$ capacity (tonne $CO_2$/h). A minimum and maximum operation output of 85% - 90% built capacity is enforced for all industrial processes. A discount rate of 4.5% is used to annualize investment costs and costs are converted to 2022 dollars.

| Technology | Lifetime (year) | Investment cost (\$/(t$CO_2$/h)) | Annualized CAPEX (\$/(t$CO_2$)/h)/y) | Fixed operation and maintenance (FOM) cost (\$/(t$CO_2$/h)/y) | Variable operation and maintenance (VOM) cost (\$/t$CO_2$) | Electricity input (MWh/t$CO_2$) | Natural gas input (MMBtu/t$CO_2$) |
|---|---|---|---|---|---|---|---|
| Solvent DAC (with NGCC CCS) | 30 | 15,579,017 | 956,420 | 422,596 | 64 | -0.13 | 10.40 |
| Sorbent DAC (Electricity) | 30 | 17,138,793 | 1,052,177 | 837,003 | 25 | 4.38 | - |

Table S27. $CO_2$ emissions and $CO_2$ captured for DAC with built-in NGCC CCS based on a natural gas fuel emission factor of 0.0536 t$CO_2$ /MMBtu [4]. Deployment of DAC with NGCC CCS will lead to additional $CO_2$ being stored due to $CO_2$ capture of NGCC emissions.

| Technology | NGCC combustion emissions (t$CO_2$ /t$CO_2$ captured by DAC) | NGCC $CO_2$ capture rate (%) | Total $CO_2$ captured (t$CO_2$ /t$CO_2$ net removal by DAC) |
|---|---|---|---|
| Solvent DAC (with NGCC CCS) | 0.01 | 99 | 1.55 |

Table S28. $CO_2$ pipeline cost assumptions obtained from 2019 National Academies report on negative emissions technologies [30]. The report also included the pipeline loss fraction, as well as operational costs and energy requirement of 3 required pumps per 10 miles that were included in the model inputs. We assumed that investment cost of pumps is negligible. The total number of $CO_2$ pipes (modeled as a continuous variable) along each line will be determined by the model based on the capacity requirement of $CO_2$ transportation respectively. Discount rate for annualization is 4.5%. Unless otherwise reported, all costs are in 2022 dollars. Candidate pipelines paths (i.e. regional source-sink pairs) and pipeline lengths are the same as the power transmission network shown in Figure S17.

| Technology | $CO_2$ throughput (Mt CO2/y) | Lifetime | Pipeline Investment Cost ($/mile) | Pipeline Annualized Investment Cost ($/mile/y) |
|---|---|---|---|---|
| $CO_2$ Pipeline | 10 | 30 | 2,975,000 | 182,640 |

Table S29. Total number of $CO_2$ storage sites, total storage availability according to high and low sequestration scenarios as described in Figure 3A of the main text, weighted average maximum $CO_2$ injection rate, and weighted average cost of $CO_2$ injection (in 2022 dollars) in each region as shown in Figure S17. $CO_2$ geological sequestration sites in each region are obtained from NREL ReEDS (Regional Energy Development Model) [15], and represented as individual sites in each region in the MACRO model, creating a supply curve in MACRO where the cheapest sites would first be utilized, as shown in Figure S18.

| | CA | NW | SW | TX | NCEN | CEN | SE | MIDAT | NE |
|---|---|---|---|---|---|---|---|---|---|
| Number of modeled $CO_2$ storage sites in region | 19 | 37 | 39 | 24 | 15 | 9 | 11 | 1 | 0 |
| Total $CO_2$ storage availability – High sequestration (HS) scenarios (Mt $CO_2$/y) | 133.7 | 121.5 | 244.6 | 163.9 | 62.4 | 99.1 | 40.4 | 0.4 | 0 |
| Total $CO_2$ storage availability – Low sequestration (LS) scenarios (Mt $CO_2$/y) | 66.9 | 60.7 | 122.3 | 81.9 | 31.2 | 50.0 | 20.2 | 0.2 | 0 |
| Wt. average maximum $CO_2$ Injection Rate (t$CO_2$/h) | 84,089 | 25,052 | 64,450 | 41,773 | 63,993 | 84,126 | 27,980 | 1,155 | 0 |
| Wt. average cost of $CO_2$ Injection ($/t$CO_2$) | 7.35 | 11.47 | 7.76 | 6.72 | 8.19 | 7.69 | 7.57 | 7.9 | 0 |

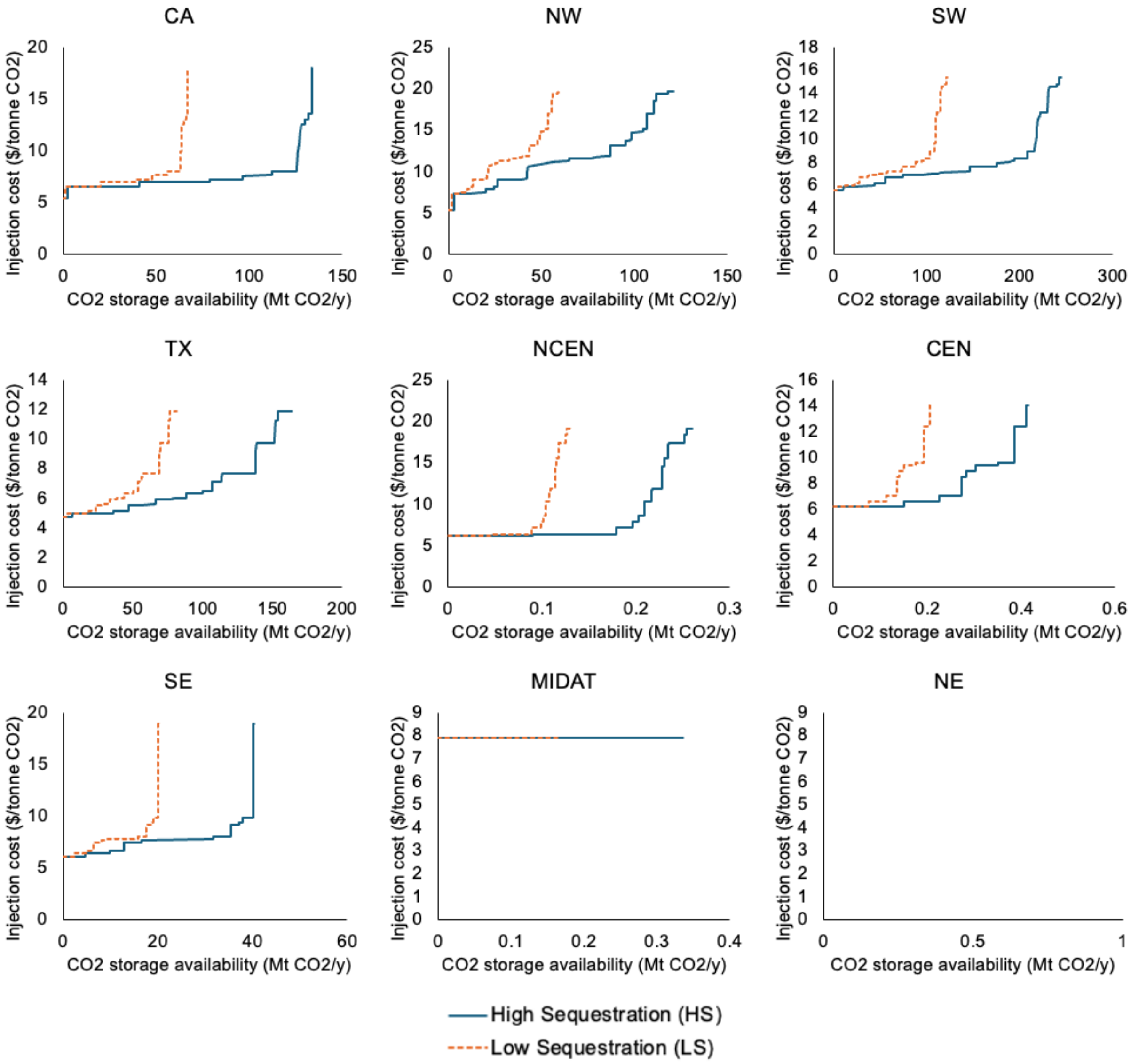

Figure S18. $CO_2$ sequestration supply curves according to regional $CO_2$ storage availability (Mt $CO_2$/y) and injection cost ($/tonne $CO_2$) using data from NREL ReEDS as described in Table S29 [15], for high sequestration (HS) and low sequestration (LS) scenarios as described in Figure 3A of the main text.

### S3.4. Liquid fuels sector

Table S30. Systemwide cost in 2022 dollars of fossil liquid fuels in 2050 based on EIA's AEO 2023 reference scenario [18].

| | Gasoline | Jet Fuel | Diesel |
|---|---|---|---|
| Fossil fuel cost ($/MMBtu) | 27.83 | 22.18 | 28.53 |

Table S31. Carbon content for fossil liquid fuels obtained from the GREET 2023 database [4].

| | Gasoline | Jet Fuel | Diesel | NG |
|---|---|---|---|---|
| Fossil fuel carbon content ($tCO_2$/MMBtu) | 0.0715 | 0.0722 | 0.0731 | 0.0536 |

Table S32. Lower and upper bounds on the fossil liquid supply distribution according to product yield of fossil gasoline, jet fuel, and diesel from U.S. refineries from 2009 – 2023 [31], used in Eq. S18 – S19 in Section S5.1 to represent realistic distributions of fossil liquid fuels in MACRO.

| | Jet Fuel/Gasoline (Min) | Diesel/Gasoline (Min) | Jet Fuel/Gasoline (Max) | Diesel/Gasoline (Max) |
|---|---|---|---|---|
| Fossil refinery constraint | 0.199 | 0.566 | 0.276 | 0.750 |

Table S33. Cost and modeling input assumptions for synthetic liquid fuel production technologies from $H_2$ and captured $CO_2$ obtained from Zang et al. as described in Section 2.2 and Figure 2B of the main text [32]. Processes are labeled as ("S" – Synthetic fuel), the dominant fuel product ( "J" – jet fuel), $CO_2$ capture rate ("CCX") if applicable, with "R" indicating the captured $CO_2$ is recycled back to the input $CO_2$ feed, effectively a high carbon conversion variant of the original process. "$CO_2$ capture rate" = rate of capture of carbon content that is not in fuels product (see Eq. S11 – S12 in Section S5.1: synthetic fuel carbon balance modeling in MACRO). Individual fuel production output is indicated (see Eq. S5 – S7 in Section S5.1: synthetic fuel energy balance modeling in MACRO). The synthetic fuel process is modeled without $CO_2$ capture in their original reference, and the approximation method elaborated in Section S2.2 was used to parameterize $CO_2$ capture variants of S-J-CC99 and S-J-CC99-R used in this study by estimating the additional costs and energy associated with capturing emissions from various flue gas streams. Input parameters might differ slightly from original references due to harmonization of energy and carbon balance for standardization for MACRO inputs as described in Section S2.3. A minimum and maximum operation output of 85% - 90% built capacity is enforced for all industrial processes. Cost assumptions are scaled according to per tonne biomass input basis and converted to 2022 dollars, a technology lifetime of 30 years and discount rate of 4.5% is used to annualize investment costs. Carbon balances of the processes are shown in Figure S28 in Section S4.

| Technology | Investment cost ($/(tonne $CO_2$ input/h)) | Annualized Investment cost ($/(tonne $CO_2$ input /h)/y) | Fixed operation and maintenance (FOM) cost ($/(tonne $CO_2$ input /h)/y) | Variable operation and maintenance (VOM) cost ($/$CO_2$ input biomass) | Electricity input (MWh/tonne $CO_2$ input) | $H_2$ input (MWh/tonne $CO_2$ input) | Gasoline output (MMbtu/ tonne $CO_2$ input) | Jet fuel output (MMbtu/ tonne $CO_2$ input) | Diesel output (MMbtu/ tonne $CO_2$ input) | $CO_2$ capture rate (%) |
|---|---|---|---|---|---|---|---|---|---|---|
| S-J | 4,522,542 | 277,646 | 240,597 | 9.66 | 0.04 | 3.68 | 1.69 | 3.06 | 1.79 | 0.0 |
| S-J-CC99 | 5,971,513 | 366,600 | 286,557 | 12.58 | 0.64 | 3.68 | 1.69 | 3.06 | 1.79 | 99.0 |
| S-J-CC99-R | 12,500,587 | 767,430 | 599,869 | 26.33 | 1.34 | 7.70 | 3.54 | 6.40 | 3.75 | 0.0 |

Table S34. Cost and modeling assumptions for biofuel technologies obtained from the literature as described in Section 2.2 and Figure 2B of the main text. Processes are labeled as ("B" - Biofuel), the dominant fuel product ("G" – gasoline [33], "J" – jet fuel [34], "D" – diesel [35]), $CO_2$ capture rate ("CCX") if applicable. "$CO_2$ capture rate" = rate of capture of biomass carbon content that is not in fuels product (see Eq. S13 – S14 in Section S5.1: biofuel carbon balance modeling in MACRO). "Biomass energy converted to fuels product" = the percentage of biomass energy converted into biofuels based on higher heating value basis with individual fuel shares indicated in the adjacent columns (see Eq. S8 – S10 in Section S5.1: biofuel energy balance modeling in MACRO). Amount of carbon captured and bioenergy produced would differ across biomass types and regions according to their respective carbon and energy content as shown in Table S38 and Table S39. Certain biofuel technologies (B-G and B-D) were not modeled with $CO_2$ capture in their original references, and the approximation method elaborated in Section S2.2 was used to parameterize $CO_2$ capture variants used in this study: B-G-CC31, B-G-CC99, B-D-CC53, B-D-CC99 by estimating the additional costs and energy associated with capturing emissions from various flue gas streams. In the case of B-J technologies, the $CO_2$ capture variants (B-J-CC75, B-J-CC84, B-J-CC99) were already parameterized in the original reference, and excess electricity is exported to grid as indicated by "Biomass energy converted to electricity" = the percentage of biomass energy converted to electricity on higher heating value basis (see Eq. S22: biofuel power balance in MACRO). Input parameters might differ slightly from original references due to harmonization of energy and carbon balance for MACRO inputs as described in Section S2.3. A minimum and maximum operation output

of 85% - 90% built capacity is enforced. Cost assumptions are scaled to per tonne biomass input basis and converted to 2022 dollars, a technology lifetime of 30 years and discount rate of 4.5% is used to annualize investment costs. Carbon balances of the processes are shown in Figure S25 to Figure S27 in Section S4.

| Technology | Investment cost ($/(tonne biomass/h)) | Annualized Investment cost ($/(tonne biomass/h)/y) | Fixed operation and maintenance (FOM) cost ($/(tonne biomass/h)/y) | Variable operation and maintenance (VOM) cost ($/tonne biomass) | Electricity input (MWh/tonne biomass) | $CO_2$ capture rate (%) | Biomass energy converted to fuels product (% HHV) | Gasoline (% fuels product) | Jet fuel (% fuels product) | Diesel (% fuels product) | Biomass energy converted to electricity (% HHV) |
|---|---|---|---|---|---|---|---|---|---|---|---|
| B-G | 5,799,717 | 356,054 | 283,503 | 22.04 | 0.00 | 0.0 | 37.90 | 100.00 | 0.00 | 0.00 | 0.00 |
| B-G-CC31 | 7,079,612 | 434,628 | 354,088 | 23.59 | 0.10 | 31.0 | 37.90 | 100.00 | 0.00 | 0.00 | 0.00 |
| B-G-CC99 | 9,618,367 | 590,486 | 434,614 | 28.74 | 1.16 | 99.3 | 37.90 | 100.00 | 0.00 | 0.00 | 0.00 |
| B-J-CC75 | 12,320,949 | 756,402 | 344,509 | 49.22 | 0.00 | 74.8 | 32.27 | 18.33 | 81.67 | 0.00 | 14.53 |
| B-J-CC84 | 12,774,703 | 784,259 | 351,017 | 50.29 | 0.00 | 83.9 | 42.47 | 18.30 | 81.70 | 0.00 | 4.63 |
| B-J-CC99 | 13,333,156 | 818,543 | 359,028 | 52.08 | 0.00 | 99.0 | 32.26 | 18.32 | 81.68 | 0.00 | 9.54 |
| B-D | 8,028,222 | 492,865 | 547,340 | 0.00 | 2.87 | 0.00 | 50.40 | 24.55 | 25.31 | 50.14 | 0.00 |
| B-D-CC53 | 8,918,107 | 547,496 | 596,415 | 1.06 | 2.94 | 52.8 | 50.40 | 24.55 | 25.31 | 50.14 | 0.00 |
| B-D-CC99 | 10,496,429 | 644,392 | 646,477 | 4.26 | 3.60 | 99.5 | 50.40 | 24.55 | 25.31 | 50.14 | 0.00 |

## S3.5. NG sector

Table S35. Regional cost in 2022 dollars of fossil natural gas based on EIA's AEO 2023 reference scenario, with modeled regions mapped to EIA's AEO fuel regions weighted by state-level demand from Net-Zero America[18].

| Region | CA | NW | SW | TX | NCEN | CEN | SE | MIDAT | NE |
|---|---|---|---|---|---|---|---|---|---|
| Fossil Natural Gas ($/MMBtu) | 8.85 | 8.11 | 6.89 | 4.89 | 6.09 | 5.62 | 6.59 | 6.33 | 7.61 |

Table S36. Cost and modeling assumptions for synthetic NG production is based on a TEA study of $CO_2$ methanation process [36], operation costs are assumed to be negligible in the methanation process, and $CO_2$ capture variant was not characterized due to high carbon conversion (96.6%). Carbon balance in Figure S30 in Section S4.

| Technology | Investment cost ($/(tonne $CO_2$ input/h)) | Annualized Investment cost ($/(tonne $CO_2$ input /h)/y) | Electricity input (MWh/tonne $CO_2$ input) | $H_2$ input (MWh/tonne $CO_2$ input) | NG output (MMbtu/ tonne $CO_2$ input) |
|---|---|---|---|---|---|
| S-NG | 1,485,773 | 91,214 | 0.29 | 7.17 | 18.01 |

Table S37. Cost and modeling input assumptions for bio-NG production technologies [37]. Processes are labeled as ("B" - Biofuel), the dominant fuel product ("NG" – natural gas), $CO_2$ capture rate ("CCX") if applicable. "$CO_2$ capture rate" = rate of capture of biomass carbon content that is not in fuels product. "Biomass energy converted to NG" = the percentage of biomass energy converted into NG based on higher heating value basis. Amount of carbon captured and bioenergy produced would differ across biomass types and regions according to their respective carbon and energy content as shown in Table S38 and Table S39. The original reference contains the B-NG and B-NG-CC40 (partial $CO_2$ capture) variant, and the approximation method elaborated in Section S2.2 was used to parameterize B-NG-CC99 by estimating the additional costs and energy associated with capturing emissions from the remaining flue gas stream in the B-NG-CC40 variant. Input parameters might differ slightly from original references due to harmonization of energy and carbon balance for standardization for MACRO inputs as described in Section S2.3. A minimum and maximum operation output of 85% - 90% built capacity is enforced for all industrial processes. Cost assumptions are scaled according to per tonne biomass input basis and converted to 2022 dollars, a technology lifetime of 30 years and discount rate of 4.5% is used to annualize investment costs. Carbon balances of the processes are shown in Figure S29 in Section S4.

| Technology | Investment cost ($/(tonne biomass/h)) | Annualized Investment cost ($/(tonne biomass/h)/y) | Fixed operation and maintenance (FOM) cost ($/(tonne biomass/h)/y) | Variable operation and maintenance (VOM) cost ($/tonne biomass) | Electricity input (MWh/tonne biomass) | $CO_2$ capture rate (%) | Biomass energy converted to NG (% HHV) |
|---|---|---|---|---|---|---|---|
| B-NG | 14,138,646 | 867,993 | 500,251 | 9.04 | 0.21 | 0.00 | 51.60 |
| B-NG-CC40 | 16,534,444 | 1,015,075 | 586,131 | 9.76 | 0.25 | 39.51 | 51.20 |
| B-NG-CC99 | 18,327,293 | 1,125,141 | 646,258 | 14.86 | 0.98 | 99.11 | 51.20 |

## S3.6. Biomass sector

Table S38. Total biomass availability weighted average costs (in 2022 dollars) and properties (energy and carbon content) for biomass types (herbaceous, woody, and agricultural residue) in each region as shown in Figure S17 for high biomass (HB) scenarios as described in Figure 3A of the main text. Biomass availability, cost, and energy content are quantified using projected county-level 2040 biomass availabilities from the high mature market scenario in the 2023 Billion Ton Study[38]. The carbon content of each biomass type in each region was estimated by mapping the county-level biomass resources to similar feedstocks primarily from the GREET 2023 database as shown in Table S40 [4]. Biomass of the same type and costs within the same region were aggregated together and modeled as individual regional-level resources in MACRO, effectively creating regional supply curves for each biomass type in MACRO as shown in Figure S19 to Figure S21, where the cheapest biomass would be first purchased. The weighted average energy and carbon content of biomass types are used as MACRO inputs to model the energy and carbon balance of bioenergy processes built for a specific biomass type of in a specific region (e.g. B-G-CC99 process consuming herbaceous biomass in NCEN region). according to Eq. S8 – S10 and Eq. S13 – S14 respectively.

| **High biomass availability (HB) scenarios** | **CA** | **NW** | **SW** | **TX** | **NCEN** | **CEN** | **SE** | **MIDAT** | **NE** |
|---|---|---|---|---|---|---|---|---|---|
| Total herbaceous biomass availability (Mt biomass /y) | 0.00 | 2.17 | 15.55 | 131.57 | 56.97 | 195.08 | 69.76 | 11.96 | 2.61 |
| Wt. avg. herbaceous biomass cost ($/t biomass) | 77.16 | 77.16 | 77.16 | 77.16 | 77.16 | 77.16 | 77.16 | 77.16 | 77.16 |
| Wt. avg. herbaceous biomass energy content (MMBtu/t biomass) | 18.19 | 18.23 | 18.11 | 18.15 | 18.34 | 18.23 | 18.20 | 18.30 | 18.37 |
| Wt. avg. herbaceous biomass carbon content ($tCO_2$/t biomass) | 1.72 | 1.73 | 1.71 | 1.71 | 1.74 | 1.72 | 1.72 | 1.74 | 1.75 |
| Total woody biomass availability (Mt/y) | 0.69 | 6.25 | 6.97 | 8.20 | 23.88 | 29.40 | 39.51 | 26.37 | 9.53 |
| Wt. average woody biomass cost ($/t) | 52.11 | 67.44 | 74.96 | 73.41 | 73.55 | 74.08 | 69.10 | 74.00 | 73.07 |
| Wt. avg. woody biomass energy content (MMBtu/t biomass) | 19.36 | 19.32 | 19.14 | 18.99 | 19.06 | 19.07 | 19.14 | 19.01 | 18.89 |
| Wt. avg. woody biomass carbon content ($tCO_2$/t biomass) | 1.84 | 1.82 | 1.80 | 1.82 | 1.80 | 1.80 | 1.82 | 1.81 | 1.83 |
| Total agricultural residue availability (Mt/y) | 4.89 | 6.30 | 2.42 | 4.01 | 111.61 | 50.03 | 7.63 | 6.19 | 1.79 |
| Wt. average agricultural residue cost ($/t) | 52.64 | 74.43 | 75.70 | 67.79 | 77.12 | 76.05 | 70.56 | 76.51 | 73.77 |
| Wt. avg. agricultural residue energy content (MMBtu/t biomass) | 18.29 | 16.72 | 17.03 | 16.59 | 17.30 | 17.17 | 17.23 | 17.28 | 17.48 |
| Wt. avg. agricultural residue carbon content ($tCO_2$/t biomass) | 1.72 | 1.66 | 1.69 | 1.63 | 1.71 | 1.70 | 1.69 | 1.71 | 1.72 |

Table S39. Total biomass availability weighted average costs (in 2022 dollars) and properties (energy and carbon content) for biomass types (herbaceous, woody, and agricultural residue) in each region as shown in Figure S17 for low biomass (LB) scenarios as described in Figure 3A of the main text. Biomass availability, cost, and energy content are quantified using projected county-level 2040 biomass availabilities from the low mature market scenario in the 2023 Billion Ton Study [38]. The carbon content of each biomass type in each region was estimated by mapping the county-level biomass resources to similar feedstocks primarily from the GREET 2023 database as shown in Table S40 [4]. Biomass of the same type and costs within the same region were aggregated together and modeled as individual regional-level resources in MACRO, effectively creating regional supply curves for each biomass type in MACRO as shown in Figure S22 to Figure S24, where the cheapest biomass would be first purchased. The weighted average energy and carbon content of biomass types are used as MACRO inputs to model the energy and carbon balance of bioenergy processes built for a specific biomass type of in a specific region according to Eq. S8 – S10 and Eq. S13 – S14 respectively.

| **Low biomass availability (LB) scenarios** | **CA** | **NW** | **SW** | **TX** | **NCEN** | **CEN** | **SE** | **MIDAT** | **NE** |
|---|---|---|---|---|---|---|---|---|---|
| Total herbaceous biomass availability (Mt biomass /y) | - | 1.11 | 7.39 | 82.94 | 14.45 | 105.02 | 38.82 | 7.04 | 0.88 |
| Wt. avg. herbaceous biomass cost ($/t biomass) | - | 77.16 | 77.16 | 77.16 | 77.16 | 77.16 | 77.16 | 77.16 | 77.16 |
| Wt. avg. herbaceous biomass energy content (MMBtu/t biomass) | - | 18.10 | 18.10 | 18.12 | 18.35 | 18.20 | 18.16 | 18.16 | 18.35 |
| Wt. avg. herbaceous biomass carbon content ($tCO_2$/t biomass) | - | 1.71 | 1.71 | 1.71 | 1.74 | 1.72 | 1.72 | 1.72 | 1.75 |
| Total woody biomass availability (Mt/y) | 0.69 | 3.65 | 2.36 | 3.72 | 11.82 | 12.38 | 28.72 | 15.92 | 8.64 |
| Wt. average woody biomass cost ($/t) | 52.11 | 60.50 | 70.68 | 68.91 | 69.88 | 69.86 | 66.07 | 71.93 | 72.64 |
| Wt. avg. woody biomass energy content (MMBtu/t biomass) | 19.36 | 19.5 | 19.3 | 19.11 | 19.06 | 19.16 | 19.18 | 18.99 | 18.87 |
| Wt. avg. woody biomass carbon content ($tCO_2$/t biomass) | 1.84 | 1.84 | 1.82 | 1.83 | 1.82 | 1.81 | 1.83 | 1.82 | 1.83 |
| Total agricultural residue availability (Mt/y) | 4.76 | 5.27 | 1.76 | 3.44 | 76.84 | 42.34 | 7.62 | 7.82 | 1.74 |
| Wt. average agricultural residue cost ($/t) | 51.95 | 73.90 | 75.15 | 66.25 | 77.10 | 75.85 | 70.56 | 76.65 | 73.69 |
| Wt. avg. agricultural residue energy content (MMBtu/t biomass) | 18.33 | 16.76 | 16.97 | 16.48 | 17.28 | 17.14 | 17.23 | 17.30 | 17.51 |
| Wt. avg. agricultural residue carbon content ($tCO_2$/t biomass) | 1.72 | 1.67 | 1.68 | 1.62 | 1.71 | 1.70 | 1.69 | 1.71 | 1.72 |

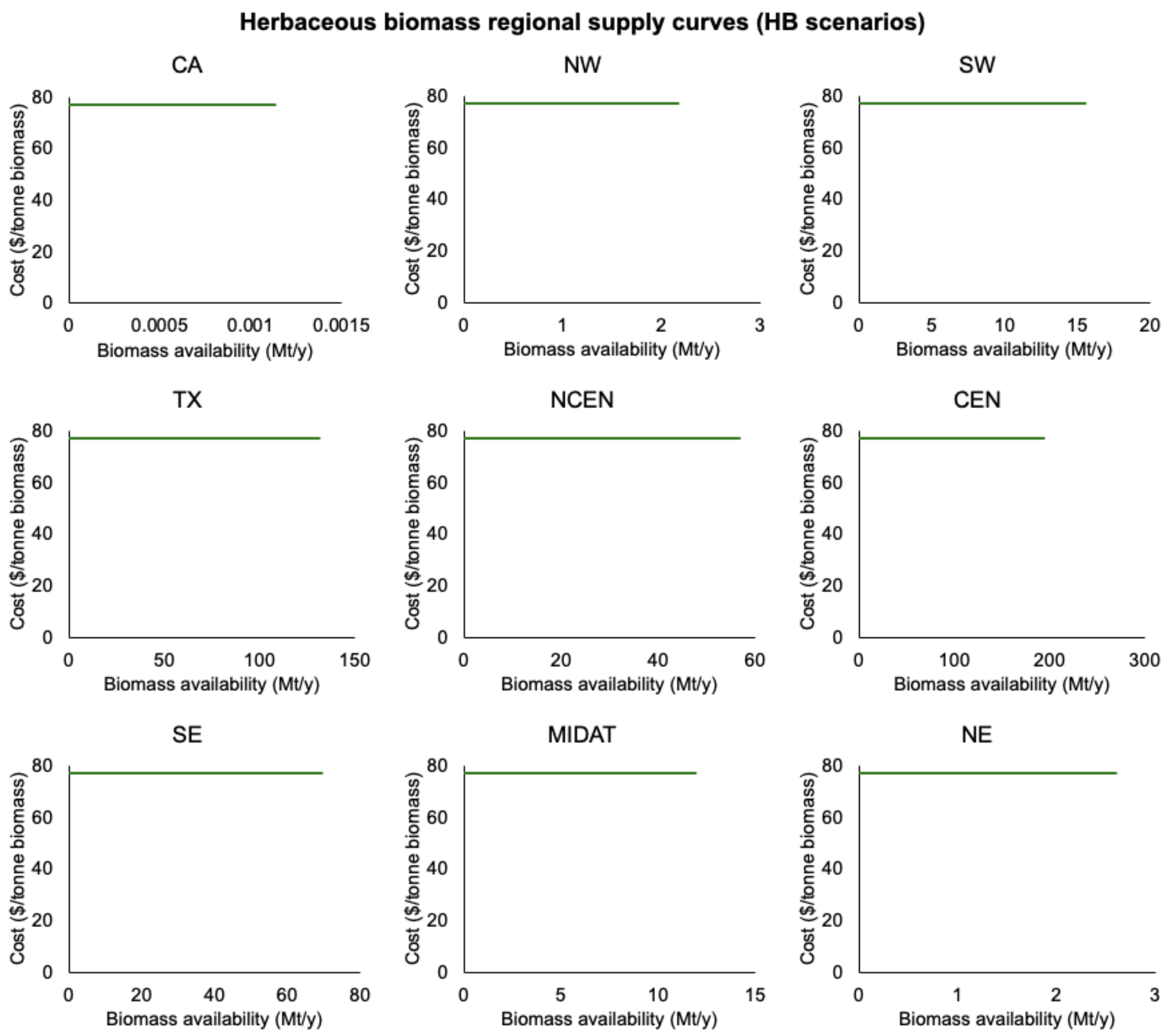

Figure S19. Supply curves for herbaceous biomass type according to biomass availability (Mt/y) and purchase cost ($/tonne biomass) as described in Table S38 for high biomass availability (HB) scenarios as described in Figure 3A of the main text, obtained from the 2023 Billion Ton Study [38].

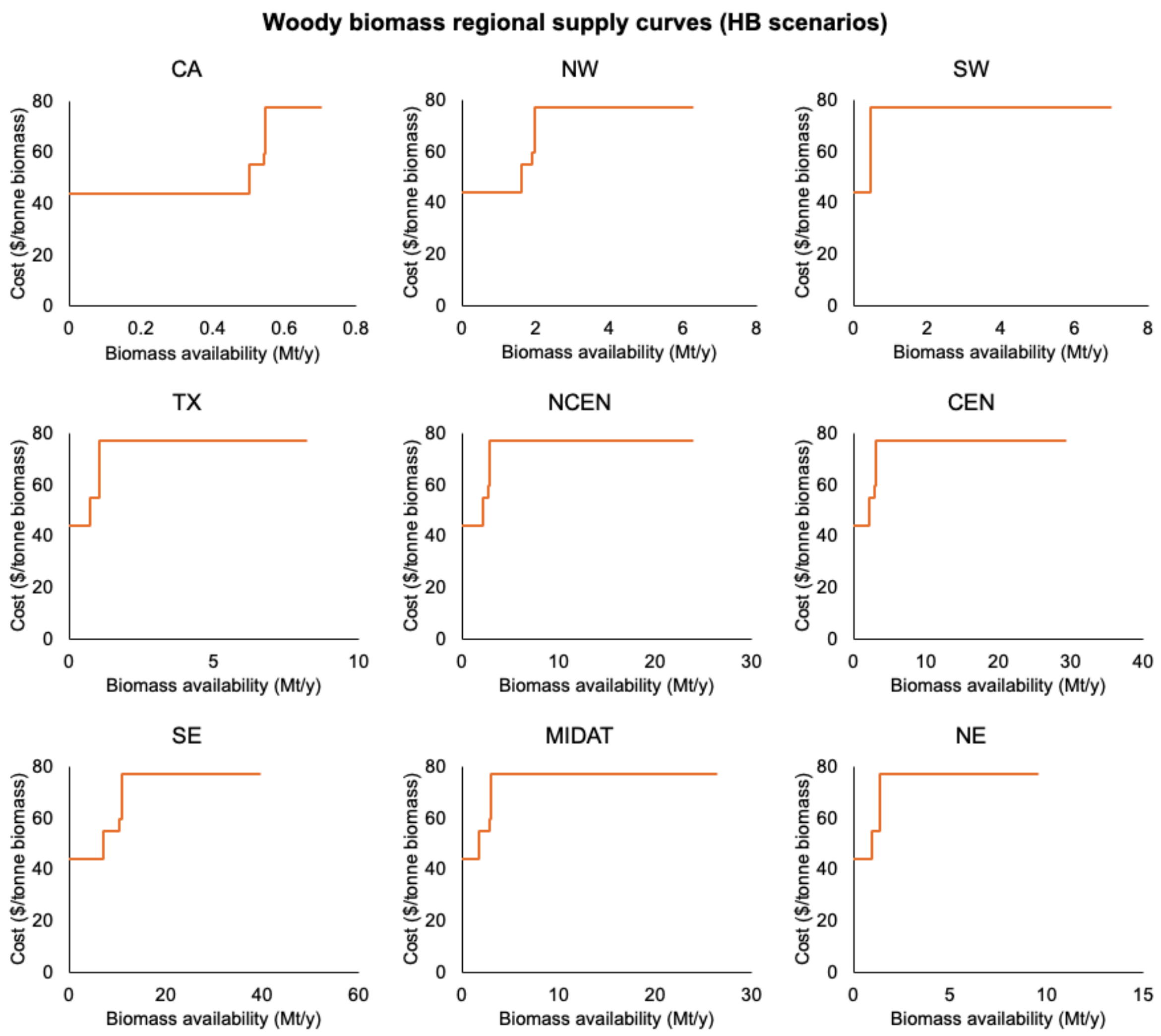

Figure S20. Supply curves for woody biomass type according to biomass availability (Mt/y) and purchase cost ($/tonne biomass) as described in Table S38 for high biomass availability (HB) scenarios as described in Figure 3A of the main text, obtained from the 2023 Billion Ton Study [38].

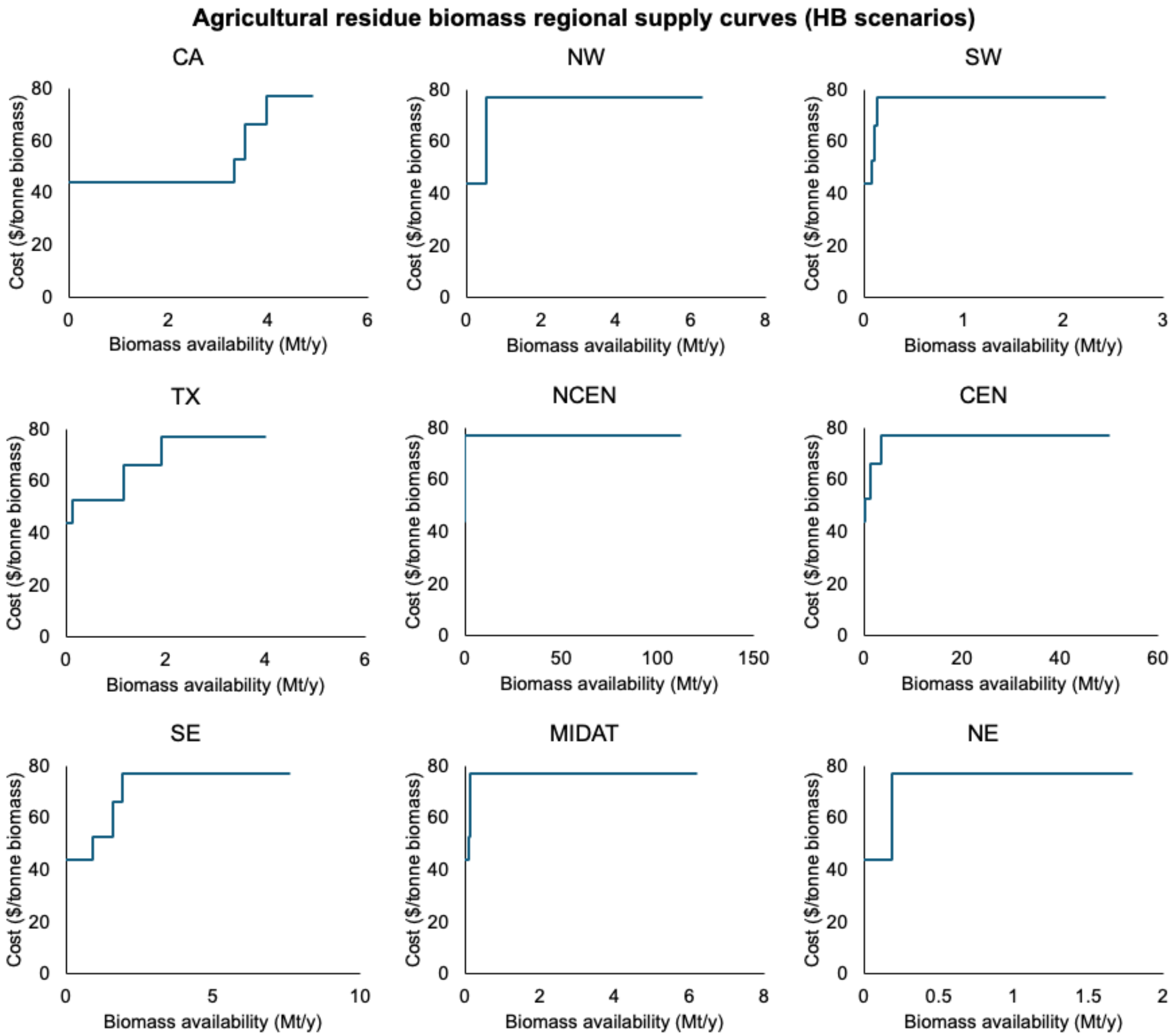

Figure S21. Supply curves for agricultural residue biomass type according to biomass availability (Mt/y) and purchase cost ($/tonne biomass) as described in Table S38 for high biomass availability (HB) scenarios as described in Figure 3A of the main text, obtained from the 2023 Billion Ton Study [38].

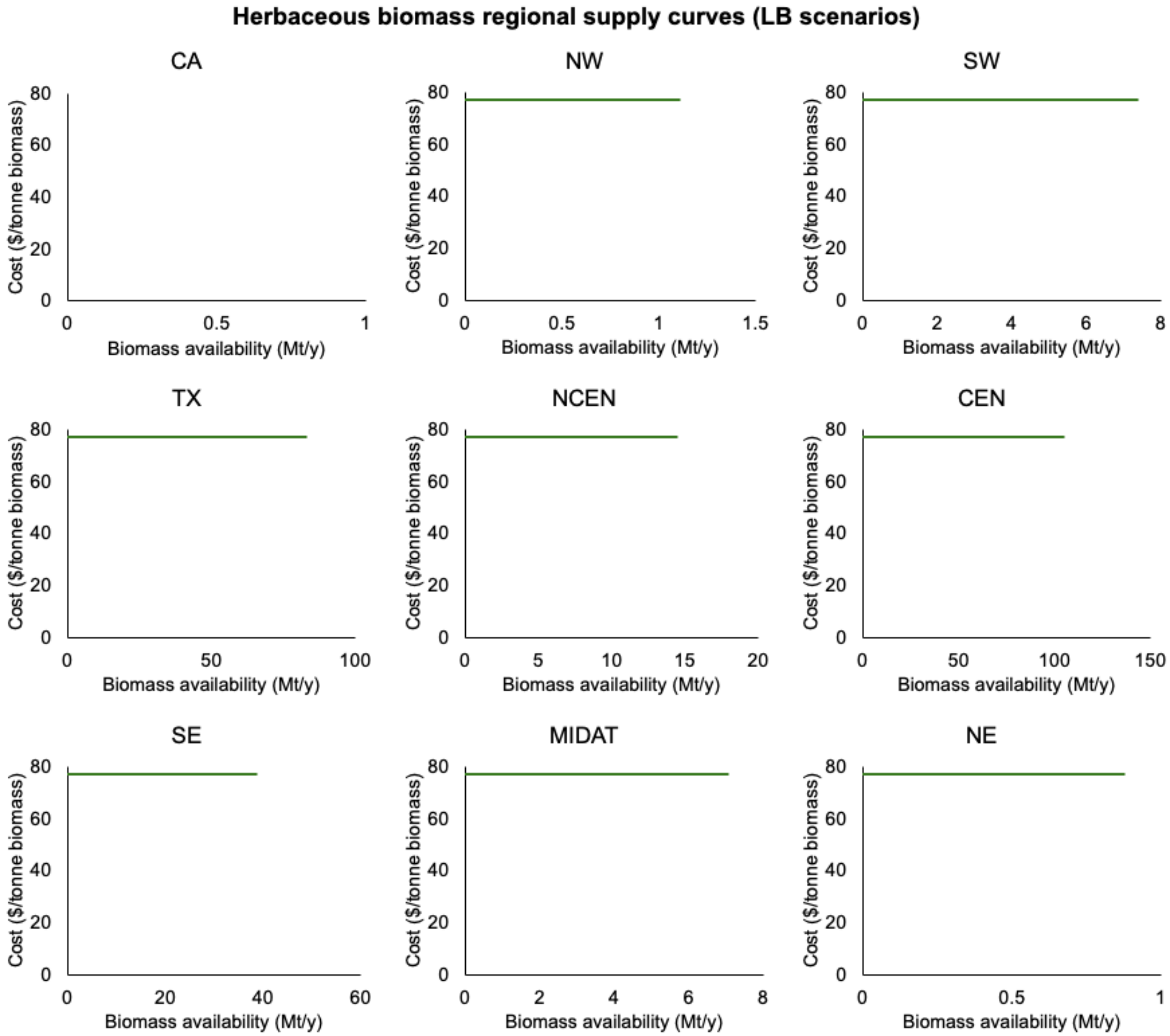

Figure S22. Supply curves for herbaceous biomass type according to biomass availability (Mt/y) and purchase cost ($/tonne biomass) as described in Table S39 for low biomass availability (LB) scenarios as described in Figure 3A of the main text, obtained from the 2023 Billion Ton Study [38].

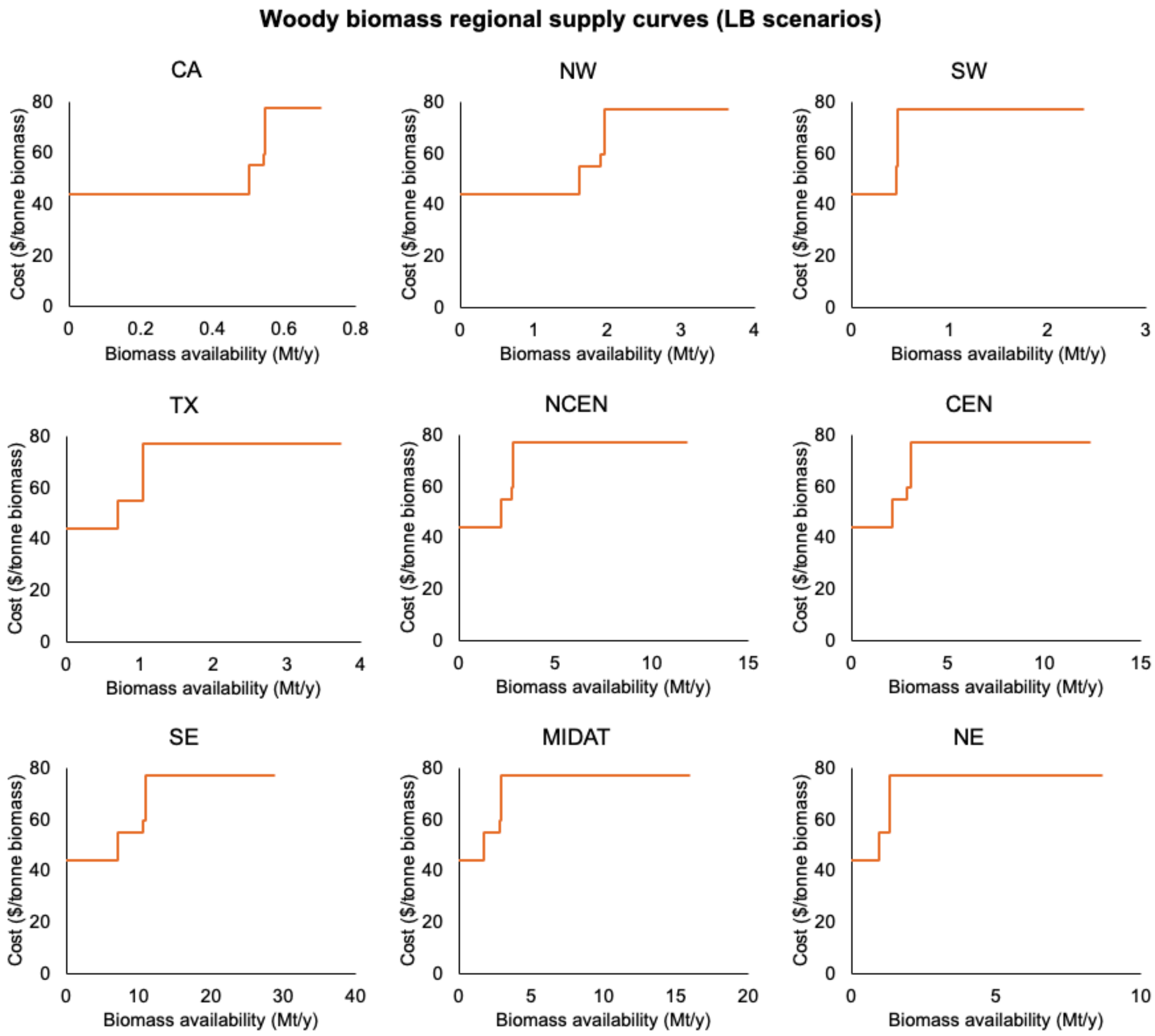

Figure S23. Supply curves for woody biomass type according to biomass availability (Mt/y) and purchase cost ($/tonne biomass) as described in Table S39 for low biomass availability (LB) scenarios as described in Figure 3A of the main text, obtained from the 2023 Billion Ton Study [38].

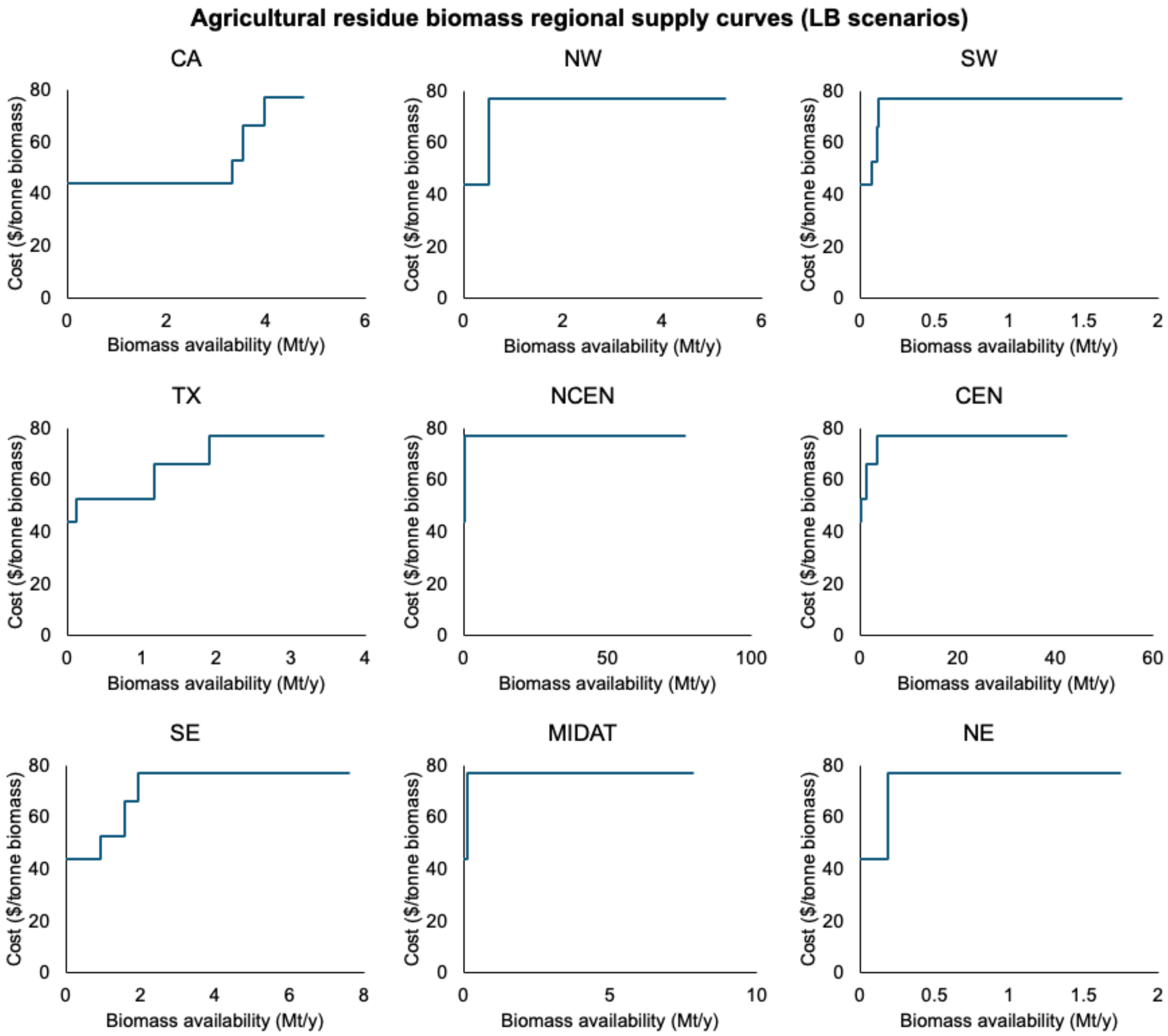

Figure S24. Supply curves for agricultural residue biomass type according to biomass availability (Mt/y) and purchase cost ($/tonne biomass) as described in Table S39 for low biomass availability (LB) scenarios as described in Figure 3A of the main text, obtained from the 2023 Billion Ton Study [38].

Table S40. Biomass carbon content (% wt.) according to species in various biomass types. County-level biomass data from the 2023 Billion-Ton Study contain information about these species-level biomass availabilities, which could be used to estimate regional weighted average carbon content of each biomass types for MACRO inputs. When possible, individual biomass species are mapped to an analogous feedstock in the GREET 2023 database to obtain their respective carbon contents as shown in the "Represented by" column. If no suitable representations are available in GREET 2023, values are taken from other cited literature in the "Reference" column.

| Biomass type | Individual biomass species | Represented by | Reference | Carbon % wt. |
|---|---|---|---|---|
| Herbaceous | Biomass sorghum | Sorghum | Bioenergy Feedstock Library [39] | 0.456 |
| | Miscanthus | Miscanthus | GREET 2023 [4] | 0.476 |
| | Energy cane | Energy cane | | 0.463 |
| | Switchgrass | Switchgrass | | 0.466 |
| Woody | Eucalyptus | Eucalyptus | | 0.492 |
| | Willow | Willow | | 0.487 |
| | Pine | Pine | | 0.501 |
| | Poplar | Poplar | | 0.501 |
| | Logging residual | Forest Residues | | 0.503 |
| | Forest processing waste | | | |
| | Fire reduction thinning | | | |
| | Other forest waste | | | |
| | Small diameter trees | Poplar | | 0.501 |
| Agricultural residue | Corn stover | Corn stover | | 0.467 |
| | Sorghum stubble | Forage sorghum bagasse | | 0.420 |
| | Pruning Residuals | Yard Trimming Waste | | 0.478 |
| | Wheat Straw | Wheat Straw | Bioenergy Feedstock Library [39] | 0.450 |
| | Oats Straw | | | |
| | Rice Straw | | | |
| | Cotton field residue | Cotton Stalk | Tao et al. [40] | 0.460 |
| | Cotten gin trash | Cotton gin | Subramani et al. [41] | 0.410 |
| | Rice hulls | Rice husk | AboDalam et al. [42] | 0.419 |

## S4. Carbon balance of biofuel and synthetic fuels technologies

This section shows carbon balances of biofuel and synthetic fuels production processes used in this study. For biofuel technologies, the carbon balances are calculated using biomass properties of 47.3% carbon content per biomass weight and 18.15 MMBtu/tonne energy content, which are the weighted average values from the high biomass (HB) scenario. The carbon content of fuel products are harmonized to those obtained from the GREET 2023 database shown in Table S31 as described in Section S2.3 [4], and might differ from carbon balances from their original reference TEAs.

Carbon conversion is not a technology input parameter in the MACRO model. Instead, biofuel production technologies are modeled using energy conversion efficiency and product distribution inputs as described in Section S5.1 (see Eq. S8 – S10) using energy conversion efficiency of individual biofuel technologies shown in Table S34. As such, the actual carbon conversion might differ according to biomass type consumed as shown in Table S38 and Table S39.

Dashed boxes in the carbon balance indicate $CO_2$ capture units estimated by applying the approximation method of matching the $CO_2$ concentrations of the flue gas stream(s) in each pathway to analogous industrial point sources $CO_2$ capture units as described in Section S2.2, and applying the corresponding capture rate to quantify the amount of $CO_2$ captured and emitted.

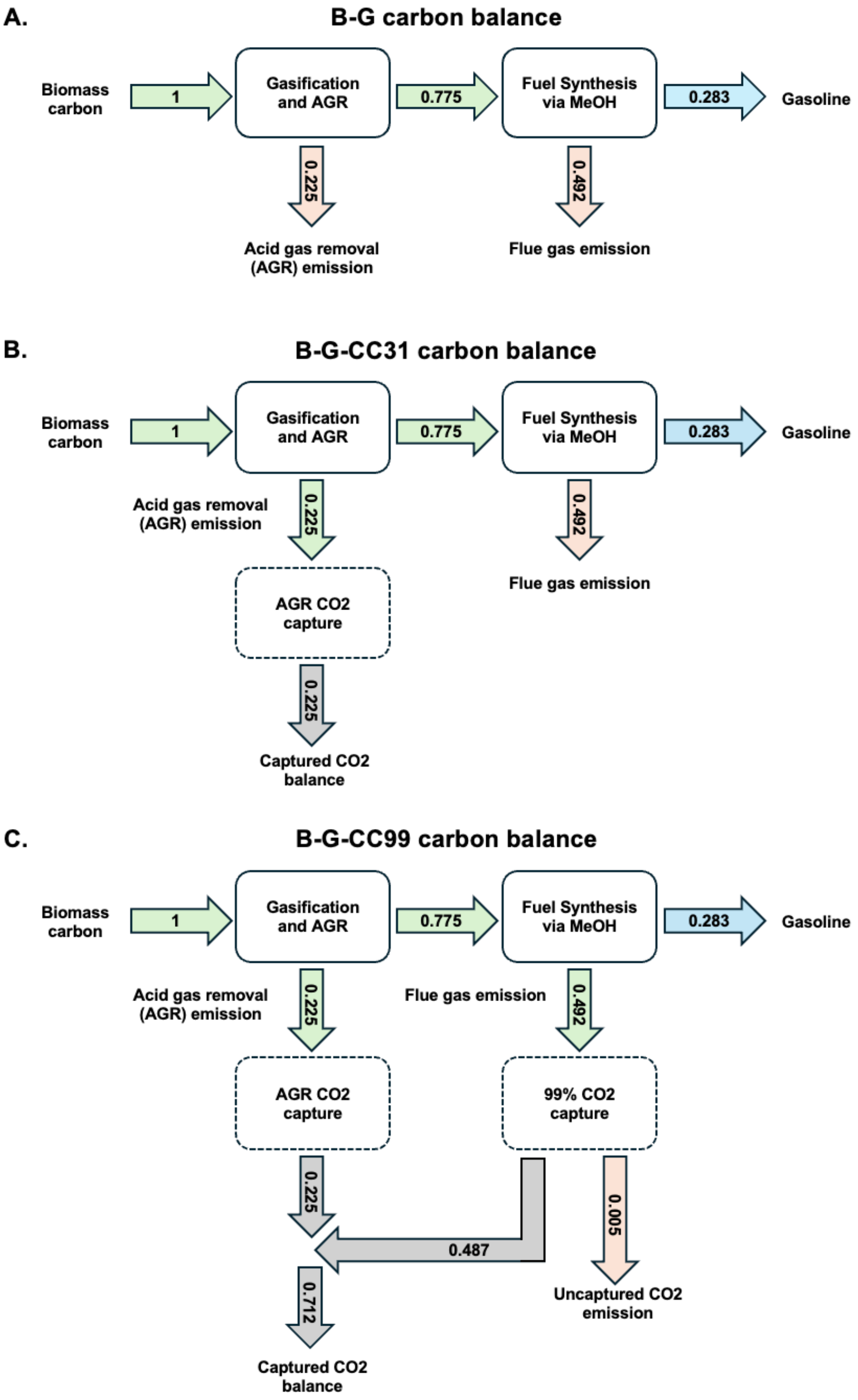

Figure S25. Carbon balance of B-G biofuel production technologies in Table S34, with box diagrams showing simplified process and carbon flows (normalized to % wt. basis). B-G is based on the SOT (state-of-technology) process configuration in the reference TEA by NREL for gasoline production from biomass gasification followed by methanol synthesis [33]. B-G-CC31 and B-G-CC99 are characterized by using the approximation method described in Section S2.2 by matching the $CO_2$ concentrations of various flue gas stream to analogous industrial point sources $CO_2$ capture units (represented by dashed boxes) from the original B-G configuration – The AGR $CO_2$ capture and 99% $CO_2$ capture units are based on ethanol and cement production $CO_2$ capture units respectively as shown in Table S11. Biomass properties of 47.3% carbon content per biomass weight and 18.15 MMBtu/tonne energy content were used to obtain this carbon balance.

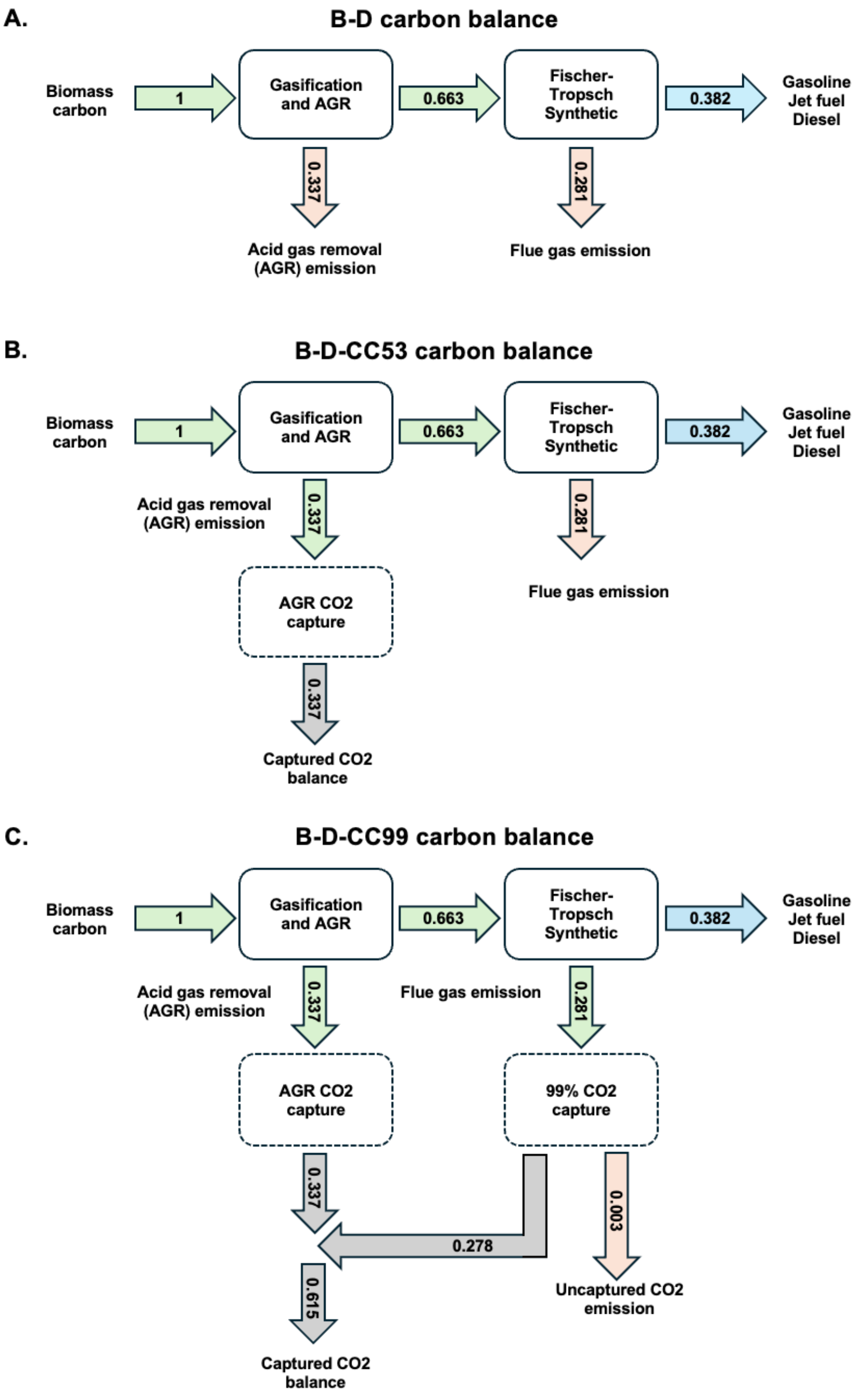

Figure S26. Carbon balance of B-D biofuel production technologies in Table S34, with box diagrams showing simplified process and carbon flows (normalized to % wt. basis). B-D is based on the CFB-FT (circulating fluidized bed Fischer-Tropsch) process configuration in the reference TEA by Dimitriou et al. for production of diesel (dominant product), jet fuel and gasoline (both side product) from biomass gasification followed by FT process [35]. B-D-CC53 and B-G-CC99 are characterized by using the approximation method described in Section S2.2 by matching the $CO_2$ concentrations of various flue gas stream to analogous industrial point sources $CO_2$ capture units (represented by dashed boxes) from the original B-D configuration – The AGR $CO_2$ capture and 99% $CO_2$ capture units are based on ethanol and cement production $CO_2$ capture units respectively as shown in Table S11. Biomass properties of 47.3% carbon content per biomass weight and 18.15 MMBtu/tonne energy content were used to obtain this carbon balance.

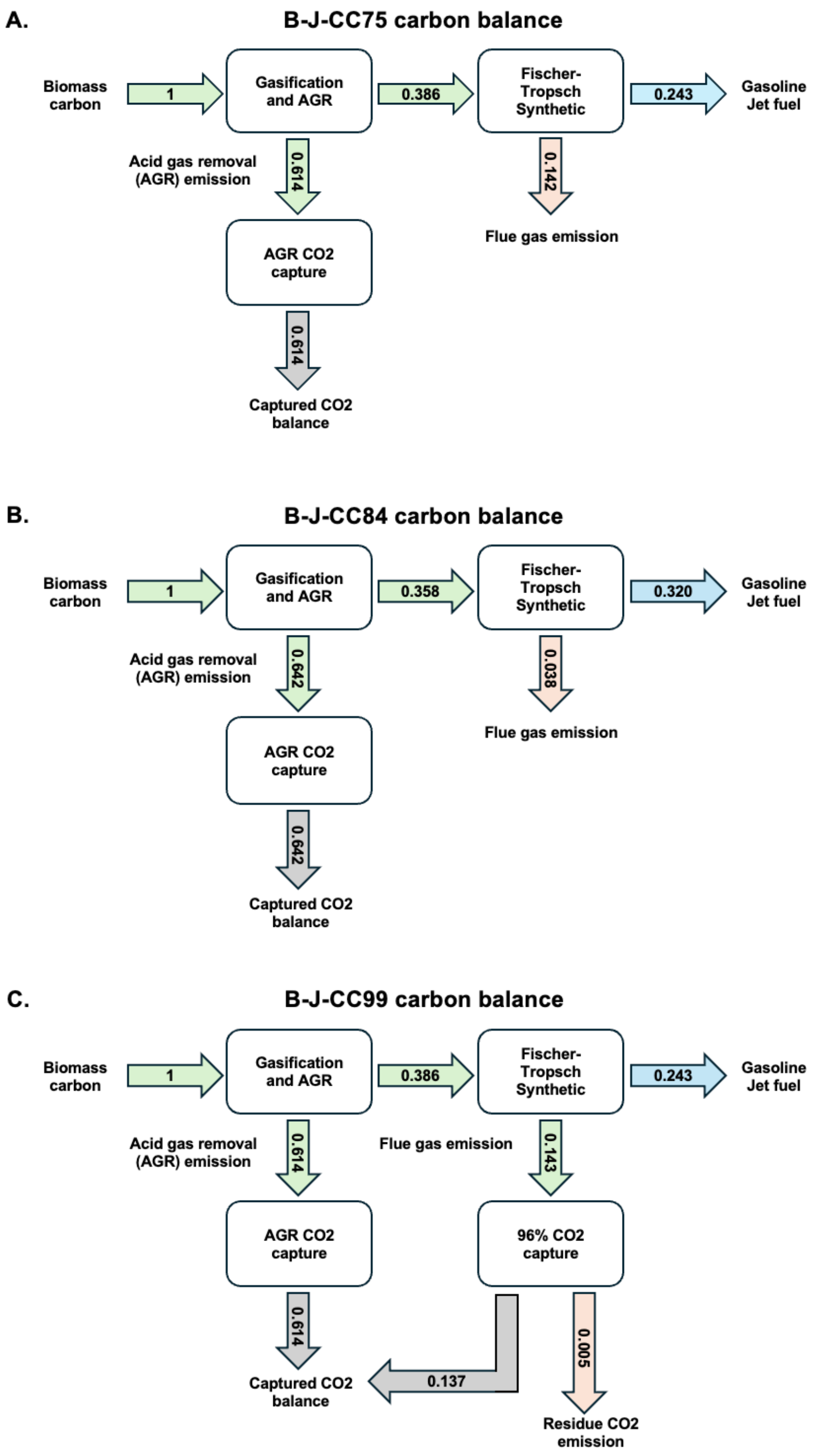

Figure S27. Carbon balance of B-J biofuel production technologies in Table S34, with box diagrams showing simplified process and carbon flows (normalized to % wt. basis). S-J-CC75, B-J-CC84 and B-J-CC99 are obtained from the original process configurations of OT-B, RC-B, and OTA-B in the reference TEA by Kreutz et al. for production of jet fuel (dominant product) and gasoline (side product) from biomass gasification followed by FT process [34]. Biomass properties of 47.3% carbon content per biomass weight and 18.15 MMBtu/tonne energy content were used to obtain this carbon balance

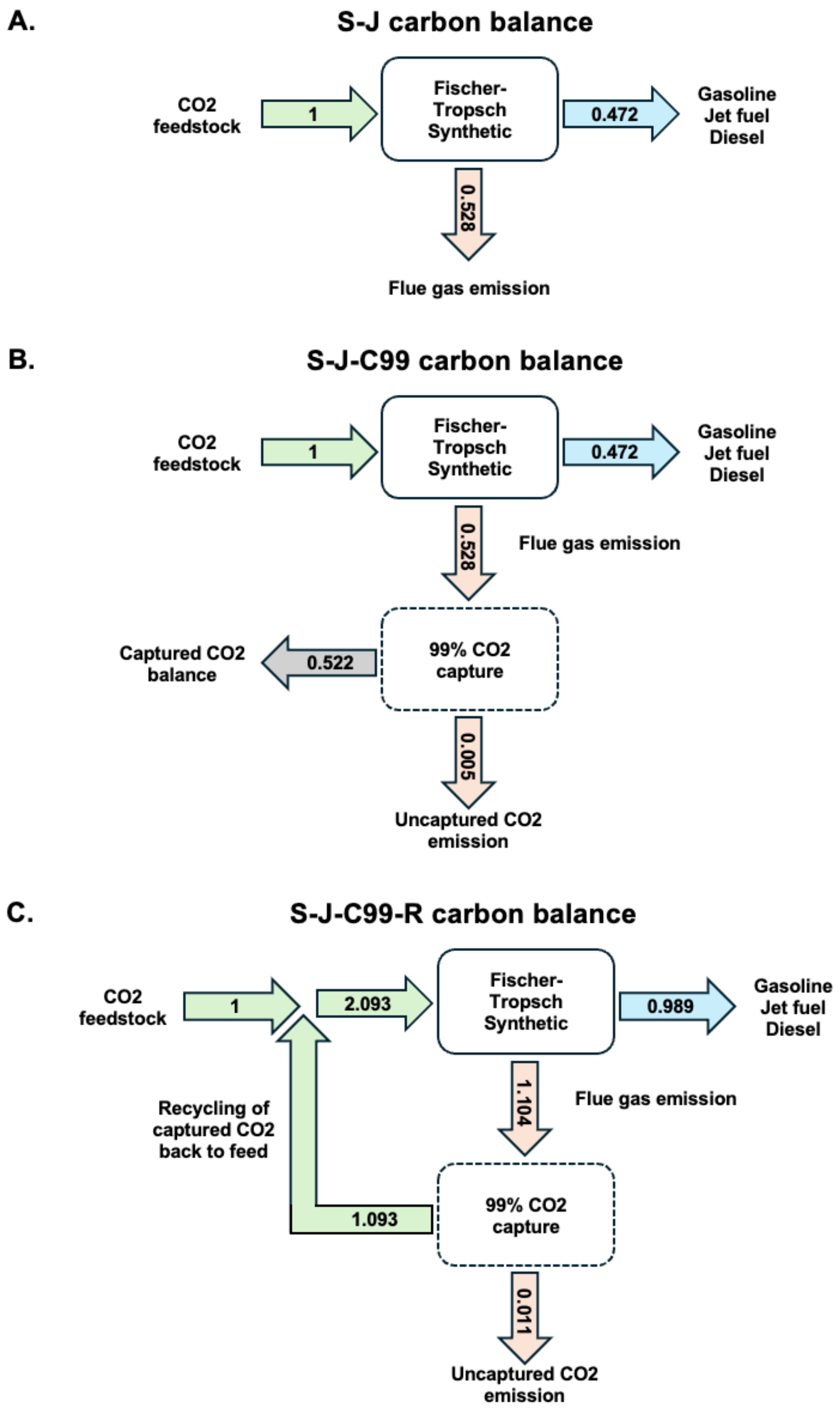

Figure S28. Carbon balance of S-J synthetic fuel production technologies in Table S33, with box diagrams showing simplified process and carbon flows (normalized to % wt. basis). S-J is based on the FT process in the reference TEA by Zang et al. for production of jet fuel (dominant product), gasoline and diesel (both side product) from $H_2$ and $CO_2$ feedstocks [32]. S-J-CC99 is characterized by using the approximation method described in Section S2.2 by matching the flue gas stream to an analogous industrial point source (represented by dashed box) from the original S-J configuration – The 99% $CO_2$ capture unit is based on a cement production $CO_2$ capture unit as shown in Table S11. The S-J-CC99-R configuration is characterized by assuming the recycling of captured $CO_2$ stream from S-J-CC99 into the input feedstock of the synthetic fuel production process, resulting in a high carbon conversion variant with additional $H_2$ and electricity inputs, and high fuels product output.

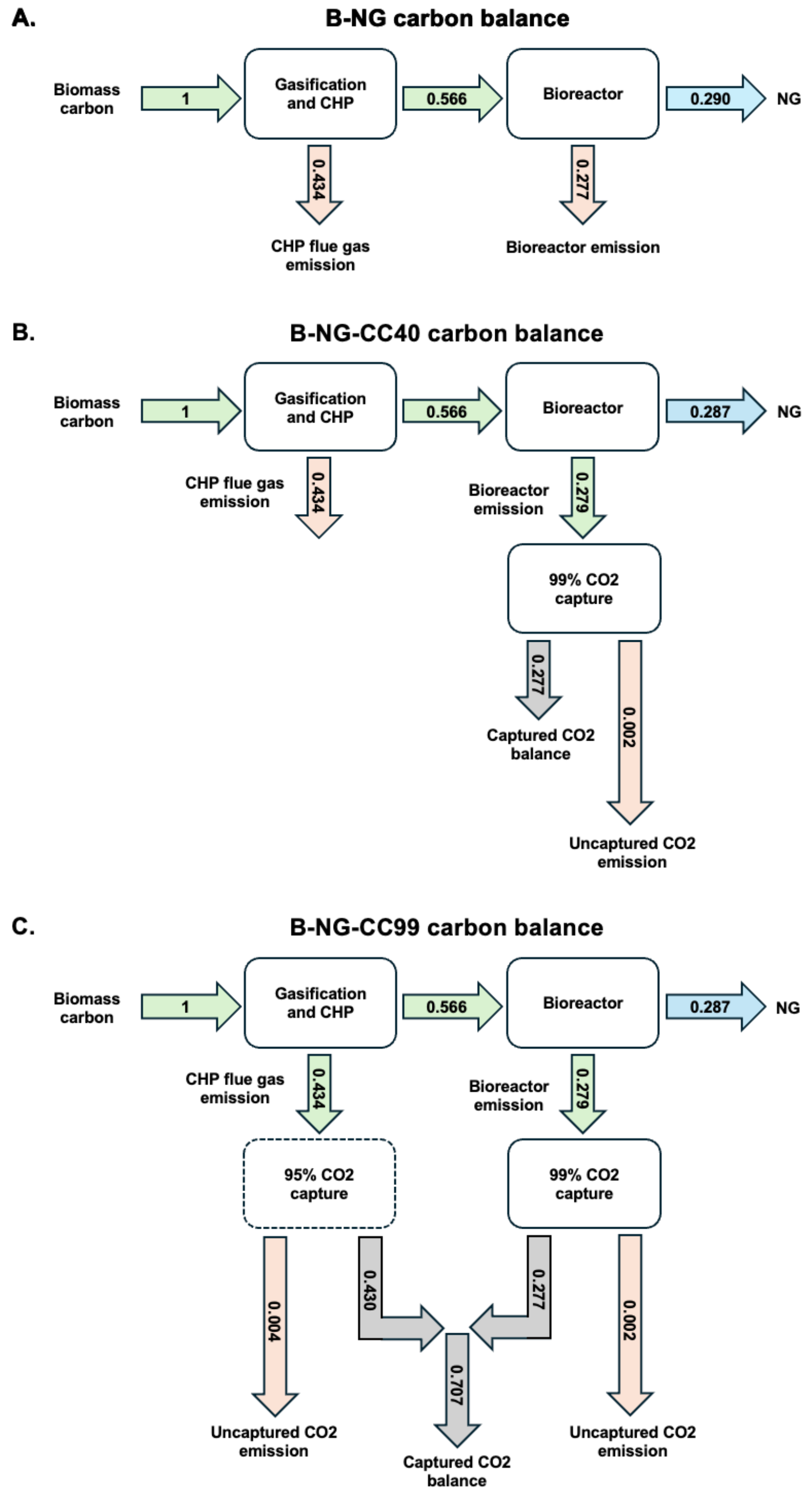

Figure S29. Carbon balance of B-NG production technologies in Table S34, with box diagrams showing simplified process and carbon flows (normalized to % wt. basis). B-NG and B-NG-CC40 is based on the process configurations with and without bioreactor flue gas $CO_2$ capture in the reference TEA by Michailos et al. for NG production from biomass gasification followed by fermentation [37]. B-NG-CC99 is characterized by using the approximation method described in Section S2.2 by matching the $CO_2$ concentrations of the CHP flue gas stream to an analogous industrial point source (represented by dashed box) from the original B-NG-CC40 configuration – The 95% $CO_2$ capture unit is based on a NGCC $CO_2$ capture unit as shown in Table S11. Biomass properties of 47.3% carbon content per biomass weight and 18.15 MMBtu/tonne energy content were used to obtain this carbon balance. to an NGCC $CO_2$ capture unit (represented by dashed boxes)

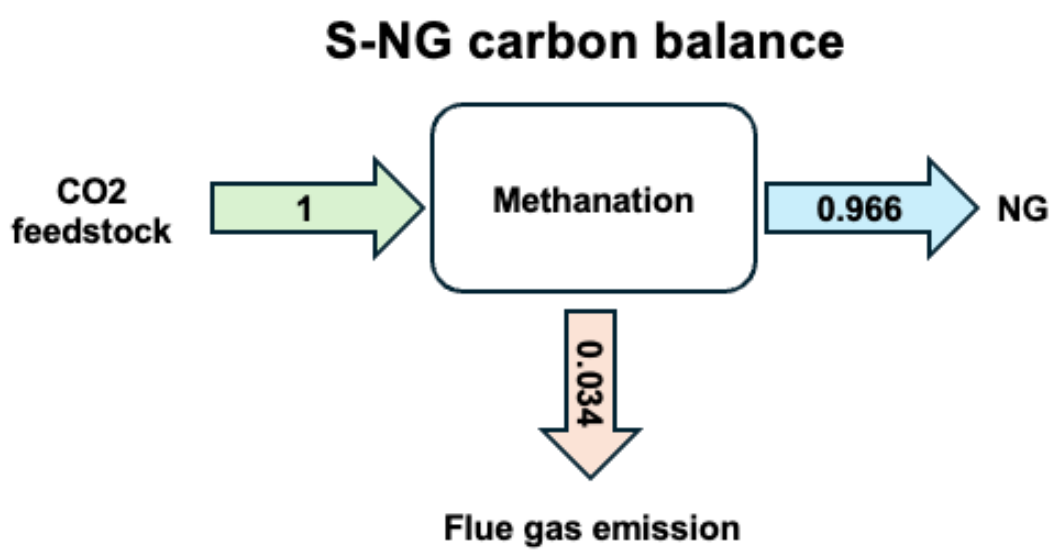

Figure S30. Carbon balance of S-NG production technology in Table S36 with box diagrams showing simplified process and carbon flows (normalized to % wt. basis). S-NG is based on the methanation process in the reference TEA by Kiani et al. for production NG from $H_2$ and $CO_2$ feedstocks [36]. Due to the high carbon conversion of 96.6%, $CO_2$ capture variants of S-NG were not necessary to be developed.

## S5. MACRO model

The documentations of the electricity, $H_2$, $CO_2$, and biomass sectors of the MACRO model is available in a public GitHub repository [22]. Existing bioenergy modeling in MACRO includes electricity generation and H2 production from biomass. The liquid fuels and natural gas supply chains add the modeling of liquid fuels (S5.1 and S5.2) and natural gas (S5.3) production from synthetic (electricity, $H_2$, and captured $CO_2$ as inputs) and biomass pathways, as well as the option of purchasing fossil fuels. S5.4 and S5.5 describes the changes made to the captured $CO_2$ balance constraint and $CO_2$ emission policy constraint in MACRO.

### S5.1 Liquid fuels supply chains modeling in MACRO

The liquid fuels supply chain in MACRO is modeled such that gasoline, jet fuel, and diesel can be produced by synthetic fuel and biofuel processes, as well as purchased fossil fuel options. Each synthetic and biofuel processes follow fixed product distribution as specified per input parameters, with the option of introducing user specified flexibility in product distribution as described in S5.2. Individual liquid fuels balance is enforced at the system level, assuming low cost of transportation of liquid fuels across regions.

### Notation

Table S41. Notation of various supply chains in MACRO.

| Notation | Description |
|---|---|
| $z \in Z$ | Zone z in set of all zones in the system |
| $t \in T$ | Time step t in set of all modeled time steps |
| $p \in P$ | Index and set of power generation resource p |
| $h \in H$ | Index and set of H2 production resource h |
| $d \in D$ | Index and set of DAC resource d |
| $s \in S$ | Index and set of synthetic fuel resources s |
| $b \in B$ | Index and set of bioenergy resources b |

### Parameters

Table S42. Input parameters of liquid fuels supply chain modeling in MACRO.

| **Liquid fuel demands** | |
|---|---|
| $d_t^G$ | Systemwide gasoline demand in MMBtu at time step t |
| $d_t^J$ | Systemwide jet fuel demand in MMBtu at time step t |
| $d_t^D$ | Systemwide diesel demand in MMBtu at time step t |
| **Synthetic fuel resources** | |
| $\eta_s^{H_2}$ | Tonne H2 per tonne $CO_2$ input to resource s |
| $\eta_s^{Elec}$ | MWh electricity per tonne $CO_2$ input to resource s |
| $\epsilon_s^{G,Syn}$ | MMBtu gasoline output per tonne $CO_2$ input from resource s |
| $\epsilon_s^{J,Syn}$ | MMBtu jet fuel output per tonne $CO_2$ input from resource s |
| $\epsilon_s^{D,Syn}$ | MMBtu diesel output per tonne $CO_2$ input from resource s |
| $\gamma_s$ | % capture of input $CO_2$ not converted to fuels in resource s |
| $\overline{\Omega_s}$ | Maximum operating factor of built capacity of resource s |
| $\underline{\Omega_s}$ | Minimum operating factor of built capacity of resource s |
| $c_s^{Inv}$ | Annualized investment cost of resource s ($/(tonne $CO_2$ input/h)/y) |
| $c_s^{FOM}$ | Annual fixed operation cost of resource s ($/(tonne $CO_2$ input/h)/y) |
| $c_s^{VOM}$ | Variable operation cost of resource s ($/tonne $CO_2$ input) |
| $\alpha^{G,Syn}$ | Emission factor of synthetic gasoline (tonne $CO_2$/MMBtu) |
| $\alpha^{J,Syn}$ | Emission factor of synthetic jet fuel (tonne $CO_2$/MMBtu) |
| $\alpha^{D,Syn}$ | Emission factor of synthetic diesel (tonne $CO_2$/MMBtu) |
| **Biofuel resources** | |
| $E_z^{Bio}$ | Energy content of biomass in zone z (MMBtu/tonne biomass) |
| $C_z^{Bio}$ | Carbon content of biomass in zone z (tonne biogenic $CO_2$/tonne biomass) |
| $\eta_b^{Elec}$ | MWh electricity per tonne biomass input to resource b |
| $\xi_b$ | % of biomass energy converted into biofuels |
| $f_b^{G,Bio}$ | % of gasoline in biofuel produced by resource b |
| $f_b^{J,Bio}$ | % of jet fuel in biofuel produced by resource b |
| $f_b^{D,Bio}$ | % of diesel in biofuel produced by resource b |
| $\gamma_b$ | % capture of biogenic CO2 not converted to fuels in resource b |
| $P_b$ | % of biomass energy converted into electricity by resource b |
| $\overline{\Omega_b}$ | Maximum operating factor of built capacity of resource b |
| $\underline{\Omega_b}$ | Minimum operating factor of built capacity of resource b |
| $c_b^{Inv}$ | Annualized investment cost of resource b ($/(tonne biomass/h)/y) |
| $c_b^{FOM}$ | Annual fixed operation cost of resource b ($/(tonne biomass/h)/y) |
| $c_b^{VOM}$ | Variable operation cost of resource b ($/tonne biomass) |
| $\alpha^{G,Bio}$ | Emission factor of bio gasoline (tonne $CO_2$/MMBtu) |
| $\alpha^{J,Bio}$ | Emission factor of bio jet fuel (tonne $CO_2$/MMBtu) |
| $\alpha^{D,Bio}$ | Emission factor of bio diesel (tonne $CO_2$/MMBtu) |
| **Fossil liquid fuels** | |
| $\min^{J:G,Fossil}$ | Min fraction of fossil jet fuel to fossil gasoline |
| $\max^{J:G,Fossil}$ | Max fraction of fossil jet fuel to fossil gasoline |
| $\min^{D:G,Fossil}$ | Min fraction of fossil diesel to fossil gasoline |
| $\max^{D:G,Fossil}$ | Max fraction of fossil diesel to fossil gasoline |
| $c^{G,Fossil}$ | Cost of purchasing fossil gasoline ($/MMBtu) |
| $c^{J,Fossil}$ | Cost of purchasing fossil jet fuel ($/MMBtu) |
| $c^{D,Fossil}$ | Cost of purchasing fossil diesel ($/MMBtu) |
| $\alpha^{G,Fossil}$ | Emission factor of fossil gasoline (tonne $CO_2$/MMBtu) |
| $\alpha^{J,Fossil}$ | Emission factor of fossil jet fuel (tonne $CO_2$/MMBtu) |
| $\alpha^{D,Fossil}$ | Emission factor of fossil diesel (tonne $CO_2$/MMBtu) |

## Decision Variables

Table S43. Decision variables in liquid fuels supply chain modeling in MACRO.

| | |
|---|---|
| **Synthetic fuel resources** | |
| $y_s^{CO_2,in}$ | Invested capacity of resource s (tonne $CO_2$ input/h) |
| $x_{s,t}^{CO_2,in}$ | Tonne $CO_2$ input to resource s in zone z in time step t |
| **Biofuel resources** | |
| $y_b^{Bio,in}$ | Invested capacity of b (tonne biomass input/h) |
| $x_{b,t}^{Bio,in}$ | Tonne biomass input to biofuel resource b in zone z in time step t |
| **Fossil fuel purchase** | |
| $\theta_t^{G,Fossil}$ | MMBtu fossil gasoline purchased in time step t |
| $\theta_t^{J,Fossil}$ | MMBtu fossil jet fuel purchased in time step t |
| $\theta_t^{D,Fossil}$ | MMBtu fossil diesel purchased in time step t |

## Expressions

Table S44. Model expressions in liquid fuels supply chain in MACRO.

| | |
|---|---|
| **Synthetic fuel resources** | |
| $\theta_{s,t}^{G,Syn}$ | MMBtu fossil gasoline produced in resource s in time step t |
| $\theta_{s,t}^{J,Syn}$ | MMBtu fossil jet fuel produced in resource s in time step t |
| $\theta_{s,t}^{D,Syn}$ | MMBtu fossil diesel produced in resource s in time step t |
| $\theta_{s,t}^{CO2,emi}$ | $CO_2$ emitted from resource s in time step t |
| $\theta_{s,t}^{CO2,capt}$ | $CO_2$ captured by resource s in time step t |
| **Biofuel resources** | |
| $\theta_{b,t}^{G,Bio}$ | MMBtu bio gasoline produced in resource b in time step t |
| $\theta_{b,t}^{J,Bio}$ | MMBtu bio jet fuel produced in resource b in time step t |
| $\theta_{b,t}^{D,Bio}$ | MMBtu bio diesel produced in resource b in time step t |
| $\theta_{b,t}^{CO2,emi}$ | $CO_2$ emitted from resource b in time step t |
| $\theta_{b,t}^{CO2,capt}$ | $CO_2$ captured by resource b in time step t |

The following expressions represent the amount of individual synthetic and biofuel types produced by respective resources.

$$\theta_{s,t}^{G,Syn} = x_{s,t}^{CO_2,in} \cdot \epsilon_s^{G,Syn} \quad \text{(Eq. S5)}$$

$$\theta_{s,t}^{J,Syn} = x_{s,t}^{CO_2,in} \cdot \epsilon_s^{J,Syn} \quad \text{(Eq. S6)}$$

$$\theta_{s,t}^{D,Syn} = x_{s,t}^{CO_2,in} \cdot \epsilon_s^{D,Syn} \quad \text{(Eq. S7)}$$

$$\theta_{b,t}^{G,Bio} = E_z^{Bio} \cdot x_{b\in z,t}^{Bio,in} \cdot \xi_b \cdot f_b^{G,Bio} \quad \text{(Eq. S8)}$$

$$\theta_{b,t}^{J,Bio} = E_z^{Bio} \cdot x_{b\in z,t}^{Bio,in} \cdot \xi_b \cdot f_b^{J,Bio} \quad \text{(Eq. S9)}$$

$$\theta_{b,t}^{D,Bio} = E_z^{Bio} \cdot x_{b\in z,t}^{Bio,in} \cdot \xi_b \cdot f_b^{D,Bio} \quad \text{(Eq. S10)}$$

The following expressions represent the $CO_2$ emissions and $CO_2$ capture by synthetic and biofuel resources, according to the carbon capture rates of carbon not converted to fuels.

$$\theta_{s,t}^{CO2,emi} = \left(x_{s,t}^{CO2,in} - \theta_{s,t}^{G,Syn} \cdot \alpha^{G,Syn} - \theta_{s,t}^{J,Syn} \cdot \alpha^{J,Syn} - \theta_{s,t}^{D,Syn} \cdot \alpha^{D,Syn}\right) \cdot (1 - \gamma_s) \quad \text{(Eq. S11)}$$

$$\theta_{s,t}^{CO2,capt} = \left(x_{s,t}^{CO2,in} - \theta_{s,t}^{G,Syn} \cdot \alpha^{G,Syn} - \theta_{s,t}^{J,Syn} \cdot \alpha^{J,Syn} - \theta_{s,t}^{D,Syn} \cdot \alpha^{D,Syn}\right) \cdot \gamma_s \quad \text{(Eq. S12)}$$

$$\theta_{b,t}^{CO2,emi} = \left(x_{b\in z,t}^{Bio,in} \cdot C_z^{Bio} - \theta_{b,t}^{G,Bio} \cdot \alpha^{G,Bio} - \theta_{b,t}^{J,Bio} \cdot \alpha^{J,Bio} - \theta_{b,t}^{D,Bio} \cdot \alpha^{D,Bio}\right) \cdot (1 - \gamma_b) \quad \text{(Eq. S13)}$$

$$\theta_{b,t}^{CO2,capt} = \left(x_{b\in z,t}^{Bio,in} \cdot C_z^{Bio} - \theta_{b,t}^{G,Bio} \cdot \alpha^{G,Bio} - \theta_{b,t}^{J,Bio} \cdot \alpha^{J,Bio} - \theta_{b,t}^{D,Bio} \cdot \alpha^{D,Bio}\right) \cdot \gamma_b \quad \text{(Eq. S14)}$$

**Constraints**

The liquid fuels balance constraints are formulated with a “greater or equal” than demand, as having fixed product distribution in synthetic and biofuel processes might result in cases with mismatch in product-demand distribution, thus causing excess production of individual fuel types. In addition, systemwide demand for each time step is implemented to reflect the interconnectedness of the liquid fuels supply chain, with assumption of low cost of transportation of liquid fuels across regions.

**Gasoline balance**

$$\sum_{z\in Z} \theta_{z,t}^{G,Fossil} + \sum_{z\in Z} \sum_{s\in z} \theta_{s,t}^{G,Syn} + \sum_{z\in Z} \sum_{b\in z} \theta_{b,t}^{G,Bio} \geq d_t^G \quad \text{(Eq. S15)}$$

**Jet fuel balance**

$$\sum_{z\in Z} \theta_{z,t}^{J,Fossil} + \sum_{z\in Z} \sum_{s\in z} \theta_{s,t}^{J,Syn} + \sum_{z\in Z} \sum_{b\in z} \theta_{b,t}^{J,Bio} \geq d_t^J \quad \text{(Eq. S16)}$$

**Diesel balance**

$$\sum_{z\in Z} \theta_{z,t}^{D,Fossil} + \sum_{z\in Z} \sum_{s\in z} \theta_{s,t}^{D,Syn} + \sum_{z\in Z} \sum_{b\in z} \theta_{b,t}^{D,Bio} \geq d_t^D \quad \text{(Eq. S17)}$$

**Fossil fuels ratio bounds**

Enforcing constraints on ratio bounds between fossil fuels allow users to represent a realistic distribution of fossil fuels produced from oil refineries, instead of allowing the model to purchase fossil liquid fuels in any distribution ratios (i.e. all fossil liquids purchased in the form of jet fuel if fossil jet fuel is the cheapest among all fossil fuels). This reflects the realistic operation of oil refineries where there is a range of distribution of various fossil fuel types. In this study, historical distribution bounds were used based on data from EIA[31] (Refer to Table S32).

$$\min^{J:G,Fossil} \cdot \theta_{z,t}^{G,Fossil} \leq \theta_{z,t}^{J,Fossil} \leq \max^{J:G,Fossil} \cdot \theta_{z,t}^{G,Fossil} \quad \text{(Eq. S18)}$$

$$\min^{D:G,Fossil} \cdot \theta_{z,t}^{G,Fossil} \leq \theta_{z,t}^{D,Fossil} \leq \max^{D:G,Fossil} \cdot \theta_{z,t}^{G,Fossil} \quad \text{(Eq. S19)}$$

**Operation constraints**

The hourly operation of each synthetic and biofuel resources is constrained to maximum and minimum operating factors of their built capacities.

$$y_s^{CO_2,in} \cdot \underline{\Omega_s} \leq x_{s,t}^{CO_2,in} \leq y_s^{CO_2,in} \cdot \overline{\Omega_s} \quad \text{(Eq. S20)}$$

$$y_b^{Bio,in} \cdot \underline{\Omega_b} \leq x_{b,t}^{Bio,in} \leq y_b^{Bio,in} \cdot \overline{\Omega_b} \quad \text{(Eq. S21)}$$

**Contribution to power balance**

The following terms are the total electricity consumption of synthetic and biofuel resources, and excess electricity produced by biofuel resources (if applicable) of each zone z, which are accounted to their respective regional power balance.

$$\text{Power Balance}_z -= \sum_{s\in z} x_{s,t}^{CO_2,in} \cdot \eta_s^{Elec} + \sum_{b\in z} x_{b,t}^{Bio,in} \cdot \eta_b^{Elec} - \sum_{b\in z} x_{b,t}^{Bio,in} \cdot E_z^{Bio} \cdot P_b \quad \text{(Eq. S22)}$$

**Contribution to H2 balance**

The following terms are the total H2 consumption of each zone, which are deducted from respective regional H2 balance.

$$H_2 \text{ Balance}_z -= \sum_{s\in z} x_{s,t}^{CO_2,in} \cdot \eta_s^{H_2} \quad \text{(Eq. S23)}$$

**Addition to objective function**

The following terms representing the total cost (investment, fixed operating costs, and variable operating costs of synthetic and biofuel resources, as well as purchase costs of fossil liquid fuels) of the liquid fuels supply chain are added to the overall objective function (Obj). $\omega_t$ is the time weight of each modeled time step that sums up to 8760 hours when representative periods are used in the model instead of modeling an entire full year for computational tractability.

$$\begin{aligned} \text{Obj} += & \sum_{z\in Z}\sum_{s\in z} y_s^{CO_2,in} \cdot (c_s^{Inv} + c_s^{FOM}) + \sum_{z\in Z}\sum_{b\in z} y_b^{Bio,in} \cdot (c_b^{Inv} + c_b^{FOM}) + \sum_{z\in Z}\sum_{t\in T}\sum_{s\in z} \omega_t \cdot x_{s,t}^{CO_2,in} \cdot c_s^{VOM} \\ & + \sum_{z\in Z}\sum_{t\in T}\sum_{b\in z} \omega_t \cdot x_{b,t}^{Bio,in} \cdot c_b^{VOM} + \sum_{t\in T} c^{G,Fossil}\theta_t^{G,Fossil} + \sum_{t\in T} c^{J,Fossil}\theta_t^{J,Fossil} \\ & + \sum_{t\in T} c^{D,Fossil}\theta_t^{D,Fossil} \quad \text{(Eq. S24)} \end{aligned}$$

## S5.2 Modeling flexibility in synthetic and biofuels production in MACRO

MACRO allows the choice of operating synthetic and biofuel resources with flexibility in fuel product distribution, instead of following strictly the specified product distribution of processes. This is done by producing a certain percentage of originally produced liquid fuel type (gasoline, jet fuel, diesel) as another type, with the assumption that all costs and energy requirement remains the same. This gives the model flexibility to produce individual synthetic and biofuel types according to the optimal distribution of gasoline, jet fuel and diesel instead of adhering to specified fixed product distributions.

These parameters and variables are only activated when synthetic and biofuels production are chosen by the user to operate flexibly.

**Parameters**

Table S45. Variables for flexible synthetic and biofuels production in MACRO.

| | |
|---|---|
| **Synthetic fuel resources** | |
| $\max^{G\to J,Syn}$ | Max % of original gasoline produced as jet fuel in resource s in time step t |
| $\max^{G\to D,Syn}$ | Max % of original gasoline produced as diesel in resource s in time step t |
| $\max^{J\to G,Syn}$ | Max % of original jet fuel produced as gasoline in resource s in time step t |
| $\max^{J\to D,Syn}$ | Max % of original jet fuel produced as diesel in resource s in time step t |
| $\max^{D\to G,Syn}$ | Max % of original diesel produced as gasoline in s in time step t |
| $\max^{D\to J,Syn}$ | Max % of original diesel produced as jet fuel in s in time step t |
| **Biofuel resources** | |
| $\max^{G\to J,Bio}$ | Max % of original gasoline produced as jet fuel in resource b in time step t |
| $\max^{G\to D,Bio}$ | Max % of original gasoline produced as diesel in resource b in time step t |
| $\max^{J\to G,Bio}$ | Max % of original jet fuel produced as gasoline in resource b in time step t |
| $\max^{J\to D,Bio}$ | Max % of original jet fuel produced as diesel in resource b in time step t |
| $\max^{D\to G,Bio}$ | Max % of original diesel produced as gasoline in resource b in time step t |
| $\max^{D\to J,Bio}$ | Max % of original diesel produced as jet fuel in resource b in time step t |

**Variables**

Table S46. Parameters for flexible synthetic and biofuels production in MACRO.

| | |
|---|---|
| **Synthetic fuel resources** | |
| $\theta_{s,t}^{G\to J,Syn}$ | MMBtu original gasoline produced as jet fuel in resource s in time stp t |
| $\theta_{s,t}^{G\to D,Syn}$ | MMBtu original gasoline produced as diesel in resource s in time step t |
| $\theta_{s,t}^{J\to G,Syn}$ | MMBtu original jet fuel produced as gasoline in resource s in time step t |
| $\theta_{s,t}^{J\to D,Syn}$ | MMBtu original jet fuel produced as diesel in resource s in time step t |
| $\theta_{s,t}^{D\to G,Syn}$ | MMBtu original diesel produced as gasoline in resource s in time step t |
| $\theta_{s,t}^{D\to J,Syn}$ | MMBtu original diesel produced as jet fuel in resource s in time step t |
| **Biofuel resources** | |
| $\theta_{b,t}^{G\to J,Bio}$ | MMBtu original gasoline produced as jet fuel in resource b in time step t |
| $\theta_{b,t}^{G\to D,Bio}$ | MMBtu original gasoline produced as diesel in resource b in time step t |
| $\theta_{b,t}^{J\to G,Bio}$ | MMBtu original jet fuel produced as gasoline in resource b in time step t |
| $\theta_{b,t}^{J\to D,Bio}$ | MMBtu original jet fuel produced as diesel in resource b in time step t |
| $\theta_{b,t}^{D\to G,Bio}$ | MMBtu original diesel produced as gasoline in resource b in time step t |
| $\theta_{b,t}^{D\to J,Bio}$ | MMBtu original diesel produced as jet fuel in resource b in time step t |

### Expressions

The expressions representing the amount of individual synthetic and biofuel types produced by respective resources are thus modified to account for flexibility in liquid fuels production.

$$\theta_{s,t}^{G,Syn} = x_{s,t}^{CO_2,in} \cdot \epsilon_s^{G,Syn} - \theta_{s,t}^{G\rightarrow J,Syn} - \theta_{s,t}^{G\rightarrow D,Syn} + \theta_{s,t}^{J\rightarrow G,Syn} + \theta_{s,t}^{D\rightarrow G,Syn} \quad \text{(Eq. S25)}$$

$$\theta_{s,t}^{J,Syn} = x_{s,t}^{CO_2,in} \cdot \epsilon_s^{J,Syn} - \theta_{s,t}^{J\rightarrow G,Syn} - \theta_{s,t}^{J\rightarrow D,Syn} + \theta_{s,t}^{G\rightarrow J,Syn} + \theta_{s,t}^{D\rightarrow J,Syn} \quad \text{(Eq. S26)}$$

$$\theta_{s,t}^{D,Syn} = x_{s,t}^{CO_2,in} \cdot \epsilon_s^{D,Syn} - \theta_{s,t}^{D\rightarrow G,Syn} - \theta_{s,t}^{D\rightarrow J,Syn} + \theta_{s,t}^{G\rightarrow D,Syn} + \theta_{s,t}^{J\rightarrow D,Syn} \quad \text{(Eq. S27)}$$

$$\theta_{b,t}^{G,Bio} = E_z^{Bio} \cdot x_{b,t}^{Bio,in} \cdot \xi_b \cdot f_b^{G,Bio} - \theta_{b,t}^{G\rightarrow J,Bio} - \theta_{b,t}^{G\rightarrow D,Bio} + \theta_{b,t}^{J\rightarrow G,Bio} + \theta_{b,t}^{D\rightarrow G,Bio} \quad \text{(Eq. S28)}$$

$$\theta_{b,t}^{J,Bio} = E_z^{Bio} \cdot x_{b,t}^{Bio,in} \cdot \xi_b \cdot f_b^{J,Bio} - \theta_{b,t}^{J\rightarrow G,Bio} - \theta_{b,t}^{J\rightarrow D,Bio} + \theta_{b,t}^{G\rightarrow J,Bio} + \theta_{b,t}^{D\rightarrow J,Bio} \quad \text{(Eq. S29)}$$

$$\theta_{b,t}^{D,Bio} = E_z^{Bio} \cdot x_{b,t}^{Bio,in} \cdot \xi_b \cdot f_b^{D,Bio} - \theta_{b,t}^{D\rightarrow G,Bio} - \theta_{b,t}^{D\rightarrow J,Bio} + \theta_{b,t}^{G\rightarrow D,Bio} + \theta_{b,t}^{J\rightarrow D,Bio} \quad \text{(Eq. S30)}$$

### Constraints

These constraints control the flexibility in liquid fuel production by synthetic or biofuel technologies according to specified maximum fraction of originally produced fuel types.

$$\theta_{s,t}^{G\rightarrow J,Syn} \leq \max^{G\rightarrow J,Syn} \cdot x_{s,t}^{CO_2,in} \cdot \epsilon_s^{G,Syn} \quad \text{(Eq. S31)}$$

$$\theta_{s,t}^{G\rightarrow D,Syn} \leq \max^{G\rightarrow D,Syn} \cdot x_{s,t}^{CO_2,in} \cdot \epsilon_s^{G,Syn} \quad \text{(Eq. S32)}$$

$$\theta_{s,t}^{J\rightarrow G,Syn} \leq \max^{J\rightarrow G,Syn} \cdot x_{s,t}^{CO_2,in} \cdot \epsilon_s^{J,Syn} \quad \text{(Eq. S33)}$$

$$\theta_{s,t}^{J\rightarrow D,Syn} \leq \max^{J\rightarrow D,Syn} \cdot x_{s,t}^{CO_2,in} \cdot \epsilon_s^{J,Syn} \quad \text{(Eq. S34)}$$

$$\theta_{s,t}^{D\rightarrow G,Syn} \leq \max^{D\rightarrow G,Syn} \cdot x_{s,t}^{CO_2,in} \cdot \epsilon_s^{D,Syn} \quad \text{(Eq. S35)}$$

$$\theta_{s,t}^{D\rightarrow J,Syn} \leq \max^{D\rightarrow J,Syn} \cdot x_{s,t}^{CO_2,in} \cdot \epsilon_s^{D,Syn} \quad \text{(Eq. S36)}$$

$$\theta_{b,t}^{G\rightarrow J,Bio} \leq \max^{G\rightarrow J,Bio} \cdot E_z^{Bio} \cdot x_{b,t}^{Bio,in} \cdot \xi_b \cdot f_b^{G,Bio} \quad \text{(Eq. S37)}$$

$$\theta_{b,t}^{G\rightarrow D,Bio} \leq \max^{G\rightarrow D,Bio} \cdot E_z^{Bio} \cdot x_{b,t}^{Bio,in} \cdot \xi_b \cdot f_b^{G,Bio} \quad \text{(Eq. S38)}$$

$$\theta_{b,t}^{J\rightarrow G,Bio} \leq \max^{J\rightarrow G,Bio} \cdot E_z^{Bio} \cdot x_{b,t}^{Bio,in} \cdot \xi_b \cdot f_b^{J,Bio} \quad \text{(Eq. S39)}$$

$$\theta_{b,t}^{J\rightarrow D,Bio} \leq \max^{J\rightarrow D,Bio} \cdot E_z^{Bio} \cdot x_{b,t}^{Bio,in} \cdot \xi_b \cdot f_b^{J,Bio} \quad \text{(Eq. S40)}$$

$$\theta_{b,t}^{D\rightarrow G,Bio} \leq \max^{D\rightarrow G,Bio} \cdot E_z^{Bio} \cdot x_{b,t}^{Bio,in} \cdot \xi_b \cdot f_b^{D,Bio} \quad \text{(Eq. S41)}$$

$$\theta_{b,t}^{D\rightarrow J,Bio} \leq \max^{D\rightarrow J,Bio} \cdot E_z^{Bio} \cdot x_{b,t}^{Bio,in} \cdot \xi_b \cdot f_b^{D,Bio} \quad \text{(Eq. S42)}$$

The following constraints control the flexibility in liquid fuel production such that it does not exceed total amount of original production of liquid fuel types.

$$\theta_{s,t}^{G\rightarrow J,Syn} + \theta_{s,t}^{G\rightarrow D,Syn} \leq x_{s,t}^{CO_2,in} \cdot \epsilon_s^{G,Syn} \quad \text{(Eq. S43)}$$

$$\theta_{s,t}^{J\to G,Syn} + \theta_{s,t}^{J\to D,Syn} \le x_{s,t}^{CO_2,in} \cdot \epsilon_s^{J,Syn} \quad \text{(Eq. S44)}$$

$$\theta_{s,t}^{D\to G,Syn} + \theta_{s,t}^{D\to J,Syn} \le x_{s,t}^{CO_2,in} \cdot \epsilon_s^{D,Syn} \quad \text{(Eq. S45)}$$

$$\theta_{b,t}^{G\to J,Bio} + \theta_{b,t}^{G\to D,Bio} \le E_z^{Bio} \cdot x_{b,t}^{Bio,in} \cdot \xi_b \cdot f_b^{G,Bio} \quad \text{(Eq. S46)}$$

$$\theta_{b,t}^{J\to G,Bio} + \theta_{b,t}^{J\to D,Bio} \le E_z^{Bio} \cdot x_{b,t}^{Bio,in} \cdot \xi_b \cdot f_b^{J,Bio} \quad \text{(Eq. S47)}$$

$$\theta_{b,t}^{D\to G,Bio} + \theta_{b,t}^{D\to J,Bio} \le E_z^{Bio} \cdot x_{b,t}^{Bio,in} \cdot \xi_b \cdot f_b^{D,Bio} \quad \text{(Eq. S48)}$$

## S5.3 Natural gas supply chains modeling in MACRO

The natural supply chain in MACRO is modeled in a similar way to the liquid fuels supply chain in S5.1 such that natural gas (NG) can be produced by synthetic and bio processes, as well as purchased fossil natural gas options. One key difference between synthetic and bio-NG technologies compared to their liquid fuels counterparts is that only one product of NG is produced. Similarly, regional electricity and $H_2$ balance are modified to account for the consumption of these commodities in the natural gas supply chain.

### Constraints

Apart from the NG balance constraint, the operational constraints are similar to that of synthetic and bio liquid fuels production, and NG production does not have any product flexibility since there is only one product.

### NG balance

The NG balance constraints are formulated as regional demand for each time step, as the prices of fossil NG in each region already include any NG transportation costs from sources to respective regions. The NG balance consists of purchased fossil NG, synthetic and bio-NG produced by respective technologies, NG consumed by the power and $H_2$ sectors, and NG utilized in direct air capture (DAC) as a fuel, that must be equal to the NG demand in each region.

$$\theta_{z,t}^{NG,Fossil} + \sum_{s\in z} \theta_{s,t}^{NG,Syn} + \sum_{b\in z} \theta_{b,t}^{NG,Bio} - \sum_{p\in P} \theta_{p,t}^{NG,Power} - \sum_{h\in H} \theta_{h,t}^{NG,H_2} - \sum_{d\in D} \theta_{d,t}^{NG,DAC} = d_{z,t}^{NG} \quad \text{(Eq. S49)}$$

## S5.4 Captured $CO_2$ balance constraint in MACRO

Captured $CO_2$ terms from the liquid fuels and natural gas supply chains are added to the existing captured $CO_2$ balance constraint in MACRO.

### Expressions

Table S47. Expressions for existing sectors in $CO_2$ balance constraint in MACRO.

| **Captured $CO_2$ balance** | |
|---|---|
| $\theta_{d,t}^{CO_2,DAC,atm}$ | Tonne atmospheric $CO_2$ captured by DAC from resource d in time step t |
| $\theta_{d,t}^{CO_2,DAC\ fuel,capt}$ | Tonne $CO_2$ captured from DAC fuel combustion from resource d in time step t |
| $\theta_{p,t}^{CO_2,capt}$ | Tonne $CO_2$ captured by power generation resource p in time step t |
| $\theta_{h,t}^{CO_2,capt}$ | Tonne $CO_2$ captured by $H_2$ generation resource p in time step t |
| $\theta_{z,t}^{CO_2,export}$ | Tonne $CO_2$ exported by pipelines from zone z in time step t |
| $\theta_{z,t}^{CO_2,seq}$ | Tonne $CO_2$ sequestered in zone z in time step t |

The captured $CO_2$ balance accounts for the capture, utilization, export, and sequestration of all captured $CO_2$ by technologies with $CO_2$ capture in each region. These technologies include NG-based power and $H_2$ generation with $CO_2$ capture, synthetic and biofuel technologies, as well as atmospheric $CO_2$ and capture of combustion emissions from fuel utilization in DAC technologies.

$$\sum_{h\in z}\theta_{h,t}^{CO_2,capt}+\sum_{p\in z}\theta_{p,t}^{CO_2,capt}+\sum_{s\in z}\theta_{s,t}^{CO_2,capt}+\sum_{b\in z}\theta_{b,t}^{CO_2,capt}+\sum_{d\in z}\left(\theta_{d,t}^{CO_2,DAC,atm}+\theta_{d,t}^{CO_2,DAC\ fuel,capt}\right)$$
$$-\sum_{s\in z}x_{s,t}^{CO_2,in}-\theta_{z,t}^{CO_2,export}=\theta_{z,t}^{CO_2,seq} \qquad (Eq.\,S50)$$

## S5.5 $CO_2$ emission policy constraint

$CO_2$ emission terms from the liquid fuels and natural gas supply chains have to be added to the existing $CO_2$ emission policy constraint in MACRO. This constraint accounts for emissions from all sectors across all regions and time steps, including emissions from fossil fuel power and $H_2$ production, uncaptured $CO_2$ from DAC fuel combustion, all liquid and gaseous fuels combustion, as well as negative emissions from biogenic $CO_2$ intake and DAC atmospheric capture, which have to be lesser than the system $CO_2$ emission cap.

### Expressions

Table S48. Expressions for $CO_2$ emission policy constraint in MACRO.

| **$CO_2$ emissions** | |
|---|---|
| $\theta_{p,t}^{CO_2,emi}$ | Tonne $CO_2$ emitted by power generation resource p in time step t |
| $\theta_{h,t}^{CO_2,emi}$ | Tonne $CO_2$ emitted by $H_2$ generation resource p in time step t |
| $\theta_{d,t}^{CO_2,DAC\ fuel,emi}$ | Tonne $CO_2$ emitted from DAC fuel combustion from resource d in time step t |

$$\omega\cdot\sum_{t\in T}\Bigg(\sum_{p\in P}\theta_{p,t}^{CO_2,emi}+\sum_{h\in H}\theta_{h,t}^{CO_2,emi}+\sum_{s\in S}\theta_{s,t}^{CO_2,emi}+\sum_{b\in B}\theta_{b,t}^{CO_2,emi}+\sum_{d\in D}\theta_{d,t}^{CO_2,DAC\ fuel,emi}+\theta_{t}^{G,Fossil}$$
$$\cdot\ \alpha^{G,Fossil}+\theta_{t}^{J,Fossil}\cdot\ \alpha^{J,Fossil}+\theta_{t}^{D,Fossil}\cdot\ \alpha^{D,Fossil}+\theta_{t}^{NG,Fossil}\cdot\ \alpha^{NG,Fossil}$$
$$+\sum_{s\in S}\left(\theta_{s,t}^{G,Syn}\cdot\ \alpha^{G,Syn}+\theta_{s,t}^{J,Syn}\cdot\ \alpha^{J,Syn}+\theta_{s,t}^{D,Syn}\cdot\ \alpha^{D,Syn}+\theta_{s,t}^{NG,Syn}\cdot\ \alpha^{NG,Syn}\right)$$
$$+\sum_{b\in B}\left(\theta_{b,t}^{G,Bio}\cdot\ \alpha^{G,Bio}+\theta_{b,t}^{J,Bio}\cdot\ \alpha^{J,Bio}+\theta_{b,t}^{D,Bio}\cdot\ \alpha^{D,Bio}+\theta_{b,t}^{NG,Bio}\cdot\ \alpha^{NG,Bio}\right)$$
$$-\sum_{d\in D}\theta_{d,t}^{CO_2,DAC,atm}-\sum_{b\in B}\sum_{b\in z}C_{z}^{Bio}x_{b,t}^{Bio,in}\Bigg)\leq CO_2\ Cap \qquad (Eq.\,S51)$$

## S6. Model implementation details

The MACRO model used in this study is formulated as a linear program (LP). Across the case studies, each instance has a problem size of approximately 2,300,000 constraints and 1,400,000 continuous decision variables for scenarios with “IA-PF” and “IA-FF” modeling assumptions, and approximately 3,350,000 constraints and 2,100,000 continuous decision variables for “FA-PF” and “FA-FF” modeling assumptions. The implementation uses Julia 1.10 and JuMP 1.20, and we solve the instances with Gurobi 10.0 on 8 CPU cores on MIT SuperCloud [43]. Typical runtimes per instance are 78–93 minutes for scenarios with “IA-PF” and “IA-FF” modeling assumptions, and 124–286 minutes for “IA-PF” and “IA-FF” modeling assumptions. The codebase for this analysis is available in the MACRO GitHub repository on the “MACRO-Liquid-Fuels-Net-Zero” branch [22].